\documentclass{article}

\usepackage[utf8]{inputenc}
\usepackage{amsmath, amsthm, amssymb, amsfonts, dsfont, mathrsfs, dirtytalk, tikz-cd, adjustbox, url, nicefrac, booktabs,array, graphicx,todonotes,longtable, tabu,bm,cite, todonotes,comment,multirow,tabularx,caption}
\usepackage[T1]{fontenc}    
\usepackage[margin=1in]{geometry}
\usepackage[english]{babel}
\usepackage[pagebackref=true]{hyperref}
\usepackage[capitalise]{cleveref}
\renewcommand*{\backref}[1]{}
\renewcommand*{\backrefalt}[4]{%
    \ifcase #1%
          \or (p.~#2.)%
          \else (p.~#2.)%
    \fi%
    }

\newcommand{\e}{\varepsilon}

\newcommand{\R}{\mathbb R}
\newcommand{\st}{\text{ s.t. }}

\newcommand{\p}{\mathbf{ P }}
\newcommand{\supp}{\operatorname{supp }}

\newcommand{\N}{\mathbb N}

\newcommand{\E}{\mathbb E}

\newcommand{\loc}{_{\mathrm{loc}}}
\newcommand{\rev}{_{\mathrm{re v}}}
\newcommand{\irr}{_{\mathrm{ir r}}}
\newcommand{\n}{^{(n)}}

\renewcommand{\H}{\operatorname{H }}

\renewcommand{\L}{\operatorname{L }}
\newcommand{\A}{\operatorname{A }}
\renewcommand{\S}{\operatorname{S }}

\renewcommand{\det}{\operatorname{d e t}}

\newcommand{\rank}{\operatorname{ran k}}
\newcommand{\tr}{\operatorname{T r}}
\newcommand{\dom}{\operatorname{D om}}

\newcommand{\im}{\operatorname{Rang e}}
\newcommand{\intr}{\int_{\mathbb R^d}}
\newcommand{\id}{\operatorname{I d}}

\renewcommand{\d}{\mathrm{ d }}
\newcommand{\dx}{\:\mathrm{ d }x}

\theoremstyle{plain}
\newtheorem{theorem}{Theorem}[section]
\newtheorem{lemma}[theorem]{Lemma}
\newtheorem{proposition}[theorem]{Proposition}
\newtheorem{corollary}[theorem]{Corollary}

\theoremstyle{definition}
\newtheorem{definition}[theorem]{Definition}

\newtheorem{example}[theorem]{Example}

\theoremstyle{remark}
\newtheorem{remark}[theorem]{Remark}

\newtheorem{assumption}[theorem]{Assumption}
\newtheorem{notation}[theorem]{Notation}

\newenvironment{claim}[1]{\par\noindent\emph{Claim:}\space#1}{}
\newenvironment{proofclaim}[1]{\par\noindent\textit{Proof of claim.}\space#1}{\hfill $\blacksquare$}
\newenvironment{sketchproof}[1]{\par\noindent\textit{Sketch proof.}\space#1}{\hfill $\square$}

\title{The entropy production of stationary diffusions}
\author{Lancelot Da Costa$^{1,2,}\thanks{Correspondence: \url{l.da-costa@imperial.ac.uk}}\:$ \& Grigorios A. Pavliotis$^{1}$\\\\
$^1$\textit{Department of Mathematics, Imperial College London} \\
$^2$\textit{Wellcome Centre for Human Neuroimaging, University College London}
}

\begin{document}

\maketitle

\begin{abstract}
The entropy production rate is a central quantity in non-equilibrium statistical physics, scoring how far a stochastic process is from being time-reversible. In this paper, we compute the entropy production of diffusion processes at non-equilibrium steady-state under the condition that the time-reversal of the diffusion remains a diffusion. We start by characterising the entropy production of both discrete and continuous-time Markov processes. We investigate the time-reversal of time-homogeneous stationary diffusions and recall the most general conditions for the reversibility of the diffusion property, which includes hypoelliptic and degenerate diffusions, and locally Lipschitz vector fields. We decompose the drift into its time-reversible and irreversible parts, or equivalently, the generator into symmetric and antisymmetric operators. We show the equivalence with a decomposition of the backward Kolmogorov equation considered in hypocoercivity theory, and a decomposition of the Fokker-Planck equation in GENERIC form. The main result shows that when the time-irreversible part of the drift is in the range of the volatility matrix (almost everywhere) the forward and time-reversed path space measures of the process are mutually equivalent, and evaluates the entropy production. When this does not hold, the measures are mutually singular and the entropy production is infinite. We verify these results using exact numerical simulations of linear diffusions. We illustrate the discrepancy between the entropy production of non-linear diffusions and their numerical simulations in several examples and illustrate how the entropy production can be used for accurate numerical simulation. Finally, we discuss the relationship between time-irreversibility and sampling efficiency, and how we can modify the definition of entropy production to score how far a process is from being generalised reversible.
\end{abstract}



\tableofcontents

\section{Introduction}

The entropy production rate $e_p$ is a central concept in statistical physics. In a nutshell, it is a measure of the time-irreversibility of a stochastic process, that is how much random motion differs statistically speaking as one plays it forward, or backward, in time.

At non-equilibrium steady-state, the $e_p$ is quantified by the relative entropy $\H$ (a.k.a Kullback-Leibler divergence) 
\begin{align*}
    e_p= \frac 1 T \H[\p_{[0,T]} \mid \bar \p_{[0,T]}]
\end{align*}
between the path-wise distributions of the forward and time-reversed processes in some time interval $[0,T]$, denoted by $\p_{[0,T]},\bar \p_{[0,T]}$, respectively.

Physically, the $e_p$ measures the minimal amount of energy needed, per unit of time, to maintain a system at non-equilibrium steady-state. Equivalently, it quantifies the heat dissipated by a physical system at non-equilibrium steady-state per unit of time \cite[p. 86]{jiangMathematicalTheoryNonequilibrium2004}. The second law of thermodynamics for open systems is the non-negativity of the entropy production. 

The $e_p$ plays a crucial role in stochastic thermodynamics. It is the central quantity in the so-called Gallavotti-Cohen fluctuation theorem, which quantifies the probability of entropy decrease along stochastic trajectories \cite{seifertStochasticThermodynamicsFluctuation2012,gallavottiDynamicalEnsemblesNonequilibrium1995a,lebowitzGallavottiCohenTypeSymmetry1999}. 
More recently, entropy production is at the heart of the so-called thermodynamic uncertainty relations, which provide estimates of $e_p$ from observations of the system \cite{dechantMultidimensionalThermodynamicUncertainty2018}.
The $e_p$ is also an important tool in biophysics, as a measure of the metabolic cost of molecular processes, such as molecular motors \cite{seifertStochasticThermodynamicsFluctuation2012}, 
and it was shown empirically that brain states associated with effortful activity correlated with a higher entropy production from neural activity \cite{lynnBrokenDetailedBalance2021}.

In this paper, we will be primarily concerned with the entropy production of diffusion processes, that is, solutions of stochastic differential equations. Consider an Itô stochastic differential equation (SDE)
\begin{equation*}
    \d x_t=b\left(x_t\right) \d t+\sigma\left( x_t\right) \d w_{t}
\end{equation*}
with drift $b: \R^d \to \R^d$ and volatility $\sigma :\R^d\to \R^{d\times m}$, and $w_t$ a standard Brownian motion on $\R^m$, whose solution is stationary at a probability measure $\mu$ with density $\rho$.
A result known as the Helmholtz decomposition, which is central in non-equilibrium statistical physics \cite{grahamCovariantFormulationNonequilibrium1977,eyinkHydrodynamicsFluctuationsOutside1996,aoPotentialStochasticDifferential2004,pavliotisStochasticProcessesApplications2014} but also in statistical sampling \cite{barpUnifyingCanonicalDescription2021,maCompleteRecipeStochastic2015,barpGeometricMethodsSampling2022a,chaudhariStochasticGradientDescent2018}, tells us that we can decompose the drift into time-reversible and time-irreversible parts: $b= b\rev+b\irr$. In particular, $b\irr \rho$ is the stationary probability current, as considered by Nelson \cite{nelsonDynamicalTheoriesBrownian1967}. Jiang and colleagues derived the entropy production rate for such systems under the constraint that the coefficients of the SDE $b, \sigma$ are globally Lipschitz, and the solution uniformly elliptic (i.e., the diffusion matrix field $D = \frac 1 2 \sigma \sigma^\top$ is uniformly positive definite). This takes the form of \cite[Chapter 4]{jiangMathematicalTheoryNonequilibrium2004}:
\begin{align*}
    e_p = \intr b\irr^\top D^{-1} b\irr \rho(x) \d x.
\end{align*}
In this paper, we extend their work by computing the entropy production for a greater range of diffusion processes, which includes non-elliptic, hypoelliptic and degenerate diffusions, and SDEs driven by locally Lipschitz coefficients. Non-elliptic diffusions are solutions to SDEs whose diffusion matrix field $D= \frac 1 2 \sigma \sigma^\top$ is not positive definite everywhere; this means that there are regions of space in which the random fluctuations cannot drive the process in every possible direction. Depending on how the volatility interacts with the drift (i.e., Hörmander's theorem \cite[Theorem 1.3]{hairerMalliavinProofOrmander2011}), solutions initialised at a point may still have a density---the hypoelliptic case---or not---the degenerate case; prominent examples are underdamped Langevin dynamics and deterministic dynamics, respectively. In our treatment, we only assume that the time-reversal of a diffusion is a diffusion (the necessary and sufficient conditions for which were first established by Millet, Nualart and Sanz \cite{milletIntegrationPartsTime1989a}) and sufficient regularity to apply Girsanov's theorem or the Stroock-Varadhan support theorem.

This extension has become important since many processes that are commonplace in non-equilibrium statistical physics or statistical machine learning are hypoelliptic. For instance, the underdamped and generalised Langevin equations in phase space \cite{pavliotisStochasticProcessesApplications2014}, which model the motion of a particle interacting with a heat bath, and which form the basis of efficient sampling schemes such as Hamiltonian Monte-Carlo \cite{barpGeometricMethodsSampling2022a}, or, stochastic gradient descent in deep neural networks \cite{chaudhariStochasticGradientDescent2018}, or, the linear diffusion process with the fastest convergence to stationary state \cite{guillinOptimalLinearDrift2021}, which informs us of the properties of efficient samplers. Much research in statistical sampling has drawn the connection between time-irreversibility and sampling efficiency \cite{ottobreMarkovChainMonte2016,barpGeometricMethodsSampling2022a,rey-belletIrreversibleLangevinSamplers2015,hwangAcceleratingDiffusions2005}, so it is informative to understand the amount of time-irreversibility associated with the most efficient samplers. Lastly, it is known that numerically integrating a diffusion processes can modify the amount of irreversibility present in the original dynamic \cite{katsoulakisMeasuringIrreversibilityNumerical2014}. The entropy production rate is thus an important indicator of the fidelity of a numerical simulation, and can serve as a guide to developing sampling schemes that preserve the statistical properties of efficient (hypoelliptic) samplers.

The outline of the paper and our contribution are detailed below.





\subsection{Paper outline and contribution}

\paragraph{Section \ref{sec: ep stationary Markov process}:} We give various characterisations and formulas for the $e_p$ of stationary Markov processes in discrete and continuous-time, and recall a crude but general recipe for numerical estimation. 

\paragraph{Section \ref{sec: time reversal of stationary diffusions}:}
We investigate the time-reversal of time-homogeneous diffusions. We give the general conditions under which the time-reversal of a time-homogeneous diffusion remains a diffusion, based on the results of Millet, Nualart and Sanz \cite{milletIntegrationPartsTime1989a}. 
We then recall how the drift vector field of an SDE can be decomposed into time-reversible and irreversible parts. We show that this decomposition is equivalent to a decomposition of the generator of the process into symmetric and antisymmetric operators on a suitable function space. Then we show that this decomposition is equivalent to two other fundamental decompositions in the study of far from equilibrium systems: the decomposition of the backward Kolmogorov equation considered in hypocoercivity theory \cite{duongNonreversibleProcessesGENERIC2021,villaniHypocoercivity2009} and the decomposition of the Fokker-Planck equation considered in the GENERIC formalism \cite{duongNonreversibleProcessesGENERIC2021,ottingerEquilibriumThermodynamics2005}. 

\paragraph{Section \ref{sec: the ep of stationary diffusions}:}
We compute the $e_p$ of stationary diffusion processes under the condition that the time-reversal of the diffusion remains a diffusion. We show that:
\begin{itemize}
    \item \textbf{Section \ref{sec: epr regularity}:} If $b_{\mathrm{irr}}(x) \in \im \sigma(x)$, for $\mu$-almost every $x \in \R^d$\footnote{$\mu$-almost everywhere: This means that the statement holds with probability $1$ when $x$ is distributed according to the probability measure $\mu$.} (in particular for elliptic or time-reversible diffusions). Then the forward and backward path-space measures are mutually equivalent, in other words, the sets of possible trajectories by forward and backward processes are equal
    ---
    and the entropy production equals $e_p = \int b_{\mathrm{irr}}^\top D^- b_{\mathrm{irr}} \d \mu$, where $\cdot^-$ denotes the Moore-Penrose matrix pseudo-inverse. See Theorems \ref{thm: epr regular case simple}, \ref{thm: epr regular case general} for details.
    \item \textbf{Section \ref{sec: ep singularity}:} When the above does not hold, the forward and backward path-space measures are mutually singular
    , in other words, there are trajectories that are taken by the forward process that cannot be taken by the backward process---and vice-versa\footnote{Precisely, two measures are mutual singular if and only if they are not mutually equivalent.}. In particular, the entropy production rate is infinite $e_p = +\infty$. See Theorem \ref{thm: epr singular} for details.
\end{itemize}

\paragraph{Section \ref{sec: illustrating results epr}:} We compute the $e_p$ of various models such as the multivariate Ornstein-Uhlenbeck and the underdamped Langevin process. We numerically simulate and verify the value of $e_p$ when the coefficients are linear. We then discuss how numerical discretisation can influence the value of $e_p$. As examples, we compute and compare the $e_p$ of Euler-Maruyama and BBK \cite{brungerStochasticBoundaryConditions1984,roussetFreeEnergyComputations2010} discretisations of the underdamped Langevin process. We summarise the usefulness of $e_p$ as a measure of the accuracy of numerical schemes in preserving the time-irreversibility properties of the underlying process, and give guidelines, in terms of $e_p$, for developing accurate simulations of underdamped Langevin dynamics.

\paragraph{Section \ref{sec: discussion}:} We give a geometric interpretation of our main results and discuss future perspectives: what this suggests about the relationship between time-irreversibility and mixing in the context of sampling and optimisation, and how we could modify the definition---and computation---of $e_p$ to quantify how a process is far from being time-reversible up to a one-to-one transformation of its phase-space.

\section{The $e_p$ of stationary Markov processes}
\label{sec: ep stationary Markov process}

In this section, $(x_t)_{t \in [0,T]}, T>0$ is a time-homogeneous Markov process on a Polish space $\mathcal X$ with almost surely (a.s.) continuous trajectories.

\begin{definition}[Time reversed process]
\label{def: Time reversed process}
The time reversed process $(\bar x_t)_{t \in [0,T]}$ is defined as $\bar x_t= x_{T-t}$.
\end{definition}
 Note that since the final state of the forward process is the initial state of the backward process this definition makes sense only on finite time intervals.

We define $\left(C([0,T], \mathcal X), d_\infty\right)$ as the space of $\mathcal X$-valued continuous paths endowed with the supremum distance $d_\infty$, defined as
    $d_\infty(f,g)= \sup_{t\in [0,T]}d(f(t),g(t))$,
where $d$ is a choice of distance on the Polish space $\mathcal X$. Naturally, when we later specialise to $\mathcal X= \mathbb R^d$, the supremum distance will be given by the $L^\infty$-norm $\|\cdot\|_\infty$.

\begin{definition}[Path space measure]
\label{def: path space measure}
Each Markov process $(x_t)_{0\leq t \leq T}$ with a.s. continuous trajectories defines probability measure $\p_{[0,T]}$ on the canonical path space $\left(C([0,T], \mathcal X), \mathscr B\right)$, where $\mathscr B$ is the Borel sigma-algebra associated with the supremum distance. This probability measure determines the probability of the process to take any (Borel set of) paths. 
\end{definition}


\begin{remark}[Cadlag Markov processes]
All definitions and results in this section hold more generally for Markov processes with \emph{cadlag} paths (i.e., right continuous with left limits), simply replacing the canonical path-space $\left(C([0,T], \mathcal X), d_\infty\right)$ with the Skorokhod space. We restrict ourselves to processes with continuous paths for simplicity.
\end{remark}

\begin{definition}[Restriction to a sub-interval of time]
Given a path space measure $\p_{[0,T]}$ and times $a<b \in [0,T]$, we define $\p_{[a,b]}$ to be the path space measure describing $\left(x_t\right)_{t \in [ a,b]}$. This is the restriction of $\p_{[0,T]}$ to the sub sigma-algebra
    $\mathscr B_{[a, b]}:= \left\{A \in \mathscr B : \left.A\right|_{[a, b]} \in \text{Borel sigma-algebra on} \left(C([a, b], \mathcal X), d_\infty\right)\right\}$.
\end{definition}

Let $\p$ be the path space measure of the Markov process $x_t$ and $\bar \p$ be that of its time-reversal $\bar x_t$, respectively. We can measure the statistical difference between the forward and time-reversed processes at time $\tau \in [0,T]$ with the \emph{entropy production rate} 
    \begin{align}
    \label{eq: time dependent EPR}
    \lim _{\e \downarrow 0} \frac{1}{\e} \H\left[\p_{[\tau, \tau+\e]}, \bar \p_{[\tau, \tau+\e]}\right],
    \end{align}
where $\H$ is the relative entropy (a.k.a. Kullback-Leibler divergence). This measures the rate at which the forward and backward path space measures differ in the relative entropy sense at time $\tau$.

The following result \cite[Theorem 2.2.4]{jiangMathematicalTheoryNonequilibrium2004} (see also \cite[Theorem 10.4]{varadhanLargeDeviationsApplications1988}) 
shows that the limit exists in stationary and time-homogeneous Markov processes. Obviously, the limit is independent of $\tau$ in this case.

\begin{theorem} 
\label{thm: ep is a rate over time}
Suppose that $\left(x_t\right)_{t \in [ 0,T]}$ is a stationary time-homogeneous Markov process on a Polish space $\mathcal X$ with continuous sample paths. Stationarity implies that we can set the time-horizon $T>0$ of the process to arbitrarily large values. Then the quantity
\begin{align*}
\frac 1 t\H\left[\p_{[\tau,\tau+t]} \mid   \bar \p_{[\tau,\tau+t]}\right] \text{ for all } \tau \in [0, +\infty), t \in (0, +\infty)
\end{align*}
is a constant $\in [0, +\infty]$.
\end{theorem}

This yields the following general definition of entropy production rate for stationary Markov processes: 

\begin{definition}[Entropy production rate of a stationary Markov process]
\label{def: epr}
Let $\left(x_t\right)_{t \in [ 0,T]}$ be a stationary time-homogeneous Markov process. Stationarity implies that we can set the time-horizon $T>0$ to be arbitrarily large. For such processes, the entropy production rate is a constant $e_p \in [0, +\infty]$ defined as
\begin{equation}
\label{eq: def epr}
    e_p := \frac{1}{t}\H\left[\p_{[0,t]} \mid  \bar \p_{[0,t]} \right]
\end{equation}
for any $t \in (0,+\infty)$. $e_p$ scores the amount to which the forward and time-reversed processes differ per unit of time. In particular, $Te_p$ is the total entropy production in a time interval of length $T$. Note that, in the literature, the $e_p$ is often defined as $e_p = \lim _{t \to +\infty}\frac{1}{t}\H\left[\p_{[0,t]} \mid  \bar \p_{[0,t]} \right]$, e.g., \cite[Definition 4.1.1]{jiangMathematicalTheoryNonequilibrium2004}; this is just \eqref{eq: def epr} in the limit of large $t$. However, Theorem \ref{thm: ep is a rate over time} showed us that \eqref{eq: def epr} is constant w.r.t. $t\in (0,+\infty)$ so we do not need to restrict ourselves to defining the $e_p$ as \eqref{eq: def epr} in the limit of large $t$. This added generality will be very helpful to compute $e_p$ later, by exploiting the fact that \eqref{eq: def epr} is often more easily analysed in the regime of finite or small $t$.
\end{definition}

\begin{remark}[Physical relevance of Definition \ref{def: epr}]
\label{rem: physical relevance}
    In some stationary processes (e.g., Hamiltonian systems), physicists define entropy production as Definition \ref{def: epr} with an additional operator applied to the path space measure of the time-reversed process; that is,
    \begin{equation}
    \label{eq: rem def gen epr}
    e_p^{\mathrm{gen}, \theta} := \lim _{\e \downarrow 0} \frac{1}{\e} \H\left[\p_{[0, \e]}, \theta_\#\bar \p_{[0, \e]}\right],
    \end{equation}
    where $\theta_\#$ is the pushforward operator associated to an involution of phase-space $\theta$ (e.g., the momentum flip \cite{vanvuUncertaintyRelationsUnderdamped2019,eckmannEntropyProductionNonlinear1999,spinneyNonequilibriumThermodynamicsStochastic2012}) that leaves the stationary distribution invariant\footnote{This generalised definition of entropy production is taken as a limit of $\e \downarrow 0$ analogously to \eqref{eq: time dependent EPR} to capture the fact that we are modelling a \emph{rate}. We cannot, a priori state that the expression is constant for any $\e \in (0,+\infty)$, as in Definition \ref{def: epr}, since we do not know whether a result analogous to Theorem \ref{thm: ep is a rate over time} holds in this case.}. In this article, we will refer to \eqref{eq: def epr} as \emph{entropy production} and to \eqref{eq: rem def gen epr} as \emph{generalised entropy production}, and proceed to derive general results for \eqref{eq: def epr}. The results we derive are informative of the process and applicable independently of whether \eqref{eq: def epr} is the physically meaningful definition of entropy production; yet, physicists looking to interpret these results should bear in mind that they are physically informative about entropy production insofar as Definition \ref{def: epr} is physically meaningful for the system at hand. We will briefly revisit generalised entropy production in the discussion (Section~\ref{sec: discussion generalised non-reversibility}).
\end{remark}


\begin{proposition}
\label{prop: aggregating local ep}
Let $\left(x_t\right)_{t \in [ 0,T]}$ be a time-homogeneous Markov process on a Polish space $\mathcal X$, stationary at the probability measure $\mu$. Then the entropy production rate equals
\begin{equation*}
    e_p = \frac 1 t \E_{x\sim \mu}\left[\H\left[\p^{x}_{[0,t]}\mid \bar  \p^{x}_{[0,t]}\right]\right]
\end{equation*}
for any $t\in (0, +\infty)$.
\end{proposition}

A proof is provided in Appendix \ref{app: aggregating local ep}.

\begin{notation}
\label{notation: path space measure deterministic initial condition}
By $\p^{x}$ we mean the path space measure of the 
process initialised (possibly out of stationarity) at a deterministic initial condition $x_0=x \in \mathcal X$.  
\end{notation}

\begin{proposition}[$e_p$ in terms of transition kernels]
\label{prop: epr transition kernels}
Let $\left(x_t\right)_{t \in [ 0,T]}$ be a time-homogeneous Markov process on a Polish space $\mathcal X$, stationary at the probability measure $\mu$. Denote by $p_t(dy,x)$ the transition kernels of the Markov semigroup, and by $\bar p_t(dy,x)$ those of the time-reversed process. Then the entropy production rate equals
\begin{equation*}
    e_p = \lim_{\e \downarrow 0} \frac{1}{\e} \E_{x\sim \mu}\left[\H\left[p_\e(\cdot,x)\mid\bar p_\e(\cdot,x)\right]\right].
\end{equation*}
\end{proposition}

The fact that the time-reversed process possesses transition kernels holds as it is also a stationary Markov process \cite[p. 113]{jiangMathematicalTheoryNonequilibrium2004}. A proof of Proposition \ref{prop: epr transition kernels} is provided in Appendix \ref{app: epr transition kernels}.

\subsection{The $e_p$ of numerical simulations}
\label{sec: ep of numerical schemes theory}

Proposition \ref{prop: epr transition kernels} entails a formula for the $e_p$ of Markov processes in discrete time:
\begin{definition}
\label{def: EPR numerical scheme}
The entropy production rate of a discrete-time Markov process with time-step $\e$ equals
\begin{equation}
\label{eq: def EPR numerical scheme}
    e_p^{\text{NS}}(\e) = \frac{1}{\e} \E_{x\sim \tilde \mu}\int p_\e(y,x) \log \frac{p_\e(y,x)}{p_\e(x,y)} \d y,
\end{equation}
where $\tilde \mu$ is the invariant measure of the process.
\end{definition}

This definition is useful, for example, to quantify the entropy production of numerical simulations of a stochastic process \cite{katsoulakisMeasuringIrreversibilityNumerical2014}.
In particular, it suggests a simple numerical estimator of the entropy production rate for numerical simulations (at stationarity). Consider a small $\delta$ (e.g., $\delta$ is the time-step of the numerical discretisation). Given samples from the process at $\delta$ time intervals, discretise the state-space into a finite partition $U_1, \ldots, U_n$, and approximate $\p_{[0,\delta]}$ and $\bar \p_{[0,\delta]}$ by the empirical transition probabilities $p_{i \to j}$ between $U_i, U_j$ from time $0$ to $\delta$.

\begin{align*}
    e_p^{\text{NS}}= \lim_{\e \to 0} \frac{1}{\e} \H\left[\p_{[0,\e]}\mid\bar \p_{[0,\e]}\right] \approx \frac{1}{\delta} \H\left[\p_{[0,\delta]}\mid\bar \p_{[0,\delta]}\right] \approx \frac{1}{\delta} \sum_{i,j}p_{i \to j} \log \frac{p_{i \to j}}{p_{j \to i}}. 
\end{align*}

Note that this method measures the entropy production rate of the numerical discretisation as opposed to that of the continuous process. This typically produces results close to $e_p$, but does not necessarily converge to $e_p$ in the continuum limit $\delta \to 0$ of the numerical discretisation. Indeed \cite{katsoulakisMeasuringIrreversibilityNumerical2014} showed that numerical discretisations can break detailed balance, so that the continuum limit of the numerical discretisation can differ from the initial process. Thus one should choose numerical schemes carefully when preserving the entropy production rate of a process is important. We will return to this in Section \ref{sec: illustrating results epr}.


\section{Time reversal of stationary diffusions}
\label{sec: time reversal of stationary diffusions}


We now specialise to diffusion processes in $\R^d$. These are Markov processes with an infinitessimal generator that is a second order linear operator without a constant part \cite[Definition 1.11.1]{bakryAnalysisGeometryMarkov2014}, 
which entails almost surely (a.s.) continuous sample paths. 
Conveniently, diffusion processes are usually expressible as solutions to stochastic differential equations.
From now on, we consider an Itô stochastic differential equation
\begin{equation}
\label{eq: Ito SDE}
    \d x_t=b\left(x_t\right) \d t+\sigma\left( x_t\right) \d w_{t}
\end{equation}
with drift $b:\mathbb{R}^{d} \rightarrow \mathbb{R}^{d}$ and volatility $\sigma: \mathbb{R}^{d}  \rightarrow \R^{d \times m}$, and $w_t$ a standard Brownian motion on $\R^m$.

\begin{notation}
Let $D=\sigma\sigma^\top/2 \in \mathbb{R}^{{d \times d}}$ be the \textit{diffusion tensor}. Denote by $\|\cdot\|$ the Euclidean distance, and, for matrices
$$
\|\sigma\|^2:=\sum_{i=1}^{d} \sum_{j=1}^{n}\left|\sigma_{i j}\right|^{2}.
$$
Throughout, $\nabla$ and $\nabla \cdot$ are the gradient and the divergence in the distributional sense. We operationally define the divergence of a matrix field $Q : \R^d \to \mathbb{R}^{{d \times d}}$ by $(\nabla \cdot Q)_i:=\sum_{j=1}^d \partial_j Q_{i j}$ for $0\leq i \leq d$.
We will denote by $\mu$ the stationary probability measure of the process $x_t$ and by $\rho$ its density with respect to the Lebesgue measure, i.e., $\mu(\d x)= \rho(x) \d x$ (assuming they exist).
\end{notation}

\subsection{On the time-reversibility of the diffusion property}

There is a substantial literature studying the time-reversal of diffusion processes. In general, the time-reversal of a diffusion need not be a diffusion \cite{milletIntegrationPartsTime1989a}, 
but Haussman and Pardoux showed that the diffusion property is preserved under some mild regularity conditions on the diffusion process \cite{haussmannTimeReversalDiffusions1986}. A few years later Millet, Nualart, Sanz derived necessary and sufficient conditions for the time-reversal of a diffusion to be a diffusion \cite[Theorem 2.2 \& p. 220]{milletIntegrationPartsTime1989a}. We provide these conditions here, with a proof of a different nature that exploits the existence of a stationary distribution.

\begin{lemma}[Conditions for the reversibility of the diffusion property]
\label{lemma: reversibility of the diffusion property}
Let an Itô SDE \eqref{eq: Ito SDE} with locally bounded, Lebesgue measurable coefficients $b: \mathbb{R}^{d} \rightarrow \mathbb{R}^{d}, \sigma : \mathbb{R}^{d}  \rightarrow \R^{d \times m}$. Consider a strong solution $(x_t)_{t\in [0,T]}$, i.e., a process satisfying
\begin{align*}
    x_t=x_0+\int_0^t b\left(x_s\right) d s+\int_0^t \sigma\left(x_s\right) d w_s,
\end{align*}
and assume that it is stationary with respect to a probability measure $\mu$ with density $\rho$. Consider the time-reversed stationary process $(\bar x_t)_{t\in [0,T]}$. Then, the following are equivalent:
\begin{itemize}
    \item $(\bar x_t)_{t\in [0,T]}$ is a Markov diffusion process. 
    \item The distributional derivative $\nabla \cdot (D \rho)$ is a function, which is then necessarily in $L^1\loc(\R^d, \R^d)$. 
\end{itemize}
\end{lemma}
A proof is provided in Appendix \ref{app: reversibility of the diffusion property}.

\subsection{Setup for the time-reversal of diffusions}

From now on, we will work under the assumption that the time-reversal of the diffusion is a diffusion. We assume that:
\begin{assumption}
\label{ass: coefficients time reversal}
\begin{enumerate}
    \item \label{ass: locally bounded and Lipschitz} The coefficients of the SDE \eqref{eq: Ito SDE} $b, \sigma$ are 
    \emph{locally Lipschitz continuous}. In other words, $\forall x \in \R^d, \exists r >0, k > 0 \st \forall y\in \R^d:$
    \begin{align*}
         \|x-y \|<r \Rightarrow \|b( x)-b( y)\|+\|\sigma(x)-\sigma(y)\|\leq k\|x-y\|,
    \end{align*}
    \item \label{ass: global solution} The solution $x_t$ to \eqref{eq: Ito SDE} is defined globally up to time $T>0$. Sufficient conditions in terms of the drift and volatility for Itô SDEs are given in Theorem \cite[Theorem 3.1.1]{prevotConciseCourseStochastic2008}. 
\end{enumerate}
\end{assumption}
Assumption \ref{ass: coefficients time reversal}.\ref{ass: locally bounded and Lipschitz} ensures the existence and uniqueness of strong solutions locally in time \cite[Chapter IV Theorem 3.1]{watanabeStochasticDifferentialEquations1981}, while Assumption \ref{ass: coefficients time reversal}.\ref{ass: global solution} ensures that this solution exists globally in time (i.e., non-explosiveness). Altogether, Assumption \ref{ass: coefficients time reversal} ensures that the SDE \eqref{eq: Ito SDE} unambiguously defines a diffusion process. 


Furthermore, we assume some regularity on the stationary distribution of the process.

\begin{assumption}
\label{ass: steady-state time reversal}
\begin{enumerate}
    \item $(x_t)_{t\in [0,T]}$ is stationary at a probability distribution $\mu$, with density $\rho$ with respect to the Lebesgue measure, i.e., $\mu(\d x)= \rho(x) \d x$.
\end{enumerate}
Then, $\rho \in L^1(\R^d)$ and, under local boundedness of the diffusion tensor (e.g., Assumption \ref{ass: coefficients time reversal}), $D\rho \in L^1\loc(\R^d, \R^{d\times d})$. Thus, we can define the distributional derivative $\nabla \cdot (D\rho)$. We assume that:
\begin{enumerate}
\setcounter{enumi}{1}
    \item $\nabla \cdot (D\rho) \in L^1\loc(\R^d, \R^{d})$, i.e., the distributional derivative $\nabla \cdot (D\rho)$ is a function.
\end{enumerate}
\end{assumption}

Assumption \ref{ass: steady-state time reversal} ensures that the time-reversal of the diffusion process remains a diffusion process, as demonstrated in Lemma \ref{lemma: reversibility of the diffusion property}.

\subsection{The time reversed diffusion}

Now that we know sufficient and necessary conditions for the time-reversibility of the diffusion property, we proceed to identify the drift and volatility of the time-reversed diffusion. This was originally done by Hausmann and Pardoux \cite[Theorem 2.1]{haussmannTimeReversalDiffusions1986}, and then by Millet, Nualart, Sanz under slightly different conditions \cite[Theorems 2.3 or 3.3]{milletIntegrationPartsTime1989a}. Inspired by these, we provide a different proof, which applies to stationary diffusions with locally Lipschitz coefficients.

\begin{theorem}[Characterisation of time-reversal of diffusion]
\label{thm: time reversal of diffusions}
Let an Itô SDE \eqref{eq: Ito SDE} with coefficients satisfying Assumption \ref{ass: coefficients time reversal}. Assume that the solution $(x_t)_{t\in [0,T]}$ is stationary with respect to a density $\rho$ satisfying Assumption \ref{ass: steady-state time reversal}. 
Then, the time-reversed process $(\bar x_t)_{t\in [0,T]}$ is a Markov diffusion process, stationary at the density $\rho$, with drift
\begin{align}
\label{eq: time reversed drift}
\bar b(x)=& \begin{cases}
        -b(x) +2\rho^{-1} \nabla \cdot \left(D \rho\right)(x)\quad \text{when } \rho(x) > 0,\\
        -b(x) \quad \text{when } \rho(x) = 0.
    \end{cases}
\end{align}
and diffusion $\bar D = D$. Furthermore, any such stationary diffusion process induces the path space measure of the time-reversed process $\bar \p_{[0,T]}$. 
\end{theorem}

A proof is provided in Appendix \ref{app: time reversed diffusion}. 
Similar time-reversal theorems exist in various settings: for more singular coefficients on the torus \cite{quastelTimeReversalDegenerate2002}, under (entropic) regularity conditions on the forward path space measure \cite{follmerTimeReversalWiener1986,cattiauxTimeReversalDiffusion2021}, for infinite-dimensional diffusions \cite{belopolskayaTimeReversalDiffusion2001,milletTimeReversalInfinitedimensional1989,follmerTimeReversalInfinitedimensional1986}, for diffusions on open time-intervals \cite{nagasawaDiffusionProcessesOpen1996}, or with boundary conditions \cite{cattiauxTimeReversalDiffusion1988}. Furthermore, we did not specify the Brownian motion driving the time-reversed diffusion but this one was identified in \cite[Remark 2.5]{pardouxTimereversalDiffusionProcesses1985}.

We illustrate the time-reversal of diffusions with a well-known example:

\begin{example}[Time reversal of underdamped Langevin dynamics]
\label{eg: Time reversal of underdamped Langevin dynamics}
Underdamped Langevin dynamics is an important model in statistical physics and sampling \cite{pavliotisStochasticProcessesApplications2014,roussetFreeEnergyComputations2010,maThereAnalogNesterov2021}. Consider a Hamiltonian $H(q,p)$ function of positions $q\in \R^n$ and momenta $ p \in \R^n$. We assume that the Hamiltonian has the form
\begin{align*}
   H(q,p) = V(q) + \frac 1 2 p^\top M^{-1} p,
\end{align*}
for some smooth potential function $V: \R^d \to \R$ and diagonal mass matrix $M \in \R^{d\times d}$. The underdamped Langevin process is given by the solution to the SDE \cite[eq 2.41]{roussetFreeEnergyComputations2010}
\begin{align}
\label{eq: underdamped Langevin dynamics}
\begin{cases}
\d q_{t} = M^{-1}p_{t} \d t \\
\d p_{t} =-\nabla V\left(q_{t}\right) \d t-\gamma M^{-1} p_{t} \d t+\sqrt{2 \gamma \beta^{-1}} \d w_{t}
\end{cases}
\end{align}
for some friction, and inverse temperature coefficients $\gamma, \beta >0$. The stationary density, assuming it exists, is the canonical density \cite[Section 2.2.3.1]{roussetFreeEnergyComputations2010}
\begin{align}
\label{eq: canonical density underdamped}
    \rho(q,p)\propto e^{-\beta H(q,p)}= e^{-\beta V(q)-\frac{\beta}{2} p^\top M^{-1} p}.
\end{align}
Thus, the time-reversal of the stationary process (Theorem \ref{thm: time reversal of diffusions}) 
is a weak solution to the SDE
\begin{align*}
&\begin{cases}
\d \bar q_{t} = -M^{-1}\bar p_{t} \d t \\
\d \bar p_{t} =\nabla V\left(\bar q_{t}\right) \d t-\gamma M^{-1} \bar p_{t} \d t+\sqrt{2 \gamma \beta^{-1}} \d  w_{t}.
\end{cases}
\end{align*}
Letting $\hat p_{t} = -\bar p_{t}$, 
the tuple $(\bar q_t,\hat p_t)$ solves the same SDE as $(q_t,p_t)$ but with a different Brownian motion $\hat w_t$
    \begin{equation*}
    \begin{cases}
    \d \bar q_{t} = M^{-1}\hat p_{t} \d t \\
    \d \hat p_{t} =-\nabla V\left(\bar q_{t}\right) \d t-\gamma M^{-1} \hat p_{t} \d t+\sqrt{2 \gamma \beta^{-1}} \d \hat w_{t}.
    \end{cases}
    \end{equation*}
Since path space measures are agnostic to changes in the Brownian motion, 
this leads to the statement that time-reversal equals momentum reversal in underdamped Langevin dynamics (with equality in law, i.e., in the sense of path space measures)
    \begin{align*}
        (\bar q_t, \bar p_t)_{t\in [0,T]}=(\bar q_t,-\hat p_t)_{t\in [0,T]}\stackrel{\ell}{=} (q_t, -p_t)_{t\in [0,T]}.
    \end{align*}
In other words, we have $\p_{[0,T]} = \theta_\# \bar \p_{[0,T]}$ where $\theta(q,p)=(q,-p)$ is the momentum flip transformation in phase space. 
\end{example}

\subsection{The Helmholtz decomposition}
\label{sec: helmholtz decomposition}

\begin{figure}[t!]
    \centering
    \includegraphics[width= 0.45\textwidth]{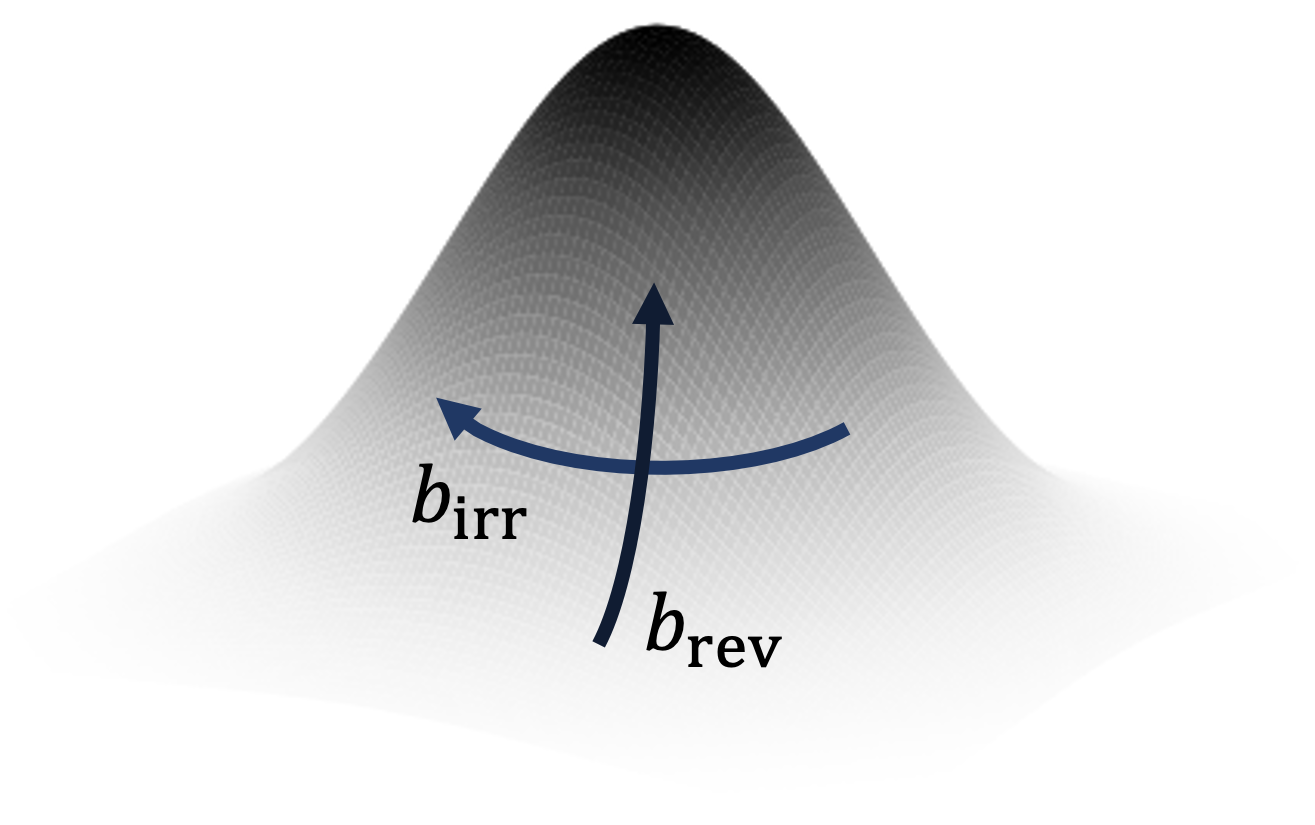}
    \includegraphics[width= 0.45\textwidth]{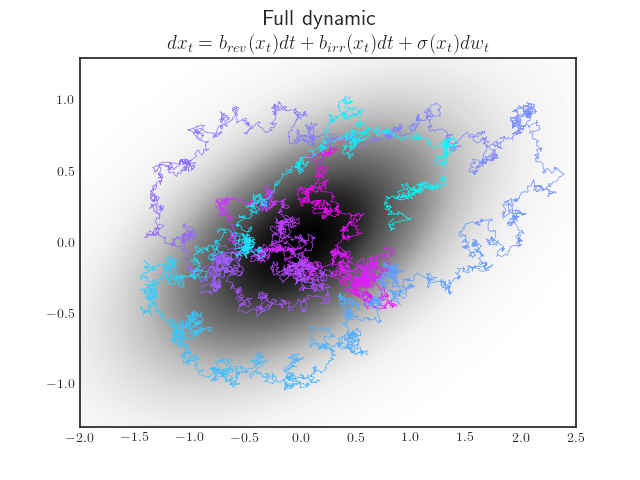}
    \includegraphics[width= 0.45\textwidth]{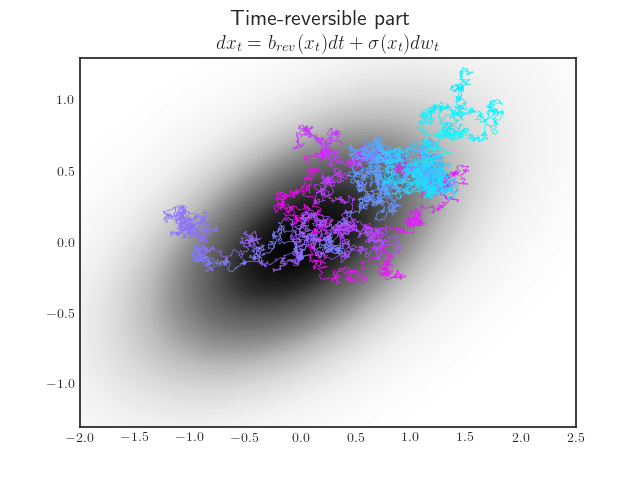}
    \includegraphics[width= 0.45\textwidth]{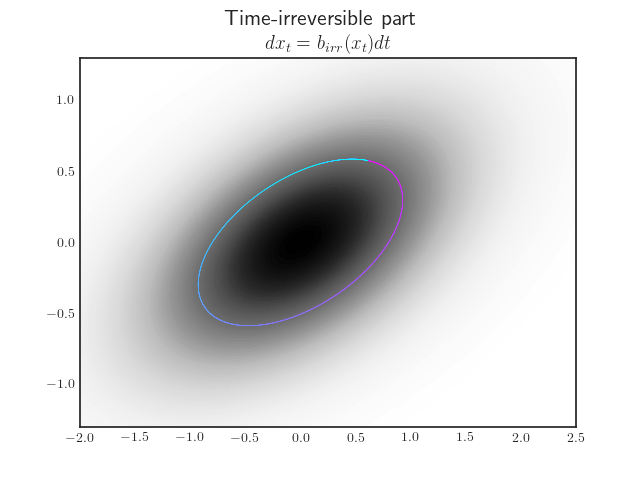}
    \caption{\footnotesize\textbf{Helmholtz decomposition.} The upper left panel illustrates the Helmholtz decomposition of the drift into time-reversible and time-irreversible parts: the time-reversible part of the drift flows towards the peak of the stationary density, while the time-irreversible part flows along its contours. The upper right panel shows a sample trajectory of a two-dimensional diffusion process stationary at a Gaussian distribution. The lower panels plot sample paths of the time-reversible (lower left) and time-irreversible (lower right) parts of the dynamic. Purely conservative dynamics (lower right) are reminiscent of the trajectories of massive bodies (e.g., planets) whose random fluctuations are negligible, as in Newtonian mechanics. Together, the lower panels illustrate time-irreversibility: If we were to reverse time, 
    the trajectories of the time-reversible process would be statistically identical, while the trajectories of the time-irreversible process be distinguishable by flow, say, clockwise instead of counterclockwise. The full process (upper right) is a combination of both time-reversible and time-irreversible dynamics. The time-irreversible part defines a non-equilibrium steady-state and induces its characteristic wandering, cyclic behaviour.}
    \label{fig: Helmholtz decomposition}
\end{figure}

Armed with the time-reversal of diffusions we proceed to decompose the SDE into its time-reversible and time-irreversible components.
This decomposition is called the Helmholtz decomposition because it can be obtained geometrically by decomposing the drift vector field $b$ into horizontal $b_{\mathrm{irr}}$ (time-irreversible, conservative) and vertical $b_{\mathrm{rev}}$ (time-reversible, non-conservative) components with respect to the stationary density \cite{barpUnifyingCanonicalDescription2021}. These vector fields are called horizontal and vertical, respectively, because the first flows along the contours of the stationary density, while the second ascends the landscape of the stationary density (see the schematic in the upper-left panel of Figure \ref{fig: Helmholtz decomposition}). For our purposes, we provide a self-contained, non-geometric proof of the Helmholtz decomposition in Appendix \ref{app: helmholtz decomposition}.

\begin{proposition}[Helmholtz decomposition]
\label{prop: helmholtz decomposition}
Consider the solution $(x_t)_{t\in [0,T]}$ of the Itô SDE \eqref{eq: Ito SDE} with coefficients satisfying Assumption \ref{ass: coefficients time reversal}. Let a probability density $\rho$ satisfying $\nabla \cdot(D\rho) \in L^1\loc (\R^d, \R^d)$. Then, the following are equivalent:
\begin{enumerate}
    \item The density $\rho$ is stationary for $(x_t)_{t\in [0,T]}$.
    \item We can write the drift as
    \begin{equation}
    \label{eq: Helmholtz decomposition of drift}
    \begin{split}
        b &= b_{\mathrm{rev}}+ b_{\mathrm{irr}}\\
        b_{\mathrm{rev}} &=
        \begin{cases}
        D\nabla \log \rho + \nabla \cdot D \quad \text{if } \rho(x) > 0\\
        0 \quad \text{if } \rho(x) = 0
        \end{cases}\\
        \nabla\cdot (b_{\mathrm{irr}} \rho)&= 0. 
    \end{split}
    \end{equation}
\end{enumerate}
Furthermore, $b_{\mathrm{rev}}$ is \emph{time-reversible}, while $b_{\mathrm{irr}}$ is \emph{time-irreversible}, i.e.,
    \begin{align*}
        b = b_{\mathrm{rev}}+ b_{\mathrm{irr}},\quad  \bar b= b_{\mathrm{rev}}- b_{\mathrm{irr}}.
    \end{align*}
\end{proposition}


The fundamental importance of the Helmholtz decomposition was originally recognised in the context of non-equilibrium thermodynamics by Graham in 1977 \cite{grahamCovariantFormulationNonequilibrium1977}, but its inception in this field dates from much earlier: for instance, the divergence free vector field $b_{\mathrm{irr}} \rho$ is precisely the stationary probability current or flux introduced by Nelson in 1967 \cite{nelsonDynamicalTheoriesBrownian1967}. More recently, the decomposition has recurrently been used in non-equilibrium statistical physics \cite{eyinkHydrodynamicsFluctuationsOutside1996,aoPotentialStochasticDifferential2004,qianDecompositionIrreversibleDiffusion2013,pavliotisStochasticProcessesApplications2014,fristonStochasticChaosMarkov2021,dacostaBayesianMechanicsStationary2021a,yangPotentialsContinuousMarkov2021}, and in statistical machine learning as the basis of Monte-Carlo sampling schemes \cite{maCompleteRecipeStochastic2015,barpUnifyingCanonicalDescription2021,lelievreOptimalNonreversibleLinear2013,pavliotisStochasticProcessesApplications2014}.

\begin{remark}[Probabilistic reversibility]
Here, time-reversible means reversibility in a probabilistic sense; that is, invariance under time reversal, also known as detailed balance \cite[Proposition 3.3.4]{jiangMathematicalTheoryNonequilibrium2004}. 
Probabilistic reversibility often leads to the non-conversation of quantities like the potential $-\log \rho(x)$
For example, the identity $\nabla\cdot (b_{\mathrm{irr}} \rho)= 0$ implies that the time-irreversible drift $b_{\mathrm{irr}}$ flows along the contours of the probability density; in other words, the probability density and the potential are conserved along the time-irreversible vector field. In contrast, none of them are conserved when flowing along the time-reversible vector field $b_{\mathrm{rev}}$. See Figure \ref{fig: Helmholtz decomposition} for an illustration.


\end{remark}

\begin{remark}[Stratonovich formulation of Helmholtz decomposition]
\label{rem: stratonovich Helmholtz}
There exists an equivalent decomposition of the drift of Stratonovich SDEs into time-reversible and irreversible parts. Assuming that $\sigma$ is differentiable, we can rewrite the Itô SDE \eqref{eq: Ito SDE} into its equivalent Stratonovich SDE 
\begin{align*}
    dx_t = b^s(x_t) dt + \sigma(x_t) \circ dw_t.
\end{align*}
where $b^s =b- \iota$ and $\iota$ is the Itô to Stratonovich correction \cite[eq. 3.31]{pavliotisStochasticProcessesApplications2014}. Note that the correction is time-reversible. It follows that
\begin{align}
\label{eq: irr drift equals ito and stratonovich}
    b^s_{\mathrm{irr}}&=b_{\mathrm{irr}},
\end{align}
and for $x \st \rho(x) >0$,
\begin{align*}
    b^s_{\mathrm{rev}}(x)&= (b_{\mathrm{rev}}-\iota)(x)=  D\nabla \log \rho(x) + \frac{1}{2} \sigma \nabla \cdot \sigma^\top(x).
\end{align*}
In particular, for $x \st \rho(x) >0$,
\begin{align}
\label{eq: stratonovich drift in image of vol equiv irr drift in image of vol}
    b^s(x) \in \im\sigma(x) &\iff b^s_{\mathrm{irr}}(x) \in \im\sigma(x)
\end{align}
as $b^s_{\mathrm{rev}}(x)\in \im\sigma(x)$. For diffusions driven by additive noise, Itô and Stratonovich formulations coincide $b^s=b$. Thus, we conclude
\begin{align}
    \sigma \text{ is constant} &\Rightarrow b(x) = \begin{cases}
        D\nabla \log \rho(x) + b_{\mathrm{irr}}(x) \quad \text{if } \rho(x) > 0\\
        b_{\mathrm{irr}}(x) \quad \text{if } \rho(x) = 0
        \end{cases} \nonumber\\
        &\Rightarrow \left( b(x) \in \im\sigma(x) \iff b_{\mathrm{irr}}(x) \in \im\sigma(x)\right) \label{eq: drift image of volatility additive noise}
\end{align}
These identities will be useful to compute the entropy production rate later on.
\end{remark}

The time-irreversible part of the drift often takes a simple form:

\begin{proposition}[Characterisation of time-irreversible drift]
\label{prop: characterisation of irreversible drift}
Consider a smooth, strictly positive probability density $\rho$ and an arbitrary smooth vector field $b_{\mathrm{irr}}$. Then
\begin{align*}
    \nabla \cdot( b_{\mathrm{irr}}\rho) = 0 \iff 
    b_{\mathrm{irr}} = Q \nabla \log \rho + \nabla \cdot Q
\end{align*}
where $Q=-Q^\top $ is a smooth antisymmetric matrix field.
\end{proposition}

A proof is provided in Appendix \ref{app: helmholtz decomposition}. We conclude this section by unpacking the Helmholtz decomposition of underdamped Langevin dynamics.

\begin{example}[Helmholtz decomposition of underdamped Langevin]
\label{eg: helmholtz decomp Langevin}
Following Example \ref{eg: Time reversal of underdamped Langevin dynamics}, it is straightforward to decompose underdamped Langevin dynamics into its time-irreversible and time-reversible parts. Indeed we just need to identify the parts of the drift whose sign changes, and remains invariant, under time reversal:
\begin{align*}
b_{\mathrm{rev}}(q,p) = \begin{bmatrix}
    0 \\
    -\gamma M^{-1} p
\end{bmatrix}, \quad b_{\mathrm{irr}}(q,p) &= \begin{bmatrix}
    M^{-1}p  \\
    -\nabla V\left(q\right)
\end{bmatrix}.   
\end{align*}
We can rewrite these in canonical form recalling the gradient of the stationary density \eqref{eq: canonical density underdamped}
\begin{align*}
    &b_{\mathrm{rev}}(q,p)= D \nabla \log \rho(q,p) , \quad b_{\mathrm{irr}}(q,p)= Q \nabla \log \rho(q,p)\\
    \nabla \log  \rho(q,p)&= -\beta \begin{bmatrix}
         \nabla V(q)\\
         M^{-1}p
    \end{bmatrix},
    \quad D=\begin{bmatrix}
        0 &0\\
        0& \gamma \beta^{-1}\id_n
    \end{bmatrix}
 , \quad Q=\beta^{-1}\begin{bmatrix}
        0 &-\id_n\\
        \id_n & 0
    \end{bmatrix}.
\end{align*}
Clearly, the time-irreversible part of the process $\d [q_t, p_t] = b_{\mathrm{irr}}(q_t,p_t)\d t$ is a Hamiltonian dynamic that preserves the energy (i.e., the Hamiltonian), while the time-reversible part is a reversible Ornstein-Uhlenbeck process. Example trajectories of the time-irreversible trajectory are exemplified in Figure \ref{fig: Helmholtz decomposition} (bottom right).
\end{example}



\subsection{Multiple perspectives on the Helmholtz decomposition}

The Helmholtz decomposition is a cornerstone of the theory of diffusion processes. In addition to being a geometric decomposition of the drift \cite{barpUnifyingCanonicalDescription2021}, it is, equivalently, a time-reversible and irreversible decomposition of the SDE \eqref{eq: Helmholtz decomp SDE}, of the generator and the (backward and forward) Kolmogorov PDEs describing the process. Briefly, the Helmholtz decomposition is equivalent to a functional analytic decomposition of the generator into symmetric and antisymmetric operators in a suitable function space. This corresponds to a decomposition of the backward Kolmogorov equation---which determines the evolution of (macroscopic) observables under the process---into a conservative and a dissipative flow. This decomposition can be used as a starting point to quantify the speed of convergence of the process to its stationary state from arbitrary initial conditions using hypocoercivity theory \cite{villaniHypocoercivity2009}. The same goes for the Fokker-Planck equation, which can be decomposed into a dissipative gradient flow, and a flow that is conservative in virtue of being orthogonal to the gradient flow in a suitable function space. This casts the Fokker-Planck equation in GENERIC form (General Equations for Non-Equilibrium Reversible-Irreversible Coupling), a general framework for analysing dynamical systems arising in non-equilibrium statistical physics  \cite{ottingerEquilibriumThermodynamics2005,duongNonreversibleProcessesGENERIC2021}.

Below we outline these different equivalent perspectives. 
This section is provided for independent interest but will not be used to derive our main results on entropy production; \emph{you may conveniently skip it on a first reading}.

\subsubsection{Helmholtz decomposition of the SDE}

Proposition \ref{prop: helmholtz decomposition} is equivalent to a Helmholtz decomposition of the SDE into its time-reversible and time-irreversible parts, noting that the volatility is invariant under time-reversal (Theorem \ref{thm: time reversal of diffusions})
\begin{align}
\label{eq: Helmholtz decomp SDE}
\mathrm{d} x_t= \underbrace{b_{\mathrm{irr}}\left(x_t\right) \mathrm{d} t}_{\text{Time-irreversible}} +\underbrace{b_{\mathrm{rev}}\left(x_t\right) \mathrm{d} t+\sigma\left(x_t\right) \mathrm{d} w_t}_{\text{Time-reversible}}.
\end{align}
Figure \ref{fig: Helmholtz decomposition} illustrates this decomposition with simulations.

\subsubsection{Helmholtz decomposition of the infinitesimal generator}
\label{sec: helmholtz decomp inf gen}

Following the differential geometric viewpoint, a deterministic flow---namely, a vector field $b$---is given by a first order differential operator $b \cdot \nabla$. Similarly, a stochastic flow given by a diffusion---namely, a vector field $b$ and a diffusion tensor $D$---is characterised by a second order differential operator
\begin{align}
\label{eq: generic generator}
    \L= b \cdot \nabla + D\nabla \cdot \nabla,
\end{align}
known as the generator. Note that the first order part is the deterministic flow given by the drift while the second order part is the stochastic flow determined by the diffusion. More precisely, the generator of a diffusion process solving the SDE \eqref{eq: Ito SDE} under Assumptions \ref{ass: coefficients time reversal} and \ref{ass: steady-state time reversal} is a linear, unbounded operator defined as
\begin{align}
&\L: C_c^\infty(\R^d) \subset  \dom \L \subset  L^p_\mu(\R^d) \to L^p_\mu(\R^d), 1 \leq p \leq \infty, \quad 
     \L f(y) := \lim_{t\downarrow 0}\frac 1 t\E[f(x_t)-f(y) \mid x_0=y], \label{eq: def generator}\\
     &f \in \dom \L = \left\{f \in L^p_\mu(\R^d) \mid \exists g \in L^p_\mu(\R^d) \st \frac 1 t\E[f(x_t)-f(y) \mid x_0=y] \xrightarrow{t \downarrow 0} g(y) \text{ in }  L^p_\mu(\R^d)\right\} \nonumber
\end{align}
Diffusions are among the simplest and most canonical Markov processes because they are characterised by generators that are second order differential operators (with no constant part). 
Indeed, starting from \eqref{eq: def generator}, a quick computation using Itô's formula yields \eqref{eq: generic generator}.

Recall that we have a duality pairing $\langle \cdot, \cdot\rangle_\mu : L^{p'}_\mu(\R^d) \otimes L^p_\mu(\R^d) \to \R$ defined by
    $\langle f,g \rangle_\mu= \intr fg\: \d \mu$,
where $\frac{1}{p'}+\frac{1}{p}=1$.

A well-known fact is that the generator $\bar \L$ of the time-reversed diffusion is the adjoint of the generator under the above duality pairing \cite{yosidaFunctionalAnalysis1995,pazySemigroupsLinearOperators2011}, \cite[Thm 4.3.2]{jiangMathematicalTheoryNonequilibrium2004}.
The adjoint $\bar \L$ is implicitly defined by the relation
\begin{align*}
    &\langle f,\L g \rangle_\mu = \langle \bar \L f, g \rangle_\mu, \forall f \in \dom \bar \L, g \in \dom \L,\\
    \dom \bar \L=\:&\{f \in L^1_\mu(\R^d) \mid  \exists h \in L^1_\mu(\R^d),  \forall g\in \dom \L: \langle f,\L g \rangle_\mu =\langle h, g \rangle_\mu   \}.
\end{align*}
(The concept of adjoint generalises the transpose of a matrix in linear algebra to operators on function spaces).
The proof of Lemma \ref{lemma: reversibility of the diffusion property} explicitly computes the adjoint and shows that it is a linear operator $\bar \L: L^1_\mu(\R^d) \to L^1_\mu(\R^d)$ which equals
\begin{align*}
    \bar \L f&= -b \cdot \nabla f + 2 \rho^{-1}\nabla \cdot (D\rho )\cdot \nabla f+ D\nabla \cdot \nabla f.
\end{align*}
Notice how the first order part of the adjoint generator is the drift of the time-reversed diffusion, while the second order part is its diffusion, as expected (cf. Theorem~\ref{thm: time reversal of diffusions}).

Much like we derived the Helmholtz decomposition of the drift by identifying the time-reversible and irreversible parts (see the proof of Proposition \ref{prop: helmholtz decomposition}), we proceed analogously at the level of the generator. Indeed, just as any matrix can be decomposed into a sum of antisymmetric and symmetric matrices, we may decompose the generator into a sum of antisymmetric and symmetric operators
\begin{equation}
\label{eq: gen symmetric and antisymmetric decomp}
\begin{split}
    \L &= \A + \S ,\quad \A := \left(\L-\bar \L\right)/2, \quad \S:= \left(\L+\bar \L\right)/2, \quad \dom \A= \dom\S= \dom\L\cap \dom\bar \L.
\end{split}
\end{equation}
By its analogous construction, this decomposition coincides with the Helmholtz decomposition; indeed, the symmetric operator recovers the time-reversible part of the dynamic while the antisymmetric operator recovers the time-irreversible part. In a nutshell, the Helmholtz decomposition of the generator is as follows
\begin{align*}
\label{eq: symmetric and antisymmetric unpacked}
   \L &= \A + \S , \quad \underbrace{\A= b_{\mathrm{irr}} \cdot \nabla}_{\text{Time-irreversible}} \quad \underbrace{ \S= b_{\mathrm{rev}} \cdot \nabla + D\nabla \cdot \nabla}_{\text{Time-reversible}},
\end{align*}
where the summands are symmetric and antisymmetric operators because they behave accordingly under the duality pairing:
\begin{align*}
    \underbrace{\langle \A f, g\rangle_\mu= -\langle f, \A g\rangle_\mu}_{\text{Antisymmetric}}, \quad \forall f,g \in \dom \A,\quad  \underbrace{\langle \S f, g\rangle_\mu= \langle f, \S g\rangle_\mu}_{\text{Symmetric}}, \quad \forall f,g \in \dom \S.
\end{align*}
Noting that $-\S$ is a positive semi-definite operator, we can go slightly further and decompose it into its square roots. To summarise:

\begin{proposition}
\label{proposition: hypocoercive decomposition of the generator}
We can rewrite the generator of the diffusion process as
$\operatorname L = \operatorname A-\operatorname \Sigma^*\operatorname \Sigma$ where $\A$ is the antisymmetric part of the generator, and $-\operatorname \Sigma^*\operatorname \Sigma$ is the symmetric part, as defined in \eqref{eq: gen symmetric and antisymmetric decomp}. Here $\cdot^*$ denotes the adjoint with respect to the duality pairing $\langle\cdot, \cdot\rangle_\mu$. The operators have the following functional forms: $\operatorname A f= b_{\mathrm{irr}}\cdot \nabla f$, $\sqrt 2 \operatorname \Sigma f =\sigma^\top \nabla f$, $\sqrt 2 \operatorname \Sigma^*g=- \nabla \log \rho \cdot \sigma g- \nabla \cdot(\sigma g) $.
\end{proposition}
A proof is provided in Appendix \ref{app: multiple perspectives on Helmholtz}.

\subsubsection{Helmholtz decomposition of the backward Kolmogorov equation}

We say that a real-valued function over the state-space of the process $f: \R^d \to \R$ is an \textit{observable}. Intuitively, this is a macroscopic quantity that can be measured or observed in a physical process (e.g., energy or pressure) when the (microscopic) process is not easily accessible. The evolution of an observable $f$ given that the process is prepared at a deterministic initial condition is given by $f_t(x) = \E [f(x_t)|x_0=x]$.

The \textit{backward Kolmogorov equation} is a fundamental equation describing a Markov process, as it encodes the motion of observables
\begin{equation*}
\label{eq: def backward Kolmogorov}
    \partial_t f_t  = \L f_t, \quad f_0 = f \in \dom \L.
\end{equation*}
In other words, $f_t= \E [f(x_t)|x_0=x]$ solves the equation. This highlights the central importance of the generator as providing a concise summary of the process.

The Helmholtz decomposition entails a decomposition of the backward Kolmogorov equation
\begin{equation}
\label{eq: helmholtz BKE}
    \partial_t u_t  = \A f_t + \S f_t = (\operatorname A - \operatorname \Sigma^*\operatorname \Sigma)f_t, \quad f_0 = f \in \dom \L.
\end{equation}
This decomposition is appealing, as it allows us to further characterise the contributions of the time-reversible and irreversible parts of the dynamic. Along the time-irreversible part of the backward Kolmogorov equation $\partial_t f_t=\operatorname A f_t$, the $L^2_\mu(\R^d)$-norm $\|\cdot\|_\mu$ is conserved. Indeed, since $\operatorname A$ is antisymmetric, $\langle \operatorname A f, f\rangle_\mu=0$ for every $f \in \dom \operatorname A$, and hence
$$
\partial_t\left\|f_t\right\|_\mu^2=2\left\langle \operatorname A f_t, f_t\right\rangle_\mu=0 .
$$
On the other hand, along the time-reversible part of the backward Kolmogorov equation generated by $-\operatorname \Sigma^*\operatorname \Sigma$, the $L^2_\mu(\R^d)$-norm is dissipated:
$$
\partial_t\left\|f_t\right\|_\mu^2=-2\left\langle \operatorname \Sigma^*\operatorname \Sigma f_t,  f_t\right\rangle_\mu=-2\left\|\operatorname \Sigma f_t\right\|_\mu^2 \leq 0.
$$
This offers another perspective on the fact that the time-irreversible part of the dynamic is conservative, while the time-reversible part is dissipative---of the $L_\mu^2(\R^d)$-norm.

Beyond this, hypocoercivity theory allows us to analyse the backward Kolmogorov equation, once one has written its Helmholtz decomposition. Hypocoercivity is a functional analytic theory developed to analyse abstract evolution equations of the form \eqref{eq: helmholtz BKE}, originally devised to systematically study the speed of convergence to stationary state of kinetic diffusion processes like the underdamped Langevin dynamics and the Boltzmann equation. As an important result, the theory provides sufficient conditions on the operators $\operatorname A,\operatorname \Sigma$ to ensure an exponentially fast convergence of the backward Kolmogorov equation to a fixed point \cite[Theorems 18 \& 24]{villaniHypocoercivity2009}. Dually, these convergence rates quantify the speed of convergence of the process to its stationary density from a given initial condition.

\subsubsection{GENERIC decomposition of the Fokker-Planck equation}


This perspective can also be examined directly from the Fokker-Planck equation. The Fokker-Planck equation is another fundamental equation describing a diffusion process: it encodes the evolution of the density of the process over time (when it exists). 
The Fokker-Planck equation is the $L^2(\R^d)$-dual to the backward Kolmogorov equation. It reads
\begin{align*}
    \partial_t \rho_t =\L'\rho_t= \nabla \cdot (-b \rho_t + \nabla \cdot(D \rho_t)),
\end{align*}
where $\L'$ is the adjoint of the generator with respect to the standard duality pairing $\langle\cdot, \cdot\rangle$; in other words $\langle \L'f , g \rangle =\langle f , \L g \rangle$ where $\langle f, g \rangle = \intr fg(x) \: \d x$.

The Helmholtz decomposition implies a decomposition of the Fokker-Planck equation into two terms: assuming for now that $\rho_t,\rho >0$ (e.g., if the diffusion is elliptic)
\begin{equation}
\label{eq: Fokker-Planck equation GENERIC form}
\begin{split}
    \partial_t \rho_t &= \nabla \cdot (-b\irr \rho_t) + \nabla \cdot(-b\rev \rho_t +\nabla \cdot(D \rho_t))\\
    &= \nabla \cdot (-b\irr \rho_t) + \nabla \cdot(-\rho_t \rho^{-1}\nabla \cdot(D \rho) +\nabla \cdot(D \rho_t))\\
    &= \nabla \cdot (-b\irr \rho_t) + \nabla \cdot\left(\rho_t D \nabla \log \frac{\rho_t}{\rho} \right).
\end{split}
\end{equation}
We will see that this decomposition casts the Fokker-Planck equation in pre-GENERIC form.

GENERIC (General Equations for Non-Equilibrium Reversible-Irreversible Coupling) is an important theory for analysing dynamical systems arising in non-equilibrium statistical physics like the Fokker-Planck equation. The framework rose to prominence through the seminal work of Ottinger \cite{ottingerEquilibriumThermodynamics2005} and was later developed by the applied physics and engineering communities. Only recently, the framework developed into a rigorous mathematical theory. We refer to \cite{duongNonreversibleProcessesGENERIC2021} for mathematical details. The following Proposition shows how we can rewrite the Fokker-Planck equation in pre-GENERIC form:

\begin{align}
\label{eq: pre-GENERIC equation}
    &\partial_t \rho_t = \underbrace{\operatorname W(\rho_t)}_{\text{time-irreversible}} - \underbrace{\operatorname M_{\rho_t}\left(\d \operatorname{H}[\rho_t \mid \rho ]\right)}_{\text{time-reversible}},
\end{align}

\begin{proposition}[GENERIC decomposition of the Fokker-Planck equation]
\label{prop: GENERIC decomposition of the Fokker-Planck equation}
The Fokker-Planck equation \eqref{eq: Fokker-Planck equation GENERIC form} is in pre-GENERIC form \eqref{eq: pre-GENERIC equation}, with
\begin{align*}
    &W(\rho_t) = \nabla \cdot (-b\irr \rho_t), \quad -M_{\rho_t}\left( \d \operatorname{H}[\rho_t \mid \rho ]\right)=\nabla \cdot\left(\rho_t D \nabla \log \frac{\rho_t}{\rho} \right),\\
     &M_{\rho_t}(\xi) = \Sigma'(\rho_t \Sigma \xi) = - \nabla \cdot (\rho_t D \nabla \xi),\quad \d \operatorname{H}[\rho_t \mid \rho ]= \log \frac{\rho_t}{\rho}+1,
\end{align*}
where $\H[\cdot \mid \cdot]$ is the relative entropy, $\cdot '$ denotes the adjoint under the standard duality pairing $\langle\cdot, \cdot\rangle$, $\d$ is the Fréchet derivative in $L^2(\R^d)$, and $W, M_{\rho_t}$ satisfy the following relations:
\begin{itemize}
    \item \emph{Orthogonality}: $\langle \operatorname W(\rho_t), \d \operatorname{H}[\rho_t \mid \rho ] \rangle =0$,
    \item \emph{Semi-positive definiteness}: $\langle \operatorname M_{\rho_t}(h), g\rangle=\langle h, \operatorname M_{\rho_t}(g)\rangle$, and $\langle \operatorname M_{\rho_t}(g), g\rangle \geq 0$.
\end{itemize}
\end{proposition}

A proof is provided in Appendix \ref{app: multiple perspectives on Helmholtz}.

Writing the Fokker-Planck equation in pre-GENERIC form \eqref{eq: pre-GENERIC equation} explicits the contributions of the time-reversible and time-irreversible parts at the level of density dynamics. Indeed, the relative entropy functional $\operatorname{H}[\rho_t \mid \rho ]$ is conserved along the time-irreversible part of the Fokker-Planck equation $\partial_t \rho_t = \operatorname W(\rho_t)$
\begin{align*}
    \frac{d \operatorname{H}[\rho_t \mid \rho ]}{d t}=\left\langle \partial_t \rho_t, \d \operatorname{H}[\rho_t \mid \rho ]\right\rangle=\langle \operatorname W(\rho_t), \d \operatorname{H}[\rho_t \mid \rho ]\rangle =0.
\end{align*}
Contrariwise, the relative entropy is dissipated along the time-reversible part of the equation
\begin{align*}
    \frac{d \operatorname{H}[\rho_t \mid \rho ]}{d t}=\left\langle \partial_t \rho_t, \d \operatorname{H}[\rho_t \mid \rho ]\right\rangle=-\left\langle\operatorname M_{\rho_t}\left( \d \operatorname{H}[\rho_t \mid \rho ]\right), \d \operatorname{H}[\rho_t \mid \rho ] \right\rangle \leq 0.
\end{align*}
Aggregating these results, we obtain the well-known fact that the relative entropy with respect to the stationary density is a Lyapunov function of the Fokker-Planck equation; a result sometimes known as de Bruijn's identity or Boltzmann's H-theorem \cite[Proposition 1.1]{chafaiEntropiesConvexityFunctional2004}.

\section{The $e_p$ of stationary diffusions}
\label{sec: the ep of stationary diffusions}

We are now ready to investigate the entropy production of stationary diffusions. First, we give sufficient conditions guaranteeing the mutual absolute continuity of the forward and time-reversed path space measures and compute the entropy production rate. Second, we demonstrate that when these conditions fail the entropy production is infinite.

\subsection{Regular case}
\label{sec: epr regularity}

\begin{figure}[t!]
    \centering
    \includegraphics[width= 0.45\textwidth]{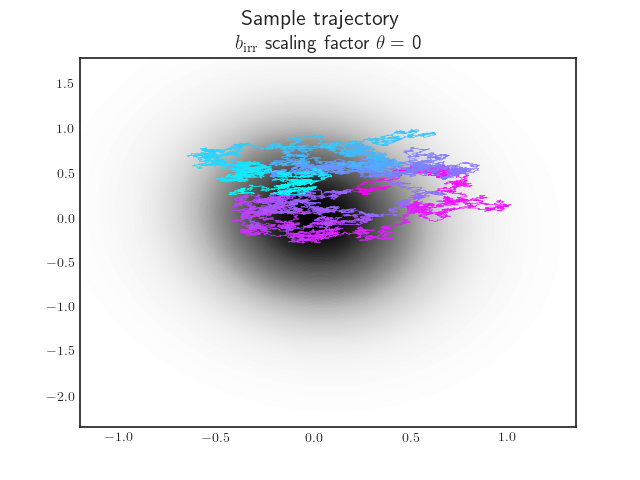}
    \includegraphics[width= 0.45\textwidth]{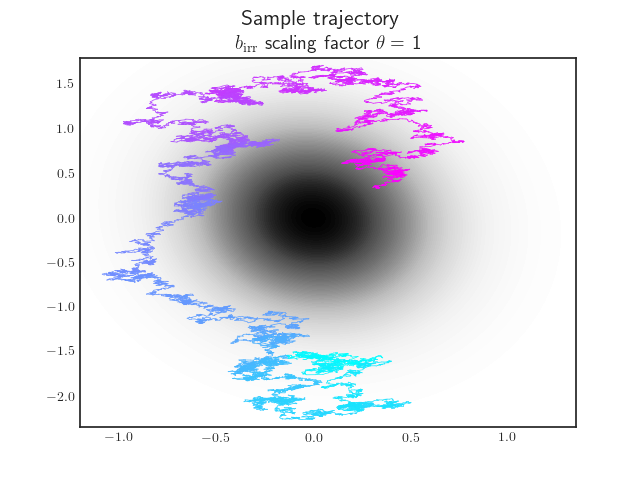}
    \includegraphics[width= 0.45\textwidth]{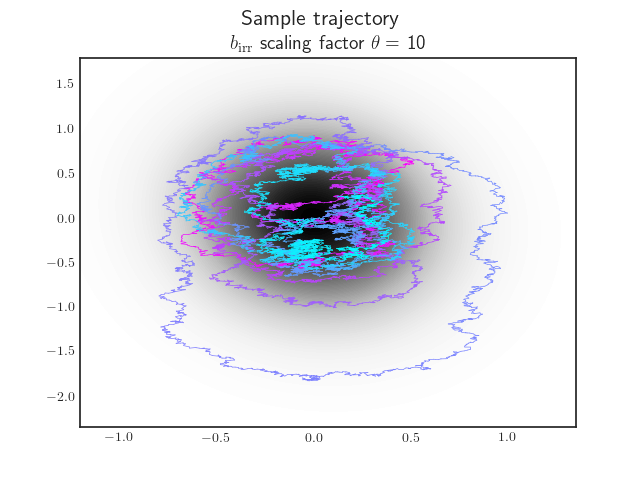}
    \includegraphics[width= 0.45\textwidth]{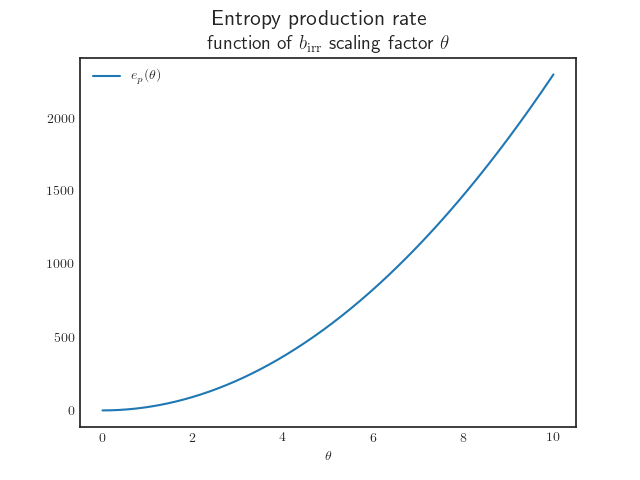}
    \caption{\footnotesize\textbf{Entropy production as a function of time-irreversible drift.} This figure illustrates the behaviour of sample paths and the entropy production rate as one scales the irreversible drift $b_{\mathrm{irr}}$ by a parameter $\theta$. The underlying process is a two-dimensional Ornstein-Uhlenbeck process, for which exact sample paths and entropy production rate are available (Section \ref{sec: OU process}). The heat map represents the density of the associated Gaussian steady-state. One sees that a non-zero irreversible drift induces circular, wandering behaviour around the contours of the steady-state, characteristic of a non-equilibrium steady-state (top right and bottom left). This is accentuated by increasing the strength of the irreversible drift. The entropy production rate measures the amount of irreversibility of the stationary process. It grows quadratically as a function of the irreversible scaling factor $\theta$ (bottom right). When there is no irreversibility (top left), we witness an equilibrium steady-state. This is characterised by a vanishing entropy production (bottom right).}
    \label{fig: EPR as a function of gamma}
\end{figure}

\begin{theorem}
\label{thm: epr regular case simple}
Let an Itô SDE \eqref{eq: Ito SDE} with coefficients satisfying Assumption \ref{ass: coefficients time reversal}. Assume that the solution $(x_t)_{t\in [0,T]}$ is stationary with respect to a density $\rho$ satisfying Assumption \ref{ass: steady-state time reversal}. Denote by $b_{\mathrm{irr}}$ the time-irreversible part of the drift (Proposition \ref{prop: helmholtz decomposition}), and by $\cdot^-$ the Moore-Penrose matrix pseudo-inverse. Suppose that:
\begin{enumerate}
    \item \label{item: epr regular drift image of volatility} For $\rho$-almost every $x \in \R^d$,
        $b_{\mathrm{irr}}(x) \in \im\sigma(x)$, and
    \item \label{item: continuity} The product $\sigma^- b_{\mathrm{irr}}: \R^d \to \R^m$ is Borel measurable (e.g., if $\sigma^- b_{\mathrm{irr}}$ is continuous), and
    \item 
         \label{item: epr bounded}
        $\int_{\R^d} b_{\mathrm{irr}}^\top D^- b_{\mathrm{irr}} \rho(x)\d x <+\infty$.
\end{enumerate}
Denote by $\p_{[0, T]}, \bar \p_{[0, T]}$ the path space measures of the forward and time-reversed diffusions, respectively, on $C([0,T], \R^d)$ (Definition \ref{def: path space measure}). Then,
\begin{enumerate}
    \item The path-space measures are equivalent $\p_{[0, T]} \sim \bar \p_{[0, T]}$, and 
    \item $e_p = \int_{\R^d} b_{\mathrm{irr}}^\top D^- b_{\mathrm{irr}} \rho(x)\d x$.
\end{enumerate} 
\end{theorem}

Under the assumptions of Theorem \ref{thm: epr regular case simple}, the $e_p$ is a quadratic form of the time-irreversible drift, see Figure~\ref{fig: EPR as a function of gamma}. 

A proof of Theorem \ref{thm: epr regular case simple} is provided in Appendix \ref{app: epr regularity}. The idea of the proof is simple: in the elliptic case, the approach follows \cite[Chapter 4]{jiangMathematicalTheoryNonequilibrium2004} with some generalisations. In the non-elliptic case, the condition $b_{\mathrm{irr}}(x) \in \im\sigma(x)$ intuitively ensures that the solution to the SDE, when initialised at any point, evolves on a sub-manifold of $\R^d$ and is elliptic on this manifold (e.g., Figure \ref{fig: OU process b in Im sigma}). The pseudo-inverse of the diffusion tensor is essentially the inverse on this sub-manifold. Thus, a proof analogous to the elliptic case, but on the sub-manifold (essentially replacing all matrix inverses by pseudo-inverses and making sure everything still holds), shows that the path space measures of the forward and backward processes initialised at a given point are equivalent---and Girsanov's theorem gives us their Radon-Nykodym derivative. Finally, Proposition \ref{prop: aggregating local ep} gives us the usual formula for the entropy production rate but with the matrix inverse replaced by the pseudo-inverse. Please see Section \ref{sec: geom} for a geometric discussion of this proof.

Suppose either of assumptions \ref{item: continuity}, \ref{item: epr bounded} of Theorem \ref{thm: epr regular case simple} do not hold. Then we have the following more general (and technical) result:

\begin{theorem}
\label{thm: epr regular case general}
Let $(\Omega, \mathscr{F}, P)$ be a probability space and $\left(w_t\right)_{t \geqslant 0}$ a standard Wiener process on $\mathbb{R}^{m}$, with respect to the filtration $(\mathscr{F}_{t})_{t\geq 0}$ \cite[Definition 2.1.12]{prevotConciseCourseStochastic2008}. 
Consider the Itô SDE \eqref{eq: Ito SDE} with coefficients satisfying Assumption \ref{ass: coefficients time reversal}. Consider its unique strong solution $(x_t)_{t\in [0,T]}$ with respect to the given Brownian motion on the filtered probability space $(\Omega, \mathscr{F}, (\mathscr{F}_{t})_{t\geq 0}, P)$. Assume that the solution is stationary with respect to a density $\rho$ satisfying Assumption \ref{ass: steady-state time reversal}. Denote by $b_{\mathrm{irr}}$ the time-irreversible part of the drift (Proposition \ref{prop: helmholtz decomposition}), and by $\cdot^-$ the Moore-Penrose matrix pseudo-inverse. Suppose that:
\begin{enumerate}
    \item \label{item: epr regular general drift image of volatility} For $\rho$-almost every $x \in \R^d$, $b_{\mathrm{irr}}(x) \in \im\sigma(x)$, and
    \item \label{item: epr regular general adapted} $\sigma^- b_{\mathrm{irr}}(x_t)$ is an $\mathscr F_t$-adapted process (e.g., $\sigma^- b_{\mathrm{irr}}: \R^d \to \R^m$ is Borel measurable), and
    \item \label{item: epr regular general expectation} The following holds 
    \begin{equation}
    \label{eq: exponential condition}
        \E_P\left[Z_T \right]=1,  Z_T :=\exp \left( -2 \int_0^T \langle \sigma^- b_{\mathrm{irr}}(x_t), dw_t \rangle + |\sigma^- b_{\mathrm{irr}}(x_t) |^2 dt\right).
    \end{equation}
\end{enumerate}
Denote by $\p_{[0, T]}, \bar \p_{[0, T]}$ the path space measures on $C([0,T], \R^d)$ of the forward and time-reversed diffusions, respectively (Definition \ref{def: path space measure}). Then,
\begin{enumerate}
    \item The path-space measures are equivalent $\p_{[0, T]} \sim \bar \p_{[0, T]}$, and 
    \item $e_p = \int_{\R^d} b_{\mathrm{irr}}^\top D^- b_{\mathrm{irr}} \rho(x)\d x$.
\end{enumerate}
\end{theorem}

A proof is provided in Appendix \ref{app: epr regularity}. The proof is similar to that of Theorem \ref{thm: epr regular case simple}, but much shorter, since \eqref{eq: exponential condition} allows one to apply Girsanov's theorem directly; in contrast, a large part of the proof of Theorem \ref{thm: epr regular case simple} is dedicated to showing that indeed, a version of Girsanov's theorem can be applied.




In relation to assumption \ref{item: epr regular general adapted} of Theorem \ref{thm: epr regular case general}, note that if a matrix field $\sigma : \R^d \to \R^{d\times m}$ is Borel measurable, then its pseudo-inverse $\sigma^- : \R^d \to \R^{m \times d}$ is also Borel measurable. We now give sufficient conditions for the exponential condition \eqref{eq: exponential condition}. 

\begin{proposition}
\label{prop: suff conds exp cond}
Consider a stochastic process $(x_t)_{t\in [0,T]}$ on the probability space $(\Omega, \mathscr{F}, (\mathscr{F}_{t})_{t\geq 0}, P)$, which is stationary at the density $\rho$. Assume that $\sigma^-b_{\mathrm{irr}}(x_t)$ is $\mathscr F_t$-adapted. Then, either of the following conditions implies \eqref{eq: exponential condition}:
\begin{enumerate}
    \item \label{item: suff cond martingale} $Z_t=\exp \left(-2{\int_{0}^{t}\langle \sigma^- b_{\mathrm{irr}}(x_s), d w_s\rangle+|\sigma^- b_{\mathrm{irr}}(x_s)|^{2} d s}\right), t\in [0,T]$ is a martingale on the probability space $(\Omega, \mathscr{F},\{\mathscr{F}_{t}\}_{t\geq 0}, P)$.
    \item \label{item: suff cond novikov} $\mathbb{E}_{P}\left(e^{2\int_{0}^{T}|\sigma^- b_{\mathrm{irr}}(x_t)|^{2} d t}\right)<+\infty$ (Novikov's condition).
    \item \label{item: suff cond delta exp condition} There exists $\delta>0$ such that $ \mathbb{E}_{\rho}\left(e^{\delta|\sigma^- b_{\mathrm{irr}}(x)|^{2}}\right)<+\infty$.
    \item \label{item: suff cond Kazamaki} $\sup_{t \in [0,T]} \mathbb{E}_P\left[\exp \left(- \int_0^t \langle\sigma^- b_{\mathrm{irr}}(x_s), d w_s\rangle\right)\right]<+\infty$ (Kazamaki's criterion).
    \item \label{item: suff cond Dellacherie meyer} There exists $K<1$ s.t. for all $t \in [0,T]$
    \begin{equation*}
        \E_P\left[\left. 2\int_t^T\left|\sigma^- b_{irr}(x_t)\right|^2 ds \:\right| \mathscr F_t\right]\leq K.
    \end{equation*}
    \item \label{item: suff cond tail} The tail of $|\sigma^- b_{\mathrm{irr}}(x)|^{2}, x\sim \rho$ decays exponentially fast, i.e., there exists positive constants $c,C, R > 0$ such that for all $r> R$ 
    \begin{align}
    \label{eq: exponential boundedness of tail}
         P(|\sigma^- b_{\mathrm{irr}}(x)|^{2}> r)\leq Ce^{-c r}.
    \end{align}
\end{enumerate}
Furthermore, (\ref{item: suff cond novikov} or \ref{item: suff cond Kazamaki}) $\Rightarrow$ \ref{item: suff cond martingale}; \ref{item: suff cond Dellacherie meyer} $\Rightarrow$ \ref{item: suff cond novikov}; and \ref{item: suff cond tail} $\Rightarrow$ \ref{item: suff cond delta exp condition}.
\end{proposition}

A proof is provided in Appendix \ref{app: epr regularity}. 

\subsection{Singular case}
\label{sec: ep singularity}

When the time-irreversible part of the drift is not always in the range of the volatility matrix field, we have a different result.

\begin{theorem}
\label{thm: epr singular}
Suppose that the Itô SDE \eqref{eq: Ito SDE} satisfies Assumption \ref{ass: coefficients time reversal} and that the volatility is twice continuously differentiable $\sigma\in C^2(\R^d, \R^{d\times m})$. Furthermore suppose that the solution $(x_t)_{t\in [0,T]}$ is stationary with respect to a density $\rho$ satisfying Assumption \ref{ass: steady-state time reversal}. Denote by $\p_{[0, T]}, \bar \p_{[0, T]}$ the path space measures on $C([0,T], \R^d)$ of the forward and time-reversed processes, respectively (Definition \ref{def: path space measure}). If $b_{\mathrm{irr}}(x) \in \im\sigma(x)$ does not hold for $\rho$-a.e. $x \in \R^d$, then
    \begin{align*}
        \p_{[0, T]} \perp \bar \p_{[0, T]} \text{ and } e_p = +\infty.
    \end{align*}
\end{theorem}

A proof is provided in Appendix \ref{app: epr singular}. The proof uses a version of the Stroock-Varadhan support theorem to show that there are paths that can be taken by the forward diffusion process that cannot be taken by the backward diffusion process---and vice-versa. Specifically, when considering the two processes initialised at a point $x \in \R^d$ where $b_{\mathrm{irr}}(x) \notin \im\sigma(x)$, we can see that the derivatives of their respective possible paths at time $0$ span different tangent sub-spaces at $x$. Thus the path space measures $\p^{x}_{[0,T]} , \bar \p^{x}_{[0,T]}$ are mutually singular. Since such $x$ occur with positive probability under the stationary density, it follows that the path space measures of the forward and time-reversed stationary processes $\p^{x}_{[0,T]} , \bar \p^{x}_{[0,T]}$ are also mutually singular. Finally, the relative entropy between two mutually singular measures is infinity, hence the $e_p$ must be infinite.

By Theorem \ref{thm: epr singular} and \eqref{eq: drift image of volatility additive noise} we can readily see that the underdamped \eqref{eq: underdamped Langevin dynamics}
and generalised Langevin \cite[eq. 8.33]{pavliotisStochasticProcessesApplications2014} processes in phase-space have infinite entropy production. Expert statistical physicists will note that this contrasts with previous results in the literature stating that these diffusions have finite entropy production. There is no contradiction as physicists usually add an additional operator to the definition of the entropy production in these systems (Remark \ref{rem: physical relevance}). While obviously informative of the underlying process, statistical physicists should take the results of Theorems \ref{thm: epr regular case simple}, \ref{thm: epr regular case general} and \ref{thm: epr singular} to be physically relevant to entropy production insofar as the definition of entropy production we adopted (Definition \ref{def: epr}) is physically meaningful for the system at hand. What if it is not? We will return to this in the discussion (Section \ref{sec: discussion generalised non-reversibility}).

In contrast to the underdamped and generalised Langevin equations, there exist hypoelliptic, non-elliptic diffusions with finite entropy production. 
For example, consider the following volatility matrix field
\begin{align*}
    \sigma(x,y,z) = \begin{bmatrix} x & 1  \\1&1\\ 0&1 
    \end{bmatrix}.
\end{align*}
By Hörmander's theorem \cite[Theorem 1.3]{hairerMalliavinProofOrmander2011}, for any smooth, confining (e.g., quadratic) potential $V: \R^3 \to \R$, the process solving the SDE
\begin{equation*}
   dx_t =- D\nabla V(x_t)dt + \nabla \cdot D(x_t)dt + \sigma(x_t) dw_t 
\end{equation*}
is hypoelliptic and non-elliptic. Furthermore, it is stationary and time-reversible at the Gibbs density $\rho(x)\propto \exp(-V(x))$.

\section{Examples and $e_p$ of numerical simulations}
\label{sec: illustrating results epr}

We illustrate these results for linear diffusion processes, underdamped Langevin dynamics and their numerical simulations.

\subsection{Linear diffusion processes}
\label{sec: OU process}

Given matrices $B\in \R^{d \times d}, \sigma \in \R^{d \times m}$, and a standard Brownian motion $(w_t)_{t\in [0, +\infty)}$ on $\R^m$, consider a linear diffusion process (i.e., a multivariate Ornstein-Uhlenbeck process)
\begin{equation}
\label{eq: OU process}
\begin{split}
    dx_t &= b(x_t)dt +\sigma(x_t) dw_t, \quad b(x)= -Bx, \quad \sigma(x) \equiv \sigma.
\end{split}
\end{equation}
This process arises, for example, in statistical physics as a model of the velocity of a massive Brownian particle subject to friction \cite{uhlenbeckTheoryBrownianMotion1930}; it covers the case of interacting particle systems when the interactions are linear in the states (e.g., the one dimensional ferromagnetic Gaussian spin model \cite{godrecheCharacterisingNonequilibriumStationary2019}); or when one linearises the equations of generic diffusion processes near the stationary density. 

By solving the linear diffusion process (e.g., \cite[Section 2.2]{godrecheCharacterisingNonequilibriumStationary2019}) one sees that the solution can be expressed as a linear operation on Brownian motion---a Gaussian process---thus the process must itself be Gaussian, and its stationary density as well (when it exists). Consider a Gaussian density $\rho$
\begin{align*}
    \rho (x) = \mathcal N (x;0, \Pi^{-1}), \quad -\log \rho(x) = \frac 1 2 x^\top \Pi x,
\end{align*}
where $\Pi \in \R^{d\times d}$ is the symmetric positive definite \emph{precision} matrix. By the Helmholtz decomposition (Propositions \ref{prop: helmholtz decomposition} \& \ref{prop: characterisation of irreversible drift}), $\rho$ is a stationary density if and only if we can decompose the drift as follows:
\begin{align*}
    b = b_{\mathrm{rev}} + b_{\mathrm{irr}}, \quad b_{\mathrm{rev}}(x)= -D\Pi x, \quad b_{\mathrm{irr}}(x)= -Q\Pi x,
\end{align*}
where $Q=-Q^\top \in \R^{d\times d}$ is an arbitrary antisymmetric matrix, and, recall $D= \sigma \sigma^\top/2\in \R^{d\times d}$ is the diffusion tensor. In particular, the drift of the forward and the time-reversed dynamic, are, respectively,
\begin{align*}
    b(x) = -Bx, \quad B = (D+ Q)\Pi,\quad 
    \bar b(x) = -Cx, \quad C:= (D-Q)\Pi.
\end{align*}



Suppose that $b_{\mathrm{irr}}(x)\in \im \sigma $ for any $x\in\R^d$. By definiteness of $\Pi$ this is equivalent to $\im Q \subseteq \im \sigma$. By Theorem \ref{thm: epr regular case simple}, and applying the trace trick to compute the Gaussian expectations of a bilinear form, we obtain\footnote{To obtain the last equality we used $\tr(D^- D \Pi Q)=\tr(\Pi Q D^- D)= - \tr(D D^- Q \Pi) $. By standard properties of the pseudo-inverse $D D^-$ is the orthogonal projector onto $\im D= \im \sigma$. Thus, $\im Q \subseteq \im \sigma$ implies $D D^-Q=Q$. Finally, the trace of a symmetric matrix $\Pi$ times an antisymmetric matrix $Q$ vanishes.}
\begin{equation}
\label{eq: ep OU regular}
\begin{split}
     e_p &= \int_{\R^d} b_{\mathrm{irr}}^\top D^- b_{\mathrm{irr}} \rho(x)\d x = -\int_{\R^d} x^\top \Pi Q D^- Q\Pi x \,\rho(x)\d x \\
     &= -\tr(\Pi Q D^- Q\Pi\Pi^{-1})
     = -\tr(D^- Q \Pi Q) \\
    &=-\tr(D^- B Q)+\underbrace{\tr(D^- D \Pi Q)}_{=0}=-\tr(D^- B Q).   
\end{split}
\end{equation}
This expression for the entropy production is nice as it generalises the usual formula to linear diffusion processes to degenerate noise, simply by replacing inverses with pseudo-inverses, cf. \cite[eqs. 2.28-2.29]{godrecheCharacterisingNonequilibriumStationary2019} and \cite{mazzoloNonequilibriumDiffusionProcesses2023,buissonDynamicalLargeDeviations2022}.

Contrariwise, suppose that $\im Q \not \subseteq \im \sigma$. Then by Theorem \ref{thm: epr singular},
\begin{align}
\label{eq: ep OU singular}
    e_p = +\infty.
\end{align}

\subsubsection{Exact numerical simulation and entropy production rate}

\begin{figure}[t!]
    \centering
    \includegraphics[width= 0.45\textwidth]{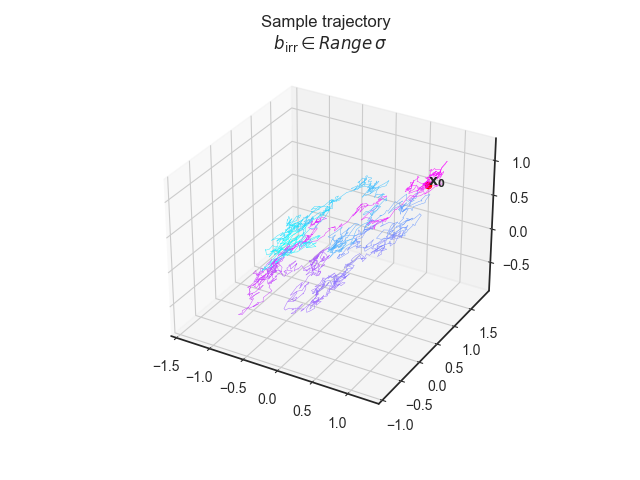}
    \includegraphics[width= 0.45\textwidth]{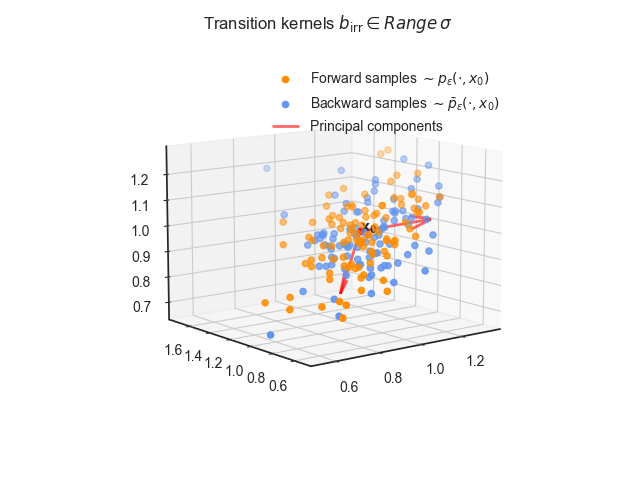}
    \includegraphics[width= 0.5\textwidth]{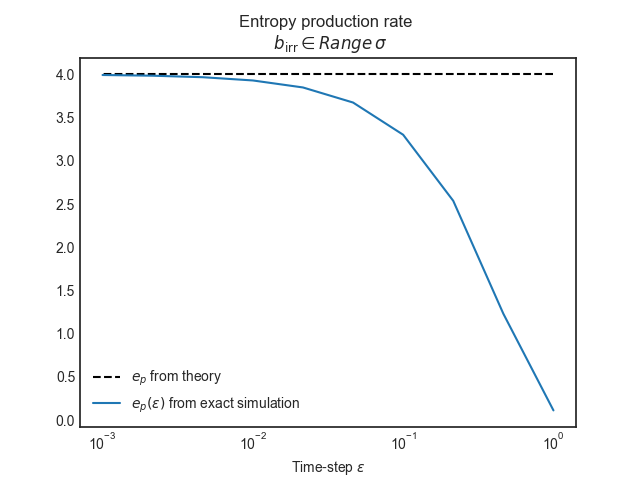}
    \caption{\footnotesize\textbf{Exact simulation of linear diffusion process with $b_{\mathrm{irr}}(x) \in \im\sigma$.} This figure considers an OU process in 3d space driven by degenerate noise, i.e., $\rank \sigma <3$. The coefficients are such that $\sigma=Q$ are of rank $2$. In particular, $b_{\mathrm{irr}}(x) \in \im\sigma$ holds for every $x$. The process is not elliptic nor hypoelliptic, but it is elliptic over the subspace in which it evolves. The upper-left panel shows a sample trajectory starting from $x_0=(1,1,1)$. The upper-right panel shows samples from different trajectories after a time-step $\e$. There are only two principal components to this point cloud as the process evolves on a two dimensional subspace. In the bottom panel, we verify the theoretically predicted value of $e_p$ by evaluating the entropy production of an exact simulation $e_p(\e)$ with time-step $\e$. As predicted, we recover the true $e_p$ in the infinitesimal limit as the time-step of the exact simulation tends to zero $\e \to 0$. Furthermore, since the process is elliptic in its subspace, the entropy production is finite.}
    \label{fig: OU process b in Im sigma}
\end{figure}

\begin{figure}[t!]
    \centering
    \includegraphics[width= 0.45\textwidth]{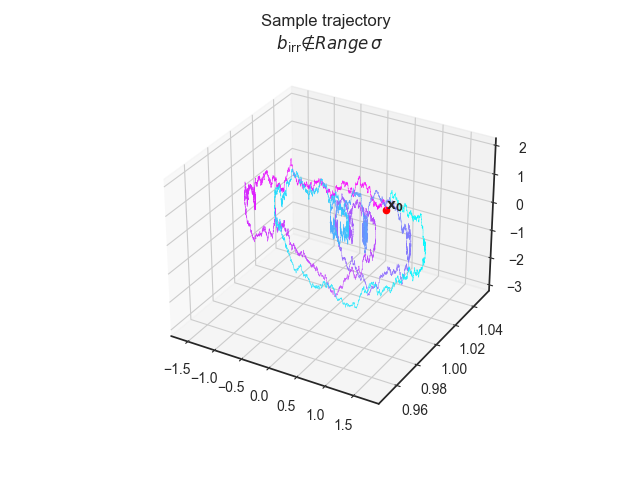}
    \includegraphics[width= 0.45\textwidth]{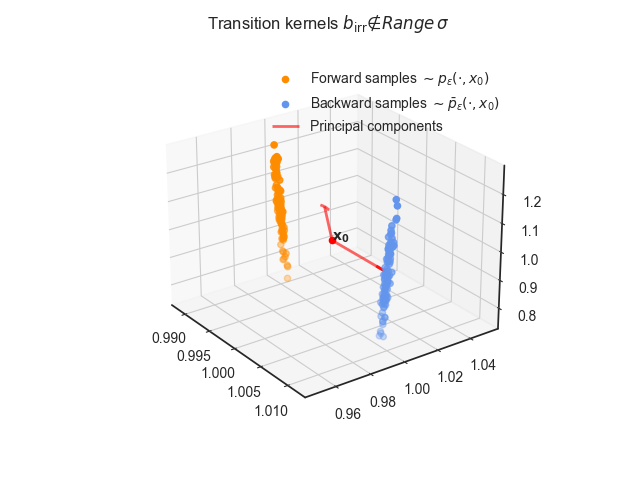}
    \includegraphics[width= 0.5\textwidth]{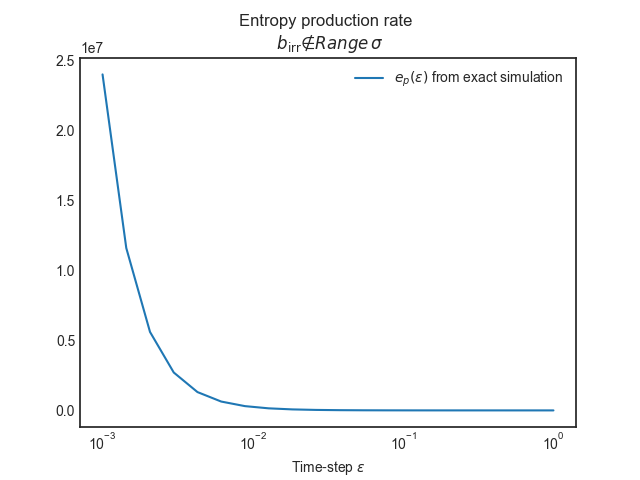}
    \caption{\footnotesize\textbf{Exact simulation of linear diffusion process with $b_{\mathrm{irr}}(x) \not\in \im\sigma$.} This figure considers an OU process in 3d space driven by degenerate noise. The coefficients are such that $\im b_{\mathrm{irr}}$ is two-dimensional while $\im\sigma$ is one-dimensional, and such that the process does not satisfy Hörmander's hypoellipticity condition. As such the process is hypoelliptic on a two-dimensional subspace; see a sample trajectory in the upper-left panel. By hypoellipticity its transition kernels are equivalent in the sense of measures, although far removed: On the upper right panel we show samples from different trajectories after a time-step $\e$. There are only two principal components to this data-cloud as the process evolves on a two dimensional subspace. In the bottom panel, we verify the theoretically predicted $e_p$ by evaluating the entropy production of an exact simulation $e_p(\e)$ with time-step $\e$. As predicted, we recover $e_p=+\infty$ in the infinitessimal limit as the time-step of the exact simulation tends to zero $\e \downarrow 0$.  This turns out to be as the transition kernels of the forward and time-reversed processes become more and more mutually singular as the time-step decreases.}
    \label{fig: OU process b not in Im sigma}
\end{figure}

Linear diffusion processes can be simulated exactly as their transition kernels are known. Indeed, by solving the process, one obtains the transition kernels of the Markov semigroup as a function of the drift and volatility \cite[Theorem 9.1.1]{lorenziAnalyticalMethodsMarkov2006}. The forward and time-reserved transition kernels are the following:
\begin{equation}
\label{eq: OU transition kernels}
\begin{split}
    p_\e(\cdot,x) &= \mathcal N(e^{-\e B}x, S_\e),\quad S_\e:= \int_0^\e e^{-tB}\sigma \sigma ^\top e^{-tB^\top} \d t, \\
    \bar p_\e(\cdot,x) 
    &= \mathcal N (e^{-\e C}x, \bar S_\e),\quad \bar S_\e
    := \int_0^\e e^{-t C}\sigma \sigma^\top e^{-tC^\top}\d t.
\end{split}
\end{equation}
Sampling from the transition kernels allows one to simulate the process exactly, and offers an alternative way to express the entropy production rate. Recall from Proposition \ref{prop: epr transition kernels} that the $e_p$ is the infinitesimal limit of the entropy production rate of an exact numerical simulation $e_p(\e)$ with time-step $\e$
    \begin{align*}
        e_p &= \lim_{\e\downarrow 0} e_p(\e), \quad e_p(\e) = \frac{1}{\e} \E_{x\sim \rho}[\H[p_\e(\cdot,x)\mid \bar p_\e(\cdot,x)]].
    \end{align*}
We can leverage the Gaussianity of the transition kernels to compute the relative entropy and obtain an alternative formula for the entropy production rate:

\begin{lemma}
\label{lemma: limit ep formula OU process}
The entropy production rate of the stationary linear diffusion process can also be expressed as 
\begin{equation}
\label{eq: epr hypo OU process lim}
\begin{split}
    e_p &= \lim_{\e \downarrow 0} e_p(\e),\\
    e_p(\e)&= \frac{1}{2\e} \left[\tr(\bar S_\e^{-}S_\e)-\rank \sigma + \log \frac{\det^*(\bar S_\e)}{\det^*(S_\e)}\right. \\
    &\left.+\tr \left( \Pi^{-1} (e^{-\e C}-e^{-\e B})^\top \bar S_\e^{-}(e^{ -\e C}-e^{-\e B})\right)\right],
\end{split}
\end{equation}
where $\cdot^-$ is the Moore-Penrose pseudo-inverse and $\det^*$ is the pseudo-determinant.
\end{lemma}
A proof is provided in Appendix \ref{app: exact simul OU process}. Computing the limit \eqref{eq: epr hypo OU process lim} analytically, gives us back \eqref{eq: ep OU regular}, \eqref{eq: ep OU singular}, however, we will omit those details here. For our purposes, this gives us a way to numerically verify the value of $e_p$ that was obtained from theory. See Figures \ref{fig: OU process b in Im sigma} and \ref{fig: OU process b not in Im sigma} for illustrations.

\subsection{Underdamped Langevin dynamics}
\label{sec: underdamped numerical simulations}

In this sub-section, we consider the entropy production rate of underdamped Langevin dynamics and its numerical simulations. Recall that the $e_p$ we compute here is defined \textit{without} an additional momentum flip operator on the path space measure of the time-reversed process (i.e., \eqref{eq: def epr} and not \eqref{eq: rem def gen epr}), and may be a distinct quantity from the entropy production that physicists usually consider in such systems (see the discussion in Section \ref{sec: discussion generalised non-reversibility}).

Consider a Hamiltonian $H(q,p)$ function of positions $q\in \R^n$ and momenta $ p \in \R^n$ of the form
\begin{equation}
H(q, p)=V(q)+\frac{1}{2} p^\top M^{-1} p
\end{equation}
for some smooth potential function $V: \R^n \to \R$ and diagonal mass matrix $M \in \R^{n\times n}$.

The underdamped Langevin process is the solution to the SDE \cite[eq 2.41]{roussetFreeEnergyComputations2010}
\begin{equation}
\begin{cases}
d q_{t} = M^{-1}p_{t} d t \\
d p_{t} =-\nabla V\left(q_{t}\right) d t-\gamma M^{-1} p_{t} d t+\sqrt{2 \gamma \beta^{-1}} d w_{t}
\end{cases}
\end{equation}
for some friction coefficient $\gamma >0$. This process arises in statistical physics, as a model of a particle coupled to a heat bath \cite{rey-belletOpenClassicalSystems2006}, \cite[Chapter 8]{pavliotisStochasticProcessesApplications2014}; in Markov chain Monte-Carlo as an accelerated sampling scheme \cite{maThereAnalogNesterov2021,barpGeometricMethodsSampling2022a}; and also as a model of interacting kinetic particles.

The invariant density, assuming it exists, is \cite[Section 2.2.3.1]{roussetFreeEnergyComputations2010}
\begin{align*}
    \rho(q,p)=\frac 1 Z e^{-\beta H(q,p)}=\frac 1 Z e^{-\beta V(q)}e^{-\frac{\beta}{2} p^\top M^{-1} p}.
\end{align*}

Since the noise is additive the Itô interpretation of the SDE coincides with the Stratonovich interpretation, thus the irreversible drift is in the range of the volatility if and only if the drift is in the range of the volatility \eqref{eq: drift image of volatility additive noise}. Observe that when the momentum is non-zero $p\neq 0$ the drift is not in the image of the volatility: in the $q$ components the drift is non-zero while the volatility vanishes. Since $p \neq 0$ has full measure under the stationary density $\rho(q,p)$ we obtain, from Theorem \ref{thm: epr singular}, 
\begin{align}
\label{eq: ep underdamped}
    e_p = +\infty.
\end{align}

\begin{figure}[t!]
    \centering
    \includegraphics[width= 0.45\textwidth]{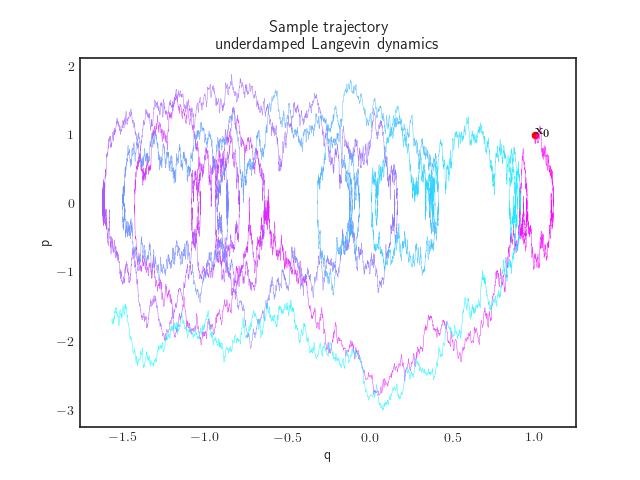}
    \includegraphics[width= 0.45\textwidth]{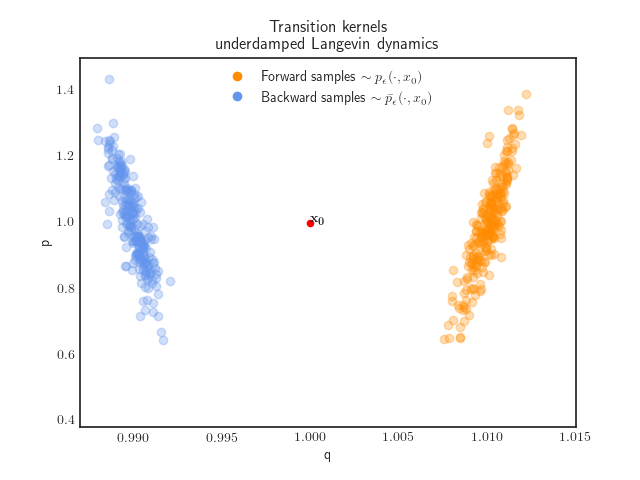}
    \includegraphics[width= 0.5\textwidth]{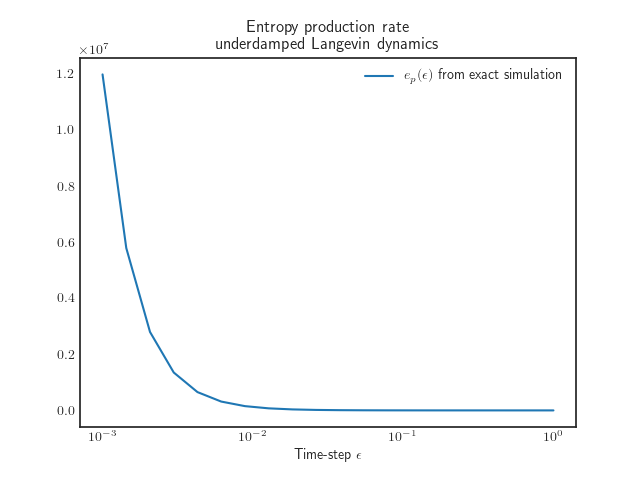}
    \caption{\footnotesize\textbf{Exact simulation of underdamped Langevin dynamics.} This figure plots underdamped Langevin dynamics in a quadratic potential. Here, the process is two dimensional, i.e., positions and momenta evolve on the real line. We exploit the fact that underdamped Langevin in a quadratic potential is an Ornstein-Uhlenbeck process to simulate sample paths exactly. The choice of parameters was: $V(q)=q^2/2, M=\gamma =1$. The upper left panel plots a sample trajectory. One observes that the process is hypoelliptic: it is not confined to a prespecified region of space, cf. Figures \ref{fig: OU process b in Im sigma}, \ref{fig: OU process b not in Im sigma}, even though random fluctuations affect the momenta only. The upper right panel plots samples of the forward and time-reversed processes after a time-step of $\e$. In the bottom panel, we verify the theoretically predicted $e_p$ by evaluating the entropy production of an exact simulation $e_p(\e)$ with time-step $\e$. As predicted, we recover $e_p=+\infty$ in the infinitessimal limit as the time-step of the exact simulation tends to zero $\e \downarrow 0$. This turns out to be because the transition kernels of the forward and time-reversed processes become more and more mutually singular as the time-step decreases.}
    \label{fig: exact underdamped}
\end{figure}

Note that the entropy production of an exact numerical simulation with time-step $\e >0$ is usually finite by hypoellipticity\footnote{\eqref{eq: finite ep for exact simulation of underdamped} always holds in a quadratic potential, whence the process is a linear diffusion and the results from Section \ref{sec: OU process} apply. We conjecture this to hold in the non-linear case as well, as hypoellipticity guarantees that the transition kernels are mutually equivalent in the sense of measures.}
\begin{align}
\label{eq: finite ep for exact simulation of underdamped}
    e_p(\e) <+\infty.
\end{align}
Figure \ref{fig: exact underdamped} illustrates this with an exact simulation of underdamped Langevin dynamics in a quadratic potential.

When the potential is non-quadratic, the underdamped process is a non-linear diffusion and one is usually unable to simulate it exactly. Instead, one resolves to numerical approximations to the solution of the process. We now turn to two common numerical discretisations of underdamped: the Euler-Maruyama and BBK discretisations. 
We will examine whether these discretisations are good approximations to the true process by computing their entropy production rate.

\subsubsection{Euler-Maruyama discretisation}

In this section, we show that an Euler-Maruyama (E-M) discretisation of underdamped Langevin dynamics at any time-step $\e>0$ has infinite entropy production
\begin{align}
\label{eq: ep Euler}
    e_p^{\text{E-M}}(\e)=+\infty.
\end{align}

To see this, we take a step back and consider an arbitrary Itô SDE in $\R^d$
\begin{align*}
    dx_t = b(x_t)dt + \sigma(x_t)dw_t
\end{align*}
The Euler-Maruyama discretisation for some time-step $\e >0$ is
\begin{align*}
    x_{i+1} = x_i + b(x_i)\e + \sigma(x_i)\omega_i, \quad \omega_i \sim \mathcal N(0, \e \operatorname{Id}_d).
\end{align*}
This is a Markov chain with the following transition kernels
\begin{equation}
\label{eq: E-M scheme definition}
\begin{split}
    p^{\text{E-M}}_\e\left(x_{i+1}, x_{i}\right) &=\mathcal N(x_{i+1}; x_i+\e b(x_i), 2\e D(x_i)),\\
    \bar p^{\text{E-M}}_\e\left(x_{i+1}, x_{i}\right)&:=p^{\text{E-M}}_\e\left(x_{i},x_{i+1}\right),
\end{split}
\end{equation}
where $\bar p^{\text{E-M}}_\e$ denotes the transition kernel of the backward chain\footnote{Caution: this is different from the E-M discretisation of the time-reversed process.}.

It turns out that when the SDE is not elliptic the transition kernels $p^{\text{E-M}}_\e\left(\cdot,x \right), \bar p^{\text{E-M}}_\e\left(\cdot,x \right)$ tend to have different supports:

\begin{lemma}
\label{lemma: support of transition kernels Euler}
For any $x\in \R^d$
\begin{align*}
    \operatorname{supp}p^{\mathrm{E}\text{-}\mathrm{M}}_\e\left( \cdot,x \right)&= \overline{\{y: y \in x + \e b(x) + \im D(x)\}} \\
  \operatorname{supp}\bar p^{\mathrm{E}\text{-}\mathrm{M}}_\e\left( \cdot,x \right)&=\overline{\{y:  x  \in y +\e b(y)+\im D(y)\}}
\end{align*}
\end{lemma}
Lemma \ref{lemma: support of transition kernels Euler} is immediate from \eqref{eq: E-M scheme definition} by noting that the support of $\bar p^{\text{E-M}}_\e\left( \cdot,x \right)$ is the closure of those elements whose successor by the forward process can be $x$.

Unpacking the result of Lemma \ref{lemma: support of transition kernels Euler} in the case of underdamped Langevin dynamics yields
\begin{align*}
   \operatorname{supp}p^{\text{E-M}}_\e\left( \cdot,x \right)=\overline{\{y : y_q = x_q +\e x_p\}}, \quad
   \operatorname{supp}\bar p^{\text{E-M}}_\e\left(\cdot,x \right) = \overline{\{y: y_q + \e y_p = x_q\}},
\end{align*}
where $x:=(x_q,x_p)$ respectively denote position and momenta. One can see that $p^{\text{E-M}}_\e\left( \cdot,x \right) \perp \bar p^{\text{E-M}}_\e\left( \cdot,x \right), \forall x\in \R^d$. From Definition \ref{def: EPR numerical scheme}, we deduce that the entropy production rate of E-M applied to the underdamped process is infinite for any time-step $\e >0$.


\begin{figure}[ht!]
    \centering
    \includegraphics[width= 0.45\textwidth]{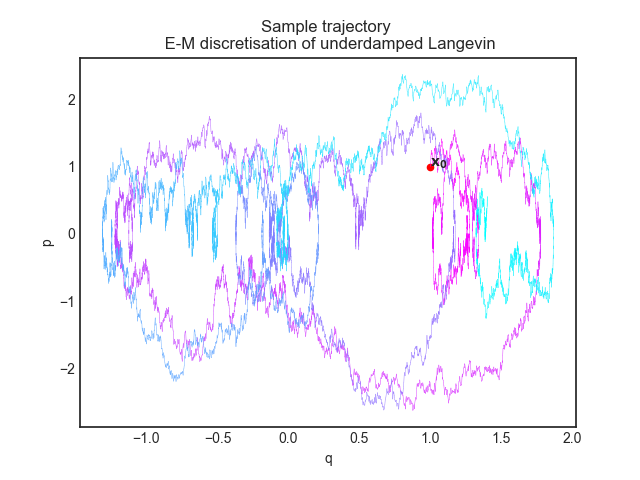}
    \includegraphics[width= 0.45\textwidth]{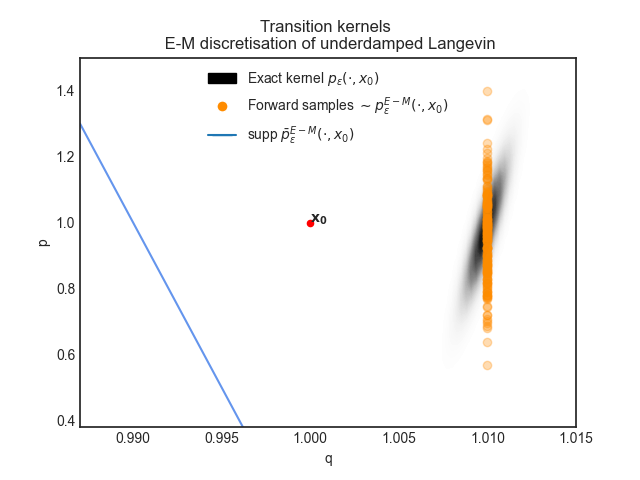}
    \caption{\footnotesize\textbf{Euler-Maruyama simulation of underdamped Langevin dynamics.} This figure compares the Euler-Maruyama simulation of underdamped Langevin dynamics with the exact simulation available in Figure \ref{fig: exact underdamped}. The choice of parameters was the same: $V(q)=q^2/2, M=\gamma =1$. The upper left panel plots a sample trajectory of the numerical scheme. One observes that the numerical scheme is not confined to a prespecified region of space like the true process. The upper right panel plots samples of the numerical scheme after a time-step of $\e$ (in orange) given an initial condition at $x_0$ (in red). This is superimposed onto a heat map of the true transition kernel (in black). We see that samples from the numerical scheme are in the right region of space, but are confined to a subspace which is not aligned with the heat map of the true transition kernel. The support of the transition kernel of the time-reversed scheme is shown in blue. One sees that the supports of forward and reverse transition kernels are mutually singular, thus the entropy production of the numerical discretisation is infinite for any time-step, which differs from the true process which has finite entropy production for any positive time-step.
    }
    \label{fig: Euler underdamped}
\end{figure}

\subsubsection{BBK discretisation}

Contrariwise to Euler, the BBK integrator \cite{brungerStochasticBoundaryConditions1984,roussetFreeEnergyComputations2010}
is a splitting scheme that when applied to underdamped Langevin yields absolutely continuous transition kernels. The numerical scheme consists of three intermediate steps
$$
\begin{aligned}
&p_{i+\frac{1}{2}}=p_{i}-\nabla V\left(q_{i}\right) \frac{\e}{2}-\gamma M^{-1} p_{i} \frac{\e}{2}+\sqrt{2 \gamma \beta^{-1}} \omega_{i} \\
&q_{i+1}=q_{i}+M^{-1} p_{i+\frac{1}{2}} \e \\
&p_{i+1}=p_{i+\frac{1}{2}}-\nabla V\left(q_{i+1}\right) \frac{\e}{2}-\gamma M^{-1} p_{i+\frac{1}{2}} \frac{\e}{2}+\sqrt{2 \gamma \beta^{-1}} \omega_{i+\frac{1}{2}}
\end{aligned}
$$
with $\omega_{i}, \omega_{i+\frac{1}{2}} \sim N\left(0, \frac{\e}{2} \operatorname{Id}\right)$. Its stability and convergence properties were studied in \cite{brungerStochasticBoundaryConditions1984,roussetFreeEnergyComputations2010} and its ergodic properties in \cite{mattinglyConvergenceNumericalTimeAveraging2010a,mattinglyErgodicitySDEsApproximations2002,talayStochasticHamiltonianSystems2002}.

\begin{figure}[ht!]
    \centering
    \includegraphics[width= 0.45\textwidth]{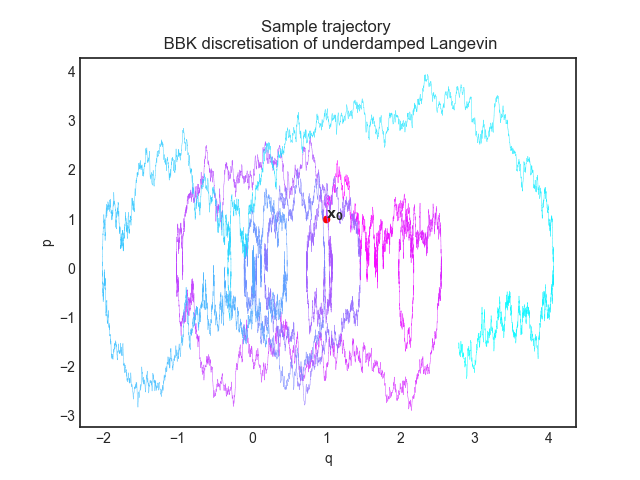}
    \includegraphics[width= 0.45\textwidth]{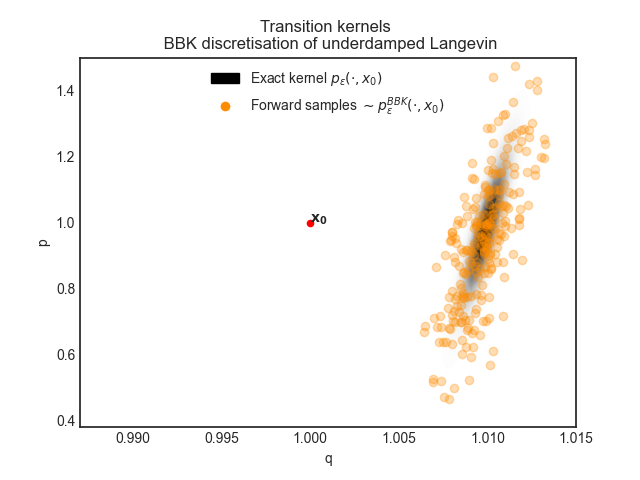}
    \includegraphics[width= 0.5\textwidth]{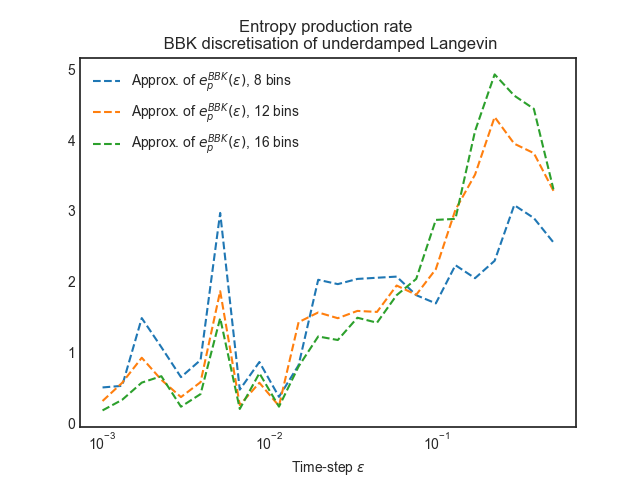}
    \caption{\footnotesize\textbf{BBK simulation of underdamped Langevin dynamics.} This figure compares the BBK simulation of underdamped Langevin dynamics with the exact simulation available in Figure \ref{fig: exact underdamped}. The choice of parameters was the same: $V(q)=q^2/2, M=\gamma =1$. The upper left panel plots a sample trajectory of the numerical scheme. One observes that the numerical scheme is not confined to a prespecified region of space like the true process. The upper right panel plots samples of the numerical scheme after a time-step of $\e$ (in orange) given an initial condition at $x_0$ (in red). This is superimposed onto a heat map of the true transition kernel (in black). We see that samples from the numerical scheme fit the true transition kernel relatively well, but have a higher variance. The bottom panel estimates the entropy production rate of the numerical scheme for several choices of time-step $\e$. This is done by discretising the state-space into a number of bins and numerically evaluating the entropy production of the resulting Markov chain using samples, see Section \ref{sec: ep of numerical schemes theory} for details. The numerical values are consistent with the theoretical result \eqref{eq: ep BBK}.
    }
    \label{fig: BBK underdamped}
\end{figure}

It was shown in \cite[Theorem 4.3]{katsoulakisMeasuringIrreversibilityNumerical2014} that the BBK discretisation of the underdamped Langevin process is quasi time-reversible, so that
\begin{align}
\label{eq: ep BBK}
    e_p^{\mathrm{BBK}}\leq O(\e).
\end{align}
One sees from Figure \ref{fig: BBK underdamped} that the BBK integrator better approximates the transition kernels than E-M, however it still is largely inaccurate from the point of view of the entropy production rate as the simulation becomes reversible when the time-step tends to zero.

\subsubsection{Summary}

In summary, the underdamped Langevin process has infinite entropy production rate in phase space\footnote{Recall that the $e_p$ we computed here is defined \textit{without} an additional momentum flip operator on the path space measure of the time-reversed process, i.e., \eqref{eq: def epr} and not \eqref{eq: rem def gen epr}. See also the discussion in Section \ref{sec: discussion generalised non-reversibility},}, but finite entropy production rate for any exact simulation with a positive time-step. When the potential is non-quadratic, the process is a non-linear diffusion that one usually cannot simulate exactly. To simulate it as accurately as possible, one should seek an approximating numerical scheme that has finite entropy production for any time-step, and whose entropy production tends to infinity for infinitesimally small time-steps.

Two well-known choices of numerical discretisation are the Euler-Maruyama and BBK schemes. By comparing their transition kernels with an exact simulation, we saw that the BBK scheme is a much better approximation to the true process than Euler-Maruyama. Analysis of the entropy production rate shows how these discretisations still far short in capturing important statistical properties of the process: the E-M discretisation has infinite entropy production for any time-step; while the BBK discretisation has finite entropy production for any time-step, and vanishing entropy production for infinitesimally small time-steps. Whenever possible, a good way to choose a time-step size for the BBK integrator might be matching its entropy production rate with that of an exact simulation. These results indicate that employing a BBK scheme with very small time-steps might be inadequate. Luckily, large step-sizes are usually preferred in practice.

In conclusion, the entropy production rate is a useful statistic of stochastic processes that can be used as a tool to devise accurate numerical schemes, particularly in a non-equilibrium statistical physics or sampling context where preserving the amount of time-irreversibility is important. Future development of numerical schemes should take entropy production into account; for example, in developing numerical schemes for underdamped Langevin, one should seek a finite entropy production rate for any positive time-step, which tends to infinity when time-steps become infinitesimally small. Other numerical schemes should be analysed in future work, such as those based on the lexicon for the approximation of the underdamped process developed by Leimkühler and Matthews \cite[p. 269 \& 271]{leimkuhlerMolecularDynamicsDeterministic2015}.

\section{Discussion}
\label{sec: discussion}

Briefly, we unpack a couple of observations and possible extensions of this work.

\subsection{$e_p$ and sampling efficiency}

A well-known criterion for efficient sampling is time-irreversibility~\cite{hwangAcceleratingDiffusions2005,rey-belletIrreversibleLangevinSamplers2015,duncanVarianceReductionUsing2016,barpGeometricMethodsSampling2022a}. Intuitively, non-reversible processes backtrack less often and thus furnish more diverse samples~\cite{nealImprovingAsymptoticVariance2004}. Furthermore, the time-irreversible part of the drift flows along the contours of the stationary probability density which yields mixing and accelerates convergence to the target measure. It is well known that removing non-reversibility worsens the spectral gap and the asymptotic variance of the MCMC estimator~\cite{rey-belletIrreversibleLangevinSamplers2015,duncanVarianceReductionUsing2016,hwangAcceleratingDiffusions2005}, which are two main indicators of the speed of convergence to stationary state \cite{barpGeometricMethodsSampling2022a}. Thus efficient samplers at non-equilibrium steady-state have positive entropy production.

In elliptic linear diffusions, one can construct the optimal time-irreversible drift to optimise the spectral gap~\cite{lelievreOptimalNonreversibleLinear2013,wuAttainingOptimalGaussian2014}
or the asymptotic variance~\cite{duncanVarianceReductionUsing2016}. 
This indicates that one cannot optimise elliptic samplers by simply increasing their entropy production at steady-state without any other constraints, as, we recall, $e_p$ is a quadratic form of the strength of the time-irreversible drift (Figure \ref{fig: EPR as a function of gamma}).


Beyond this, the entropy production rate of general diffusions (Theorems \ref{thm: epr regular case simple}, \ref{thm: epr regular case general}) 
bears a formal resemblance to the Donsker-Varadhan functional \cite[Theorem 2.2]{rey-belletIrreversibleLangevinSamplers2015}, %
from which the asymptotic variance of MCMC estimators is derived \cite{rey-belletIrreversibleLangevinSamplers2015}. It is entirely possible that one might be able to relate the non-stationary entropy production rate (\eqref{eq: time dependent EPR} or \cite[eq. 3.19]{jiangMathematicalTheoryNonequilibrium2004}) 
to the Donsker-Varadhan functional, and thus give a more complete characterisation of sampling efficiency in terms of entropy production.

Many diffusion models of efficient sampling (the underdamped \eqref{eq: underdamped Langevin dynamics} and generalised \cite[eq. 8.33]{pavliotisStochasticProcessesApplications2014} Langevin dynamics, the fastest converging linear diffusion \cite{guillinOptimalLinearDrift2021}), and stochastic optimisation (stochastic gradient descent in deep neural networks \cite{chaudhariStochasticGradientDescent2018}) are not elliptic; that is, they are driven by less Brownian drivers than there are dimensions to their phase space. In particular, these processes have their forward and backward path space measures which are mutually singular, and infinite entropy production\footnote{\cite[Section 5]{chaudhariStochasticGradientDescent2018} shows that stochastic gradient descent is out of equilibrium. Furthermore, it shows empirically that the rank of the diffusion matrix is about 1\% of its dimension in deep neural networks. The sparsity of the noise with respect to the highly out-of-equilibrium behaviour they observe conjectures $b\irr(x) \not \in \operatorname{Range} \sigma(x)$ and thus, mutual singularity of forward and backward path space measures.}. In light of this, we conjecture that mutual singularity of the forward and backward path space measures is an important facet of sampling efficiency (provided the process is ergodic). Mutual singularity apparently exacerbates the mixing effect that time-irreversibility introduces in the elliptic case. Heuristically, if some paths can be taken by the forward process and not by the backward process, these trajectories cannot be reversed, thus the process is constantly forced to visit new regions of phase space, which contributes to the (non-reversible) convergence to steady-state. 

If the above intuition holds, a useful statistic of sampling efficiency might be the probability that the forward process takes paths that cannot be taken by the backward process. By the Lebesgue decomposition theorem we can decompose the forward path space measure $\p$ into $\p_{\textrm{reg}}+\p_{\textrm{sing}}$ such that $\p_{\textrm{reg}} \ll \bar \p$ and $\p_{\textrm{sing}} \perp \bar \p$.  
 This statistic is the non-negative real number 
 \begin{align*}
    \p(\{\gamma \in C([0,T], \R^d): d\p/d \bar \p(\gamma)= +\infty\})= \p_{\textrm{sing}}\left(C([0,T], \R^d)\right),
 \end{align*}
where $d\p/d \bar \p$ is the Lebesgue derivative between forward and backward path space measures. 
Note that the linear diffusion that converges fastest to steady-state maximises the latter (under the constraint that the process remains ergodic) since it has only one Brownian driver \cite{guillinOptimalLinearDrift2021}. However, this statistic does not tell us all since the direction of the Brownian driver with respect to the drift and the stationary density is important to determine sampling efficiency. Yet, these observations indicate that employing diffusions with less Brownian drivers might be an advantage for sampling and optimisation (provided ergodicity is maintained). 
A careful investigation of these relationships is left to future work.

\subsection{Generalised non-reversibility and entropy production rate}
\label{sec: discussion generalised non-reversibility}

Many diffusions studied in statistical physics are not time-reversible but they are generalised reversible; that is, they are time-reversible up to a one-to-one transformation $\theta$ of phase-space which leaves the stationary measure invariant \cite[Section 5.1]{duongNonreversibleProcessesGENERIC2021}, \cite[eq. 2.32]{roussetFreeEnergyComputations2010}. For example, the underdamped langevin equation is generalised reversible---it is reversible up to momentum reversal (Example \ref{eg: Time reversal of underdamped Langevin dynamics}); the generalised Langevin equation is also generalised reversible. 


The entropy production, as defined in Definition \ref{def: epr}, measures time-irreversibility as opposed to generalised non-reversibility. However, as pointed out in Remark \ref{rem: physical relevance}, the physically meaningful definition of entropy production rate sometimes comprises additional operators applied to the path-space measure of the time-reversed process. This modified notion of $e_p$, which we refer to as generalised entropy production, usually takes the form of

\begin{equation}
\label{eq: def gen epr}
    e_p^{\mathrm{gen}, \theta} := \lim _{\e \downarrow 0} \frac{1}{\e} \H\left[\p_{[0, \e]}, \theta_\#\bar \p_{[0, \e]}\right],
\end{equation}
where $\theta_\#$ is the pushforward operator associated to an involution of phase-space $\theta$ that leaves the stationary distribution invariant. The generalised entropy production rate measures the generalised non-reversibility of the process; that is, the extent to which the process is time-irreversible up to the one-to-one transformation $\theta$.
Of course, generalised entropy production reduces to entropy production, as defined in Definition \ref{def: epr}, when $\theta \equiv \operatorname{Id}$.

Since generalised entropy production can sometimes be more physically meaningful, we spend the rest of this section computing it in a couple of examples.

It seems to be a general consensus in statistical physics that the physically relevant notion of entropy production for the underdamped Langevin process is the generalised entropy production when $\theta$ is the momentum reversal \cite{vanvuUncertaintyRelationsUnderdamped2019,eckmannEntropyProductionNonlinear1999,spinneyNonequilibriumThermodynamicsStochastic2012,luposchainskyEntropyProductionContinuous2013}. It is then a by-product of Example \eqref{eg: Time reversal of underdamped Langevin dynamics} that underdamped Langevin dynamics has zero (generalised) entropy production $e_p^{\mathrm{gen}, \theta}=0$, which contrasts with the infinite entropy production one obtains in the non-generalised case (Section \ref{sec: ep singularity}) when one sets $\theta \equiv \operatorname{Id}$.

Beyond this, generalised entropy production could be a useful construct to quantify how far certain diffusion processes are from being generalised reversible. 
For example we can quantify to what extent certain time-irreversible perturbations of underdamped Langevin dynamics are far from being generalised reversible up to momentum reversal.

\begin{example}[$e_p^{\mathrm{gen}, \theta}$ of perturbed underdamped Langevin dynamics]
Consider the following perturbations of underdamped Langevin dynamics \cite[eq. 8]{duncanUsingPerturbedUnderdamped2017}
\begin{equation}
\label{eq: perturbation of underdamped}
    \begin{cases}
        \mathrm{d} q_t =M^{-1} p_t \mathrm{~d} t- Q_1 \nabla V\left(q_t\right) \mathrm{d} t \\
\mathrm{d} p_t =-\nabla V\left(q_t\right) \mathrm{d} t- Q_2 M^{-1} p_t \mathrm{~d} t-\gamma M^{-1} p_t \mathrm{~d} t+\sqrt{2 \gamma \beta^{-1}} \mathrm{d} w_t,
    \end{cases}
\end{equation}
where $Q_1, Q_2 \in \mathbb{R}^{d \times d}$ are constant antisymmetric matrices. By inspection this equation has a Helmholtz decomposition that is similar to underdamped Langevin dynamics (cf. Example \ref{eg: helmholtz decomp Langevin})
\begin{align*}
    &b_{\mathrm{rev}}(q,p)= D \nabla \log \rho(q,p) , \quad b_{\mathrm{irr}}(q,p)= Q \nabla \log \rho(q,p)\\
    \nabla \log  \rho(q,p)&= -\beta \begin{bmatrix}
         \nabla V(q)\\
         M^{-1}p
    \end{bmatrix},
    \quad D=\begin{bmatrix}
        0 &0\\
        0& \gamma \beta^{-1}\id_n
    \end{bmatrix}
 , \quad Q=\beta^{-1}\begin{bmatrix}
        Q_1 &-\id_n\\
        \id_n & Q_2
    \end{bmatrix}.
\end{align*}
The time-reversed process solves the following SDE (Section \ref{sec: helmholtz decomposition})
\begin{align*}
\begin{cases}
\d \bar q_{t} = -M^{-1}\bar p_{t} \d t+ Q_1 \nabla V\left(\bar q_t\right)\mathrm{d} t \\
\d \bar p_{t} =\nabla V\left(\bar q_t\right) \mathrm{d} t+ Q_2 M^{-1} \bar p_t \mathrm{~d} t-\gamma M^{-1} \bar p_{t} \d t+\sqrt{2 \gamma \beta^{-1}} \d  w_{t}.
\end{cases}
\end{align*}
Define $\theta(q,p)=(q,-p)$ to be the momentum reversal transformation of phase space (that leaves underdamped Langevin dynamics invariant as shown in Example \ref{eg: Time reversal of underdamped Langevin dynamics}). Letting $\hat p_{t} = -\bar p_{t}$, 
the time-reversed momentum-flipped equation looks like
\begin{align}
\label{eq: time-reversed momentum flipped}
\begin{cases}
\d \bar q_{t} = M^{-1}\hat p_{t} \d t+ Q_1 \nabla V\left(\bar q_t\right)\mathrm{d} t \\
\d \hat p_{t} =-\nabla V\left(\bar q_t\right) \mathrm{d} t+ Q_2 M^{-1} \hat p_t \mathrm{~d} t-\gamma M^{-1} \hat p_{t} \d t+\sqrt{2 \gamma \beta^{-1}} \d \hat w_{t}.
\end{cases}
\end{align}
Denote by $b^{\mathrm{gen}, \theta}_{\mathrm{irr}}$ the vector field whose sign changes after successively applying these two transformations:
\begin{align*}
   b^{\mathrm{gen}, \theta}_{\mathrm{irr}}
        (q,p)
  = \begin{bmatrix}
        -Q_1 \nabla V\left( q\right)\\
        - Q_2 M^{-1} p
    \end{bmatrix}.
\end{align*}
It follows that the time-reversed, momentum-flipped equation \eqref{eq: time-reversed momentum flipped} does not induce the same path space measure as the initial equation \eqref{eq: perturbation of underdamped} unless $Q_1=Q_2=0$. To see this,  
we follow the proofs of Theorems \ref{thm: epr singular} and \ref{thm: epr regular case simple} to compute the generalised entropy production rate
\begin{align*}
     Q_1 \neq 0 \Rightarrow \p \perp \theta_\# \bar \p \Rightarrow e_p^{\mathrm{gen}, \theta}&= +\infty,\\
    Q_1 =0 \Rightarrow \p \sim \theta_\# \bar\p \Rightarrow e_p^{\mathrm{gen}, \theta}&= \iint_{\R^n}b^{\mathrm{gen}, \theta}_{\mathrm{irr}}\cdot  D^- b^{\mathrm{gen}, \theta}_{\mathrm{irr}} \rho(q,p) \,\d p \,\d q\\
    &= \gamma^{-1}\beta \int_{\R^n} (Q_2 M^{-1}p)^2\rho(p)\d p \\
    &=-\gamma^{-1}\beta \tr\left(Q_2 M^{-1}Q_2\right)< +\infty.
\end{align*}
The last line equality follows from a standard result about expectations of bilinear forms under Gaussian distributions, since $\rho(p)$ is Gaussian with covariance matrix $M$. As usual, the generalised entropy production rate is a quadratic form of the (generalised) irreversible drift.
\end{example}

\subsection{Geometric interpretation of results}
\label{sec: geom}

Our main results concerning the value of entropy production have a straightforward geometric interpretation. The Stratonovich interpretation of the SDE 
\begin{align*}
	dx_t = b^s(x_t)+ \sigma(x_t)\circ dw_t
\end{align*}
is the natural one to consider in a geometric context, when looking at the directions of the drift $b^s$ and volatility vector fields $\sigma_{\cdot i },i=1, \ldots, m$ (i.e., the columns of the volatility matrix field).

Recall from Remark \ref{rem: stratonovich Helmholtz} that the Stratonovich SDE also admits a Helmholtz decomposition $b^s= b^s\rev + b^s\irr$ with $b^s\irr= b\irr$, so that 
\begin{equation}
\label{eq: sumary}
b^s(x) \in \im \sigma(x) \iff b^s\irr (x) \in \im \sigma(x)\iff \bar b^s(x) \in \im \sigma(x) \text{ for any } x\in \operatorname{supp}\mu,
\end{equation}
where $\bar b^s$ is the drift of the time-reversed Stratonovich SDE. In particular, time-reversal is a transformation that sends $b^s$ to $\bar b^s$, $b\irr^s$ to $-b\irr^s$, or, equivalently, adds $-2 b\irr^s$ to the drift.

Our main results can be summarised in a nutshell:
\begin{equation}
\label{eq: main results stratonovich interpretation summary}
\begin{split}
  	\mu\left( \left\{x \in \R^d: b^s(x)\in \im \sigma(x) \right\}\right)= 1 &\Rightarrow e_p = \intr b\irr^s\cdot  D^- b\irr^s \d \mu \quad \text{(see Theorem \ref{thm: epr regular case simple} or \ref{thm: epr regular case general} for details)},\\
	\mu\left( \left\{x \in \R^d: b^s(x)\in \im \sigma(x) \right\}\right)< 1 &\Rightarrow e_p = +\infty \quad \text{(see Theorem \ref{thm: epr singular} for details)}.  
\end{split}
\end{equation}
We derived our main results using the Itô interpretation of an SDE because this allowed us to make more general statements, notably in the context of the general existence and uniqueness theorem of strong solutions to Itô SDEs; it turns out, however, that these results are more naturally interpreted in the Stratonovich context.

Consider the case where there is noise in the direction of the vector field $b^s$, (almost every-) where the process is; in other words, assume that $\mu\left( \left\{x \in \R^d: b^s (x)\in \im \sigma(x) \right\}\right)= 1$. Consider the process at any point $x\in \supp \mu$. In virtue of \eqref{eq: sumary}, the drifts of the forward and time reversed processes both live in $\im \sigma(x)$, the subset of the tangent space that is spanned by the volatility vector fields. Since the driving fluctuations are Gaussian on $\im \sigma(x)$, the time-reversal transformation will be reversed by the random fluctuations with positive probability. Thus, the forward and time-reversed Markov transition kernels (for an infinitesimally small time-step) have the same support---they are mutually equivalent. Under sufficient regularity, made explicit in Theorems \ref{thm: epr regular case simple} or \ref{thm: epr regular case general}, their relative entropy is finite. The $e_p$ is the relative entropy between such Markov kernels on an infinitesimally small time-step (Proposition \ref{prop: epr transition kernels}), so it too will be finite.

On the other hand, if there exists $x \in \supp \mu$ such that there is no noise in the direction of the vector field $b^s$, that is $b^s (x)\not \in \im \sigma(x)$, then the direction of the forward and time-reversed dynamics in an infinitesimal time-step lie on different tangent spaces, $b^s (x) + \im \sigma(x)$ and $  \bar b^s (x) + \im \sigma(x)$, respectively. This means that the forward and time-reversed transition kernels (for an infinitesimally small time-step) are mutually singular and their relative entropy is infinite; thus, the $e_p$ is also infinite.

In particular, it should be straightforward to extend these observations and calculations to diffusions on manifolds.

\section*{Additional information}

\subsection*{Data accessibility}

All data and numerical simulations can be reproduced with code freely available at \url{https://github.com/lancelotdacosta/entropy_production_stationary_diffusions}.

\subsection*{Funding statement}

LD is supported by the Fonds National de la Recherche, Luxembourg (Project code: 13568875). This publication is based on work partially supported by the EPSRC Centre for Doctoral Training in Mathematics of Random Systems: Analysis, Modelling and Simulation (EP/S023925/1). The work of GAP was partially funded by the EPSRC, grant number EP/P031587/1, and by JPMorgan Chase \& Co. Any views or opinions expressed herein are solely those of the authors listed, and may differ from the views and opinions expressed by JPMorgan Chase \& Co. or its affiliates. This material is not a product of the Research Department of J.P. Morgan Securities LLC. This material does not constitute a solicitation or offer in any jurisdiction.

\subsection*{Authors' contributions}

All authors made substantial contributions to conception and design, and writing of the article; and approved publication of the final version.

\subsection*{Competing interests}

We have no competing interests.

\subsection*{Acknowledgements}
We thank Dalton A. R. Sakthivadivel for interesting discussions. We are grateful to our anonymous reviewers for many helpful comments that improved the manuscript.

\appendix

\section{Proofs}

Here we provide proofs supporting the main text.

\subsection{The $e_p$ of stationary Markov processes}
\label{app: proofs ep stationary Markov processes}

\subsubsection{$e_p$ in terms of path space measures with deterministic initial condition}
\label{app: aggregating local ep}

We prove Proposition \ref{prop: aggregating local ep}:
\begin{proof}
The proof is straightforward
\begin{equation*}
\begin{split}
     e_p 
     &= \frac 1 t \H\left[\p_{[0,t]}\mid\bar \p_{[0,t]}\right] =  \frac 1 t\E_{x_\bullet \sim \p}\left [\log \frac{d \p_{[0,t]}}{d \bar \p_{[0,t]}}(x_\bullet)\right] \\
     &=  \frac 1 t\E_{x \sim \mu}\left[ \E_{x_\bullet \sim \p^{x}_{[0,t]}}\left[\log \frac{d \p_{[0,t]}}{d \bar \p_{[0,t]}}(x_\bullet)\right]\right]\\
     &=  \frac 1 t\E_{x \sim \mu}\left[ \E_{x_\bullet \sim \p^{x}_{[0,t]}}\left[\log \frac{d \p^{x}_{[0,t]}}{d \bar \p^{x}_{[0,t]}}(x_\bullet)\right]\right] =  \frac 1 t\E_{x \sim \mu}\left[\H\left[\p^{x}_{[0,t]}\mid\bar \p^{x}_{[0,t]}\right]\right].
\end{split}
\end{equation*}
\end{proof}

\subsubsection{$e_p$ in terms of transition kernels}
\label{app: epr transition kernels}

We prove Proposition \ref{prop: epr transition kernels}:
\begin{proof}
By Proposition \ref{prop: aggregating local ep},
\begin{equation*}
\begin{split}
     e_p 
     &= \lim_{\e \downarrow 0}\frac 1 \e \E_{x \sim \mu}\left[ \E_{x_\bullet \sim \p^{x}_{[0,\e]}}\left[\log \frac{d \p^{x}_{[0,\e ]}}{d \bar \p^{x}_{[0,\e]}}(x_\bullet)\right]\right]\\
     &= \lim_{\e \downarrow 0} \frac{1}{\e} \E_{x \sim \mu}\left[\E_{y \sim p_\e(\cdot,x)}\left[ \log \frac{d p_\e(\cdot,x)}{d \bar p_\e(\cdot,x)}(y)\right]\right]\\
     &= \lim_{\e \downarrow 0} \frac{1}{\e} \E_{x\sim \mu}\left[\H\left[p_\e(\cdot,x)\mid\bar p_\e(\cdot,x)\right]\right].
\end{split}
\end{equation*}
\end{proof}

\subsection{Time-reversal of stationary diffusions}

\subsubsection{Conditions for the reversibility of the diffusion property}
\label{app: reversibility of the diffusion property}

We prove Lemma \ref{lemma: reversibility of the diffusion property}:
\begin{proof}
Recall the following facts:
\begin{itemize}
    \item $(x_t)_{t\in [0,T]}$ is a Markov diffusion process. Its generator is an unbounded, linear operator given by
\begin{equation}
\label{eq: generator of diffusion process}
\begin{split}
    \L : C_c^\infty(\R^d) &\subset  \dom \L \subset L^p_\mu(\R^d) \to L^p_\mu(\R^d) , \quad 1 \leq p \leq \infty, \quad \L f = b \cdot \nabla f + D \nabla \cdot \nabla f.
\end{split}
\end{equation}
    \item The time-reversal of a Markov process is also a Markov process. Let $\bar \L$ be the generator of the time-reversed process $(\bar x_t)_{t\in [0,T]}$. It is known that $\bar \L$ is the adjoint of $\L$. In other words, we have the identity
\begin{align}
\label{eq: adjoint generator}
    \intr f \L g \:\d \mu = \intr g \bar \L f \: \d \mu, \quad \forall f \in \dom \bar \L, g \in \dom \L,
\end{align}
where
    $\dom \bar \L =\left\{f \in L^1_\mu(\R^d) \mid  \exists h \in L^1_\mu(\R^d),  \forall g\in \dom \L: \intr f \L g \d \mu =\intr h g \d \mu  \right\}$. 
     This follows from the fact that the Markov semigroup of the time-reversed process is the adjoint semigroup \cite[p. 113]{jiangMathematicalTheoryNonequilibrium2004}, 
    and thus the infinitesimal generator is the adjoint generator
    \cite{yosidaFunctionalAnalysis1995,pazySemigroupsLinearOperators2011}, \cite[Thm 4.3.2]{jiangMathematicalTheoryNonequilibrium2004}. 
    \item $L^1\loc$-functions define distributions, and hence admit distributional derivatives (which need not be functions).
\end{itemize}

We identify the generator of the time-reversed process by computing the adjoint of the generator. In the following, all integrals are with respect to the $d$-dimensional Lebesgue measure. Let $f,g \in C_c^\infty(\R^d)$.  Noting that $f\rho b \cdot \nabla g, f\rho D\nabla \cdot \nabla g \in L^1(\R^d)$, 
we have 
\begin{align*}
    \intr f \L g \rho = \intr f \rho b \cdot \nabla g +\intr f\rho D\nabla \cdot \nabla g.
\end{align*}
On the one hand, noting that $f\rho b ,\rho b \in L^1\loc(\R^d, \R^{d})$, we have
\begin{align*}
    \intr f \rho b \cdot \nabla g &= - \intr g \nabla \cdot (f\rho b) 
    =-\intr g \left( \rho b \cdot \nabla f + f  \nabla \cdot (\rho b) \right)\\
    &= -\intr g\left( \rho b \cdot \nabla f + f \nabla \cdot \nabla \cdot (\rho D) \right),
\end{align*}
where the last equality follows from the stationary Fokker-Planck equation. 
(Recall that local boundedness of coefficients $b, \sigma$, and Itô's formula imply that the stationary density $\rho$ satisfies $\nabla \cdot (-b \rho + \nabla \cdot (D\rho))=0$
where the equality is in a distributional sense).

On the other hand, noting that $f\rho D, \rho D \in L^1\loc(\R^d, \R^{d \times d})$
, we have
\begin{align*}
    \intr f\rho D\nabla \cdot \nabla g &= \intr f\rho D\cdot \nabla \cdot \nabla g= \intr g \nabla \cdot \nabla \cdot (f \rho D)\\&= \intr g \nabla \cdot \left(\rho D \nabla f + f\nabla \cdot (\rho D) \right)\\
    &=\intr g \left( 2\nabla \cdot (\rho D)\cdot \nabla f + \rho D \nabla \cdot \nabla f +  f \nabla \cdot\nabla \cdot (\rho D) \right).
\end{align*}

Finally, summing the previous two equations yields:
\begin{align*}
  \intr f \L g \rho =\intr g \left(-\rho b \cdot \nabla f + 2 \nabla \cdot (D\rho )\cdot \nabla f+ \rho D\nabla \cdot \nabla f \right)  = \intr g \bar \L f \rho.
\end{align*}
And thus, the generator of the time-reversed process satisfies $\rho \bar \L f  = -\rho b \cdot \nabla f + 2 \nabla \cdot (D\rho )\cdot \nabla f+ \rho D\nabla \cdot \nabla f$ for all $ f\in C_c^\infty(\R^d)$. The time-reversed process is a diffusion if its generator is a second order differential operator with no constant part. This is the case here, except for the fact that the generator outputs distributions as opposed to functions. For the generator to be a diffusion operator we need to assume that the distributional derivative $\nabla \cdot (D\rho )$ is indeed a function (which is then necessarily in $L^1\loc(\R^d, \R^d)$). Thus, the following are equivalent:
\begin{itemize}
    \item $\nabla \cdot (D \rho) \in L^1\loc(\R^d, \R^d)$,
    \item $\bar \L f \in L^1_\mu(\R^d)$ for any $ f\in C_c^\infty(\R^d)$, where $\bar \L f= -b \cdot \nabla f + 2 \rho^{-1}\nabla \cdot (D\rho )\cdot \nabla f+ D\nabla \cdot \nabla f$,
    \item $(\bar x_t)_{t\in [0,T]}$ is a Markov diffusion process. 
\end{itemize}
\end{proof}

\subsubsection{The time-reversed diffusion}
\label{app: time reversed diffusion}

We prove Theorem \ref{thm: time reversal of diffusions}.

\begin{proof}
Since $(\bar x_t)_{t\in [0,T]}$ is a Markov diffusion process with generator $\bar \L$, we have shown that its drift and diffusion are indeed $\bar b, D$, in the proof of Lemma \ref{lemma: reversibility of the diffusion property}.

To show that any such diffusion process induces the path space measure of the time-reversed process, it suffices to show that the martingale problem associated to $(\bar \L, \rho)$ is well-posed. First note that, by Assumption \ref{ass: coefficients time reversal}, the Itô SDE \eqref{eq: Ito SDE} has a unique strong solution. Therefore it also has a unique weak solution. 
Therefore, $(x_t)_{t\in [0,T]}$ is the unique solution to the martingale problem associated to the generator $\L= b \cdot \nabla + D \nabla \cdot \nabla $ 
\cite[Theorem 1.1]{kurtzEquivalenceStochasticEquations2011}. 
In other words, the martingale problem associated to $(\L, \rho)$ is well-posed. It remains to show that there is a one-to-one correspondence between stationary solutions to the martingale problem associated to $\L$ and $\bar \L$. 

Consider Markov processes $(y_t)_{t\in [0,T]}$, $(\bar y_t)_{t\in [0,T]}$, $\bar y_t= y_{T-t}$ stationary at the density $\rho$. We show that $(\bar y_t)_{t\in [0,T]}$ solves the martingale problem wrt $\bar \L$ if and only if $(y_t)_{t\in [0,T]}$ solves the martingale problem wrt $\L$.

\begin{itemize}
    \item $(\bar y_t)_{t\in [0,T]}$ solves the martingale problem wrt $\bar \L$ if and only if for arbitrary $0\leq s\leq t \leq T$, $f,g \in C_c^\infty(\R^d)$
\begin{equation}
\label{eq: equivalent formulations of martingale problem}
\begin{split}
    &\E\left[f(\bar y_t)- \int_0^t\bar \L f(\bar y_r) \d r \mid \bar y_\theta, 0 \leq \theta \leq s\right] = f(\bar y_s)- \int_0^s \bar \L f(x_r) \d r\\
   \stackrel{\text{Markov}}{\iff} &\E\left[f(\bar y_t)- \int_0^t\bar \L f(\bar y_r) \d r \mid \bar y_s\right] = f(\bar y_s)- \int_0^s \bar \L f(x_r) \d r\\
    \iff & \E\left[f(\bar y_t)-f(\bar y_s)- \int_s^t\bar \L f(\bar y_r) \d r \mid \bar y_s\right]=0\\
    \iff & \E\left[\left(f(\bar y_t)-f(\bar y_s)- \int_s^t\bar \L f(\bar y_r) \d r \right) g(\bar y_s)\right]=0.
\end{split}
\end{equation}
If we make the change of variable $t \leftarrow T-s, s \leftarrow T-t$, so that $0 \leq s \leq t \leq T$, this is equivalent to:
\begin{align*}
    \iff & \E\left[\left(f(y_s)-f( y_t)- \int_{T-t}^{T-s} \bar \L f( y_{T-r}) \d r \right) g(y_t)\right]=0\\
    \iff & \E\left[\left(f(y_s)-f( y_t)- \int_{s}^t \bar \L f(y_r) \d r \right) g(y_t)\right]=0
\end{align*}
\item Repeating the equivalences in \eqref{eq: equivalent formulations of martingale problem}, $(y_t)_{t\in [0,T]}$ solves the martingale problem wrt $\L$ if and only if for arbitrary $0\leq s\leq t \leq T$, $f,g \in C_c^\infty(\R^d)$
\begin{align*}
    \E\left[\left(g(y_t)-g(y_s)- \int_s^t\L g( y_r) \d r \right) f( y_s)\right]=0.
\end{align*}
\item Thus, it suffices to show that the two last expressions are equal, i.e.,
\begin{align*}
    \E\left[\left(f(y_s)-f( y_t)- \int_{s}^t \bar \L f(y_r) \d r \right) g(y_t)\right]= \E\left[\left(g(y_t)-g(y_s)- \int_s^t\L g( y_r) \d r \right) f( y_s)\right]
\end{align*}

By stationarity, we have 
\begin{align*}
    \E\left[\left(f(y_s)-f( y_t)\right) g(y_t)\right]&=\E\left[f(y_s)g(y_t)-f( y_t)g(y_t)\right]\\
    =\E\left[f(y_s)g(y_t)-f( y_s)g(y_s)\right]&=\E\left[\left(g(y_t)-g(y_s)\right) f( y_s)\right].
\end{align*}
Thus, it remains to show that
\begin{align*}
    \E\left[\int_{s}^t g(y_t) \bar \L f(y_r) \d r \right]= \E\left[\int_s^tf( y_s)\L g( y_r) \d r \right]
\end{align*}
We proceed to do this. On the one hand: 
\begin{align*}
    \E\left[\int_s^tf( y_s)\L g( y_r) \d r \right] &=\int_s^t \E\left[f( y_s)\L g( y_r) \right]\d r \\
    &= \int_s^t \E \left[\E\left[f( y_s)\L g( y_r) \mid y_s\right]\right]\d r\\
    &=\int_s^t \E \left[f( y_s)\E\left[\L g( y_r) \mid y_s\right]\right]\d r\\
    &= \int_s^t \E \left[f( y_s)\operatorname{P}_{r-s}\L g(y_s)\right]\d r\\
    &=\int_s^t \intr f( y)\operatorname{P}_{r-s}\L g(y) \rho(y) \d y\d r \quad \text{(stationarity)}\\
    &= \intr f( y)\int_s^t \operatorname{P}_{r-s}\L g(y) \d r \rho(y) \d y \\
    &=\intr f(y) \left(\operatorname{P}_{t-s} -\operatorname{P}_0\right)g(y)\rho(y) \d y \quad \text{($\partial_t \operatorname{P}_{t}= \operatorname{P}_{t} \L
    $)}\\
    &=\intr g(y) \left(\bar{\operatorname{P}}_{t-s} -\bar{\operatorname{P}}_0\right)f(y)\rho(y) \d y
\end{align*}

On the other hand: 
\begin{align*}
    \E\left[\int_s^t g( y_t)\bar \L f( y_r) \d r \right] &= \int_s^t \E\left[g( y_t)\bar \L g( y_r) \right]\d r\\
    &= \int_s^t \E \left[\E\left[g( y_t)\bar \L f( y_r) \mid y_t\right]\right]\d r\\
    &= \int_s^t \E \left[g( y_t)\E\left[\bar \L f( y_r) \mid y_t\right]\right]\d r\\
    &= \int_s^t \E \left[g( y_t)\bar {\operatorname{P}}_{t-r}\bar \L f( y_t)\right]\d r\\
    &= \int_s^t \intr g( y)\bar {\operatorname{P}}_{t-r}\bar \L f( y)\rho(y) \d y\d r \quad \text{(stationarity)}\\
    &=  \intr g( y) \int_s^t\bar {\operatorname{P}}_{t-r}\bar \L f( y) \d r\rho(y) \d y\\
    &= \intr g( y) \left(\bar{\operatorname{P}}_{t-s}-\bar{\operatorname{P}}_0\right)f(y) \rho(y) \d y \quad \text{($\partial_t \bar{\operatorname{P}}_{t}= \bar{\operatorname{P}}_{t} \bar \L
    $)}
\end{align*}
\end{itemize}
This shows the one-to-one correspondence and completes the proof.
\end{proof}

\subsubsection{The Helmholtz decomposition}
\label{app: helmholtz decomposition}

We prove Proposition \ref{prop: helmholtz decomposition}.


\begin{proof}
\begin{itemize}
    \item["$\Rightarrow$"]
    We define the time-reversible and time-irreversible parts of the drift
    \begin{align*}
        b_{\mathrm{rev}} :=\frac {b + \bar b}{2}, \quad  b_{\mathrm{irr}} :=\frac {b - \bar b}{2}.
    \end{align*}
    We now show that the have the predicted functional form. For $x$ such that $\rho(x) =0$, $b_{\mathrm{rev}} =\left(b + \bar b\right)/{2}=0$. For $x$ such that $\rho(x) >0$ 
    \begin{align}
    \label{eq: time reversible drift}
        b_{\mathrm{rev}} &=\frac {b + \bar b}{2}= \rho^{-1}\nabla \cdot (D \rho) = \rho^{-1}D \nabla \rho +\rho^{-1}\rho \nabla \cdot D = D \nabla \log \rho + \nabla \cdot D.
    \end{align}
    For the time-irreversible drift, first note that the stationary density $\rho$ solves the stationary Fokker-Planck equation \cite{pavliotisStochasticProcessesApplications2014,riskenFokkerPlanckEquationMethods1996}
    \begin{equation*}
        \nabla \cdot (-b\rho + \nabla \cdot (D \rho))=0.
    \end{equation*}
    Decomposing the drift into time-reversible and time-irreversible parts, from \eqref{eq: time reversible drift}
    \begin{align*}
        -b\rho + \nabla \cdot (D \rho)= -b_{\mathrm{rev}}\rho- b_{\mathrm{irr}}\rho + \nabla \cdot (D \rho)=- b_{\mathrm{irr}}\rho,
    \end{align*}
    we obtain that the time-irreversible part produces divergence-free (i.e., conservative) flow w.r.t. the steady-state density
    \begin{equation*}
        \nabla \cdot (b_{\mathrm{irr}}\rho)=0.
    \end{equation*}
    \item["$\Leftarrow$"] From \eqref{eq: time reversible drift} the time-reversible part of the drift satisfies the following identity 
    \begin{align}
        b_{\mathrm{rev}}\rho &= \nabla \cdot (D \rho).
    \end{align}
    
    It follows that the density $\rho$ solves the stationary Fokker-Planck equation
    \begin{align*}
        \nabla \cdot (-b\rho + \nabla \cdot (D \rho)) &= \nabla \cdot (-b_{\mathrm{rev}}\rho-b_{\mathrm{irr}}\rho + \nabla \cdot (D \rho))= \nabla \cdot (-b_{\mathrm{irr}}\rho) =0.
    \end{align*}
\end{itemize}
\end{proof}


We prove Proposition \ref{prop: characterisation of irreversible drift}:

\begin{proof}
\begin{itemize}
    \item["$\Rightarrow$"] Recall that any smooth divergence-free vector field is the divergence of a smooth antisymmetric matrix field $A=-A^\top$ \cite{RealAnalysisEvery,yangBivectorialNonequilibriumThermodynamics2021,grahamCovariantFormulationNonequilibrium1977, eyinkHydrodynamicsFluctuationsOutside1996}
    \begin{equation*}
        b_{\mathrm{irr}}\rho = \nabla \cdot A.
    \end{equation*}
    This result holds most generally a consequence of Poincaré duality in de Rham cohomology \cite[Appendix D]{yangBivectorialNonequilibriumThermodynamics2021}. We define a new antisymmetric matrix field $Q:= \rho^{-1}A$. From the product rule for divergences we can rewrite the time-irreversible drift as required
    \begin{equation*}
       b_{\mathrm{irr}} = Q \nabla \log \rho + \nabla \cdot Q.
    \end{equation*}
    \item["$\Leftarrow$"] Conversely, we define the auxiliary antisymmetric matrix field $A := \rho Q$. Using the product rule for divergences it follows that
    \begin{equation*}
        b_{\mathrm{irr}} = \rho^{-1}\nabla \cdot A.
    \end{equation*}
    Finally, 
    \begin{equation*}
       \nabla \cdot( b_{\mathrm{irr}}\rho) = \nabla \cdot (\nabla \cdot A)= 0
    \end{equation*}
    as the matrix field $A$ is smooth and antisymmetric.
\end{itemize}
\end{proof}

\subsubsection{Multiple perspectives on the Helmholtz decomposition}
\label{app: multiple perspectives on Helmholtz}

We prove Proposition \ref{proposition: hypocoercive decomposition of the generator}:

\begin{proof}[Proof of Proposition \ref{proposition: hypocoercive decomposition of the generator}]
The proof is analogous to \cite[Proposition 3]{villaniHypocoercivity2009}. In view of Section \ref{sec: helmholtz decomp inf gen}, we only need to check that: 1) if $\sqrt 2 \operatorname \Sigma f =\sigma^\top \nabla f$, then $\sqrt 2 \operatorname \Sigma^*g=- \nabla \cdot(\sigma g) - \nabla \log \rho \cdot \sigma g$; 2) the symmetric part of the generator factorises as $\operatorname S = - \Sigma^* \Sigma$. 
\begin{enumerate}
    \item[1)] For any $f,g \in C_c^\infty (\R^d):$
    \begin{align*}
        \langle f, \sqrt 2\Sigma^*g \rangle_{L^2_\mu(\R^d)}&=
        \langle \sqrt 2\Sigma f, g \rangle_{L^2_\mu(\R^d)} = \int g \sigma^\top \nabla f \rho(x) \dx \\
        &= \int \sigma g \rho \cdot \nabla f (x) \dx
        =-\int f \nabla \cdot (\sigma g \rho)(x) \dx \\
        &= \int -f \nabla \log \rho \cdot \sigma g \rho(x)-f \nabla \cdot (\sigma g) \rho(x)  \dx \\
        &= \langle f,  - \nabla \log \rho \cdot \sigma g-\nabla \cdot (\sigma g)  \rangle_{L^2_\mu(\R^d)}
    \end{align*}
    This implies $\sqrt 2 \Sigma^* g= - \nabla \log \rho \cdot \sigma g-\nabla \cdot (\sigma g)  $.
    \item[2)] For any $f \in C_c^\infty(\R^d)$:
    \begin{align*}
        -\Sigma^*\Sigma f&= \nabla \log \rho \cdot D \nabla f+\nabla \cdot ( D\nabla f)\\
        &= \nabla \log \rho \cdot D \nabla f + (\nabla \cdot D)\cdot \nabla f + D\nabla \cdot \nabla f\\
        &= b_{\mathrm{rev}} \cdot \nabla f + D\nabla \cdot \nabla f= \operatorname S f
    \end{align*}
    where the penultimate equality follows since $b_{\mathrm{rev}}= D \nabla \log \rho + \nabla \cdot D$, $\mu$-a.e.
\end{enumerate}
\end{proof}

We now prove Proposition \ref{prop: GENERIC decomposition of the Fokker-Planck equation}:

\begin{proof}[Proof of Proposition \ref{prop: GENERIC decomposition of the Fokker-Planck equation}]
\begin{itemize}
    \item We compute the Fréchet derivative of $\H[\cdot \mid \rho]$. First of all, we compute its Gâteaux derivative in the direction of $\eta$.
    \begin{align*}
        \frac{d}{d\e}\operatorname{H}[\rho_t + \e \eta \mid \rho ]&= \frac{d}{d\e}\intr (\rho_t + \e \eta) \log \frac{\rho_t + \e \eta}{\rho}\d x
        = \intr \eta \log \frac{\rho_t + \e \eta}{\rho} +\eta \d x
        = \intr \eta \left( \log \frac{\rho_t + \e \eta}{\rho} +1\right)\d x.
    \end{align*}
    By definition of the Fréchet derivative, we have $\frac{d}{d\e}\left.\operatorname{H}[\rho_t + \e \eta \mid \rho ]\right|_{\e=0}= \langle \d \operatorname{H}[\rho_t \mid \rho ], \eta \rangle$. This implies $ \d \operatorname{H}[\rho_t \mid \rho ]=  \log \frac{\rho_t }{\rho} +1$ by the Riesz representation theorem.
    \item Recall, from Proposition \ref{proposition: hypocoercive decomposition of the generator}, that
 $\sqrt 2\Sigma= \sigma^\top\nabla$. We identify $\Sigma'$. For any $f,g \in C_c^\infty(\R^d)$
 \begin{align*}
     \langle g, \sqrt 2\Sigma f\rangle= \intr g \sigma^\top\nabla f \d x= \intr \sigma g \cdot \nabla f \d x = - \intr f \nabla \cdot (\sigma g) \d x.
 \end{align*}
 This yields $\sqrt 2\Sigma'g =-\nabla \cdot (\sigma g)$. And in particular, $\Sigma'(\rho_t \Sigma \xi)= -\nabla \cdot (\rho_t D\nabla \xi)$.
 \item We define $M_{\rho_t}(\xi) := \Sigma'(\rho_t \Sigma \xi)=-\nabla \cdot (\rho_t D\nabla \xi)$ and verify that this is a symmetric semi-positive definite operator. For any $g,h \in C_c^\infty(\R^d)$:
 \begin{align*}
   \langle \operatorname M_{\rho_t}h, g\rangle= \langle \Sigma h, \Sigma g \rangle = \langle h, \operatorname M_{\rho_t}g\rangle, \quad
   \langle \operatorname M_{\rho_t}g, g\rangle=\langle \Sigma g, \Sigma g \rangle_{\rho_t}\geq 0.
 \end{align*}
 Also, $-M_{\rho_t}\left( \d \operatorname{H}[\rho_t \mid \rho ]\right)=\nabla \cdot(\rho_t D \nabla \log \frac{\rho_t}{\rho} )$ is immediate.
 \item We define $W(\rho_t) = \nabla \cdot (-b\irr \rho_t)$ and verify the orthogonality relation:
 \begin{align*}
    \langle \operatorname W(\rho_t), \d \operatorname{H}[\rho_t \mid \rho ]\rangle &= \intr \left(\log \frac{\rho_t }{\rho} +1\right)\nabla \cdot (-b\irr \rho_t)\d x= \intr b\irr \rho_t \nabla \left(\log \frac{\rho_t }{\rho} +1\right) \d x \\
    &= \intr b\irr \rho_t \frac{\rho }{\rho_t}\nabla \left(\frac{\rho_t }{\rho}\right) \d x 
    = -\intr \nabla\cdot (b\irr \rho)\frac{\rho_t }{\rho}\d x =0, 
 \end{align*}
 where the last equality holds by Proposition \ref{prop: helmholtz decomposition}.
\end{itemize}
\end{proof}

\subsection{The $e_p$ of stationary diffusions}

\subsubsection{Regular case}
\label{app: epr regularity}

We prove Theorem \ref{thm: epr regular case simple}:

\begin{proof}
By Assumption \ref{ass: coefficients time reversal} the Itô SDE \eqref{eq: Ito SDE} has a unique non-explosive strong solution $\left(x_{t}\right)_{t \geq 0}$ with respect to the given Brownian motion $\left(w_{t}\right)_{t \geq 0}$ on a filtered probability space $(\Omega, \mathscr{F},\left\{\mathscr{F}_{t}\right\}_{t \geq 0}, P)$. (Even though Assumption \ref{ass: coefficients time reversal} ensures non-explosiveness of the solution on a finite time-interval, stationarity implies that we may prolong the solution up to arbitrary large times).


By Theorem \ref{thm: time reversal of diffusions} we know that a solution to the following SDE
\begin{align}
\label{eq: reversed SDE}
d\bar x_t= \bar b(\bar x_t) dt + \sigma(\bar x_t) dw_t, \quad \bar x_0 = 
x_0, 
\end{align}
induces the path space measure of the time-reversed process. By Proposition \ref{prop: helmholtz decomposition}, we can rewrite the (forward and time-reversed) drifts as $b = b_{\mathrm{rev}}+ b_{\mathrm{irr}}$ and $\bar b = b_{\mathrm{rev}}- b_{\mathrm{irr}}$.

We define the \textit{localised} coefficients
\begin{align*}
    b\n(x) &:= b\left( \left(1 \land \frac{n}{|x|}\right) x \right)= \begin{cases}
        b(x) \text{ if } |x| \leq n\\
        b\left(n \frac{x}{|x|}\right) \text{ if } |x| > n,
    \end{cases}
\end{align*}
and analogously for $\bar b\n, \sigma\n, b_{\mathrm{rev}}\n,b_{\mathrm{irr}}\n$. Note that the assignment $\cdot\n$ respects sums and products, in particular 
\begin{equation}
\label{eq: localised drift respects helmholtz decomposition}
\begin{split}
    b\n &= (b_{\mathrm{rev}}+ b_{\mathrm{irr}})\n = b_{\mathrm{rev}}\n+ b_{\mathrm{irr}}\n,\\
    \bar b\n &= (b_{\mathrm{rev}}- b_{\mathrm{irr}})\n = b_{\mathrm{rev}}\n- b_{\mathrm{irr}}\n.
\end{split}
\end{equation}

It is easy to see that the localised SDE
\begin{align*}
        dx_t\n = b\n (x_t\n) dt + \sigma\n (x_t\n) dw_t, \quad x_0\n= x_0
\end{align*}
also has a unique strong solution $x\n=(x_t\n)_{t \geq 0}$ with respect to the given Brownian motion $(w_t)_{t\geq 0}$ on the probability space $(\Omega, \mathscr{F},\left\{\mathscr{F}_{t}\right\}_{t \geq 0}, P)$.
This follows from the fact that the localised SDE has locally Lipschitz continuous and bounded coefficients that satisfy the assumptions of Theorem \cite[Theorem 3.1.1]{prevotConciseCourseStochastic2008}.

From assumption \ref{item: epr regular drift image of volatility}, we obtain that for $\rho$-a.e. $x \in \R^d$
\begin{equation}
\label{eq: localised drift in image of the volatility}
\begin{split}
    b_{\mathrm{irr}}(x) \in \im \sigma(x)
    &\Rightarrow b_{\mathrm{irr}}(x) = \sigma \sigma^- b_{\mathrm{irr}}(x)\\
    b_{\mathrm{irr}}\n(x) \in \im \sigma\n(x)
    &\Rightarrow b_{\mathrm{irr}}\n(x) = \sigma\n \sigma^{(n)^-} b_{\mathrm{irr}}\n(x).
\end{split}
\end{equation}
Then, \eqref{eq: localised drift respects helmholtz decomposition} and \eqref{eq: localised drift in image of the volatility} imply that we can rewrite the localised SDE as
\begin{align*}
        dx_t\n = b_{\mathrm{rev}}\n(x_t\n) dt+ \sigma\n(x_t\n)\left[\sigma^{(n)^-}b_{\mathrm{irr}}\n\left(x_t\n\right)dt+dw_t\right], \quad x_0\n= x_0.
\end{align*}
By the definition of Itô's stochastic calculus, $x_t\n$ is an $\mathscr F_t$-adapted process.
By assumption \ref{item: continuity}, $\sigma^{-}b_{\mathrm{irr}}$ is Borel measurable and thus it follows that the localised map $\sigma^{(n)^-}b_{\mathrm{irr}}\n$ is Borel measurable. 
Thus $-2\sigma^{(n)^-}b_{\mathrm{irr}}\n\left(x_t\n\right)$ is also an $\mathscr F_t$-adapted process. In addition, by continuity and localisation, $\sigma^{(n)^-}b_{\mathrm{irr}}\n$ is bounded. Therefore, by \cite[Proposition 10.17 (i)]{pratoStochasticEquationsInfinite2014} 
applied to $-2\sigma^{(n)^-} b_{\mathrm{irr}}\n\left(x_s\n\right)$,
\begin{align*}
    Z_t\n &= \exp \left[-2 \int_0^t \left\langle\sigma^{(n)^-} b_{\mathrm{irr}}\n\left(x_s\n\right), dw_s\right\rangle + \left|\sigma^{(n)^-} b_{\mathrm{irr}}\n\left(x_s\n\right) \right|^2 ds\right], t \geq 0,
\end{align*}
is a martingale on the probability space $\left(\Omega, \mathscr{F},\left\{\mathscr{F}_{t}\right\}_{t \geq 0}, P\right)$.
We define a new probability measure $\bar P_n$ on the sample space $\Omega$ through
\begin{equation}
\left.\frac{d \bar P_n}{d P}\right|_{\mathscr{F}_{t}}=Z_{t}^{(n)}, \quad \forall t \geq 0.
\end{equation}
By Girsanov's theorem \cite[Theorem 10.14]{pratoStochasticEquationsInfinite2014}, 
$x_t\n$ solves the SDE
\begin{align*}
        d x_t\n &= b_{\mathrm{rev}}\n(x_t\n) dt- \sigma\n(x_t\n)\left[\sigma^{(n)^-}b_{\mathrm{irr}}\n\left(x_t\n\right)dt+dw_t\right], \quad  x_0\n=  x_0.
\end{align*}
on the probability space $\left(\Omega, \mathscr{F},\left\{\mathscr{F}_{t}\right\}_{t \geq 0}, \bar P_{n}\right)$.
Using \eqref{eq: localised drift in image of the volatility}, $x_t\n$ solves
\begin{align*}
        d x_t\n &= \bar b\n ( x_t\n) dt + \sigma\n ( x_t\n) dw_t, \quad  x_0\n=  x_0
\end{align*}
on said probability space.

We define a sequence of stopping times $\tau_0=0$ and $\tau_{n}=\inf \left\{t \geq 0:\left|x_{t}\right|>n\right\}$ for $n>0$. Since $\left(x_{t}\right)_{t \geq 0}$ is non-explosive, $P$-a.s. $\lim_{n \to \infty} \tau_n = +\infty$.
As $x_t = x\n_t$ when $t \leq \tau_n$, we have $P$-a.s.
\begin{align*}
    Z_{t \land \tau_n}\n &= \exp \left[-2 \int_0^{t \land \tau_n} \left\langle \sigma^- b_{\mathrm{irr}}(x_s), dw_s\right\rangle + \left|\sigma^- b_{\mathrm{irr}}(x_s) \right|^2 ds\right].
\end{align*}
As $Z_{t}^{(n+1)} \mathds 1_{\left\{t<\tau_{n}\right\}}=Z_{t}^{(n)} \mathds 1_{\left\{t<\tau_{n}\right\}}$, we define the limit as $n \to +\infty$
\begin{align*}
Z_{t} \equiv \sum_{n=1}^{+\infty} Z_{t}^{(n)} \mathds 1_{\left\{\tau_{n-1} \leq t<\tau_{n}\right\}}=\lim _{n \rightarrow+\infty} Z_{t}^{(n)} \mathds1_{\left\{t<\tau_{n}\right\}}=\lim _{n \rightarrow+\infty} Z_{t \wedge \tau_{n}}^{(n)}.
\end{align*}
By definition, $Z_{t}$ is a continuous local martingale on $
\left(\Omega, \mathscr{F},\left\{\mathscr{F}_{t}\right\}_{t \geq 0}, P\right)$. 

We compute $Z_t$. Let's write $-\log Z_{t \wedge \tau_{n}}^{(n)}=M_{t}^{(n)}+Y_{t}^{(n)}$, where
\begin{align*}
    M_{t}^{(n)}=2 \int_0^{t \land \tau_n} \left\langle \sigma^- b_{\mathrm{irr}}(x_s), dw_s\right\rangle, \text{ and, } Y_{t}^{(n)}=2 \int_0^{t \land \tau_n}  \left|\sigma^- b_{\mathrm{irr}}(x_s) \right|^2 ds.
\end{align*}
We also define $M_{t}=2 \int_0^{t} \left\langle \sigma^- b_{\mathrm{irr}}(x_s), dw_s\right\rangle$ and $Y_t= 2\int_0^{t}  \left|\sigma^- b_{\mathrm{irr}}(x_s) \right|^2 ds$.

From assumption \ref{item: epr bounded}, we have
\begin{align*}
    \int_{\R^d} \left|\sigma^- b_{\mathrm{irr}}(x) \right|^2\rho(x) dx = \frac 1 2\int_{\R^d} b_{\mathrm{irr}}^\top D^- b_{\mathrm{irr}} \rho(x)\d x  <+\infty.
\end{align*}
Thus, we obtain that 
\begin{align*}
\E_P\left|M_{t}^{(n)}-M_{t}\right|^{2} &=4 \E_P\left|\int_{0}^{t} \left\langle \sigma^- b_{\mathrm{irr}}(x_s)\mathds 1_{\left\{s>\tau_{n}\right\}}, dw_s\right\rangle  \right|^{2} \\
&=4 \E_P \int_{0}^{t} \left|\sigma^- b_{\mathrm{irr}}(x_s) \right|^2  \mathds 1_{\left\{s>\tau_{n}\right\}} d s \quad \text{(Itô's isometry)}\\
&\xrightarrow{ n \to+\infty} 0, \\
\E_P\left|Y_{t}^{(n)}-Y_{t}\right| &=2 \E_P \int_{0}^{t} \left|\sigma^- b_{\mathrm{irr}}(x_s) \right|^2\mathds 1_{\left\{s>\tau_{n}\right\}} d s \xrightarrow{ n \to+\infty} 0.
\end{align*}
Thus, $-\log Z_{t}=M_{t}+Y_{t}$. By Itô calculus, $M_{t}$ is a martingale on the probability space $\left(\Omega, \mathscr{F},\left\{\mathscr{F}_{t}\right\}_{t \geq 0}, P\right)$, and in particular $\E_P[M_t]=0$.

Let $T> 0$. Let $\left(C([0,T], \R^d), \mathscr{B}\right)$ denote the path space, where $\mathscr{B}$ is the Borel sigma-algebra generated by the sup norm $\|\cdot\|_\infty$. Denote trajectories of the process by $x_\bullet:=(x_t)_{t \in [0,T]}: \Omega \to C([0,T], \R^d)$. By definition of Itô calculus, $Z_{T}$ is measurable with respect to $\langle x_{s}: 0 \leq t \leq T\rangle$, so there exists a positive measurable function $Z_T^C$ on the path space, such that $P$-a.s. $Z^C_T (x_\bullet(\omega))=Z_T(\omega)$ for $\omega \in \Omega$, i.e., the following diagram commutes
\begin{equation*}
  \begin{tikzcd}
    &C([0,T], \R^d) \arrow[dr, "Z_T^C"] &  \\
    \Omega \arrow[ur, "x_\bullet"] \arrow[rr, "Z_T"']&&\R_{>0}.
    \end{tikzcd}
\end{equation*}

Note that the path space $\left(C([0,T], \R^d), \mathscr{B}\right)$ admits a canonical filtration $(\mathscr{B}_{t})_{t \in [0,T]}$, where
\begin{align*}
    \mathscr{B}_{t} = \left\{A \subset C([0,T], \R^d) : \left.A\right|_{[0,t]} \in \text{Borel sigma-algebra on} \left(C([0,t], \R^d), \|\cdot \|_\infty\right)\right\}.
\end{align*}
For any path $\gamma \in C([0,T], \R^d)$ and $n \in \mathbb N$, we define the hitting time $t_n(\gamma) =\inf \left\{t \geq 0:\left|\gamma_{t}\right|>n\right\}$.

\begin{claim}
\label{claim: hitting time is stopping time}
These hitting times are stopping times wrt to the canonical filtration, i.e., $\{\gamma \in C([0,T], \R^d): t_n(\gamma) \leq t\} \in \mathscr{B}_{t}$ for any $t\in [0,T]$.
\end{claim}
\begin{proofclaim}
Let $A:=\{\gamma \in C([0,T], \R^d): t_n(\gamma) \leq t\}$. Then, 
\begin{align*}
    \left.A\right|_{[0,t]} &= \{\gamma \in C([0,t], \R^d): t_n(\gamma) \leq t\}\\
    &= \left\{\gamma \in C([0,t], \R^d): \min\{s\geq 0: \left|\gamma_{s}\right| > n\} \leq t\right\} \text{ (continuity of $\gamma$)}\\
    &=\left\{\gamma \in C([0,t], \R^d): \|\gamma \|_\infty > n \right\},
\end{align*}
which is clearly a Borel set in $\left(C([0,t], \R^d), \|\cdot \|_\infty\right)$.
\end{proofclaim}

Thus we can define stopping time sigma-algebras in the usual way
\begin{align*}
    \mathscr{B}_{T\land t_n} = \left\{A \in \mathscr{B}_{T}: A \cap\{\gamma \in C([0,T], \R^d): T\land t_n(\gamma) \leq t \} \in \mathscr{B}_{t}, \forall t \in [0,T]\right\}.
\end{align*}

We showed above that, under $P, \bar P_n$, the distributions of $x$ restricted to $(C([0,T], \R^d),\mathscr{B}_{T\land t_n})$ are $\p_{\left[0, T \wedge t_{n}\right]}:=\left.\p_{\left[0, T \right]}\right|_{\mathscr{B}_{T\land t_n}}$ and $ \bar{\p}_{\left[0, T \wedge t_{n}\right]}:=\left.\bar{\p}_{\left[0, T \right]}\right|_{\mathscr{B}_{T\land t_n}}$, respectively.


By inspection, we have, for any $t \geq 0$,
\begin{align*}
    \{T <\tau_n \}\cap \{T \land \tau_n \leq t\} = \begin{cases}
        \{T <\tau_n\} \in \mathscr F_T \subset \mathscr F_t \text{ if } T\leq t\\
        \emptyset \subset \mathscr F_t \text{ if } T> t.
    \end{cases}
\end{align*}
Setting $t= T \land \tau_n$, we have $\{T <\tau_n\} \in \mathscr F_{T \land \tau_n}$, which also yields $\{T \geq \tau_n\} \in \mathscr F_{T \land \tau_n}$ and $\{\tau_{n-1} \leq T < \tau_{n}\} \in \mathcal F_{T \land \tau_n}$.

Fix $i \geq 0$ and $A \in \mathscr{B}_{T \wedge t_{i}}$. Then $x^{-1}_\bullet A \in \mathscr F$ as $x_\bullet$ is measurable. Thus $x^{-1}_\bullet A \cap \{T <\tau_i \} \subset \mathscr F_{T \land \tau_i}$ and $x^{-1}_\bullet A \cap \{\tau_{n-1} \leq T < \tau_{n}\} \subset \mathscr F_{T \land \tau_n}$ for any $n >i$. Finally,
\begin{align*}
    &\E_{\p}\left[Z_{T}^{C}  \mathds 1_{A}\right]=\E_{P}\left[ Z_{T} \mathds 1_{x^{-1}_\bullet A}\right]\\
&=\E_{P} \left[Z_{T}^{(i)} \mathds 1_{x^{-1}_\bullet A \cap\left\{T<\tau_{i}\right\}}\right]+\sum_{n=i+1}^{+\infty} \E_{P}\left[ Z_{T}^{(n)} \mathds 1_{x^{-1}_\bullet A \cap\left\{\tau_{n-1} \leq T<\tau_{n}\right\}}\right]\\
&=\E_{\bar{P}_{i}}\left[ \mathds 1_{x^{-1}_\bullet A \cap\left\{T<\tau_{i}\right\}}\right]+\sum_{n=i+1}^{+\infty} \E_{\bar{P}_{n}}\left[\mathds 1_{x^{-1}_\bullet A \cap\left\{\tau_{n-1} \leq T<\tau_{n}\right\}}\right]\\
&=\E_{\bar{\p}} \left[\mathds 1_{A \cap\left\{T<t_{i}\right\}}\right]+\sum_{n=i+1}^{+\infty} \E_{\bar{\p}} \left[\mathds 1_{A \cap\left\{t_{n-1} \leq T<t_{n}\right\}}\right]=\E_{\bar{\p}} \left[\mathds 1_{A}\right].
\end{align*}
From the arbitrariness of $i$ and that $\lim_{n \to \infty} \tau_n = +\infty$ $P$-a.s., it follows that
\begin{align*}
    \frac{d \bar{\p}_{[0, T]}}{d \p_{[0, T]}}=Z_{T}^{C}.
\end{align*}

Finally, we compute the relative entropy between the forward and backward path-space measures
\begin{align*}
\H\left[\p_{[0, T]}, \bar \p_{[0, T]}\right] &=\E_{\p}\left[\log \frac{d \p_{[0, T]}}{d \bar \p_{[0, T]}}(\gamma)\right] \\
&=\E_{P}\left[\log \frac{d \p_{[0, T]}}{d \bar \p_{[0, T]}}(x(\omega))\right] \\
&= \E_{P}\left[-\log Z_T^C(x(\omega))\right]=\E_{P}\left[-\log Z_{T}\right] = \overbrace{\E_{P}\left[M_T\right]}^{=0} + \E_P[Y_T]\\
&=\E_{P}\left[2\int_0^{T}  \left|\sigma^- b_{\mathrm{irr}}(x_s) \right|^2 ds\right] =2T \int_{\R^d}  \left|\sigma^- b_{\mathrm{irr}}(x) \right|^2\rho(x) dx\\
&= T\int_{\R^d} b_{\mathrm{irr}}^\top D^- b_{\mathrm{irr}} \rho(x) dx,
\end{align*}
where we used Tonelli's theorem and stationarity for the penultimate equality. By Theorem \ref{thm: ep is a rate over time}, we obtain the entropy production rate
\begin{align*}
    e_p = \frac 1 T \H\left[\p_{[0, T]}, \bar \p_{[0, T]}\right] = \int_{\R^d} b_{\mathrm{irr}}^\top D^- b_{\mathrm{irr}} \rho(x) \dx. 
\end{align*}
\end{proof}

We now prove Theorem \ref{thm: epr regular case general}:

\begin{proof}
By assumption \ref{item: epr regular general drift image of volatility}, for $\rho$-a.e. $x \in \R^d$
\begin{equation}
\label{eq: localised drift in image of the volatility general}
\begin{split}
    b_{\mathrm{irr}}(x) \in \im \sigma(x)
    &\Rightarrow b_{\mathrm{irr}}(x) = \sigma \sigma^- b_{\mathrm{irr}}(x).
\end{split}
\end{equation}

Then, \eqref{eq: localised drift in image of the volatility general} implies that we can rewrite the SDE \eqref{eq: Ito SDE} as
\begin{align*}
        dx_t = b_{\mathrm{rev}}(x_t) dt+ \sigma(x_t)\left[\sigma^{-}b_{\mathrm{irr}}\left(x_t\right)dt+dw_t\right], \quad x_0 \sim \rho.
\end{align*}

By assumptions \ref{item: epr regular general adapted}, \ref{item: epr regular general expectation}, we may define a new probability measure $\bar P$ on the sample space $\Omega$ through the relation
\begin{equation}
\left.\frac{d \bar P}{d P}\right|_{\mathscr{F}_{T}}=Z_{T},
\end{equation}
and it follows by Girsanov's theorem \cite[Theorem 10.14]{pratoStochasticEquationsInfinite2014} 
applied to the process $-2\sigma^{-} b_{\mathrm{irr}}\left(x_t\right)$, that $\left(x_{t}\right)_{t \in [0,T]}$ solves the SDE
\begin{align*}
        d x_t &= b_{\mathrm{rev}}(x_t) dt- \sigma(x_t)\left[\sigma^{-}b_{\mathrm{irr}}\left(x_t\right)dt+dw_t\right], \quad  x_0\sim \rho.
\end{align*}
on the probability space $(\Omega, \mathscr{F},\left\{\mathscr{F}_{t}\right\}_{t \geq 0}, \bar P)$.
Using \eqref{eq: localised drift in image of the volatility general}, $\left(x_{t}\right)_{t \in [0,T]}$ solves
\begin{align*}
        d x_t &= \bar b ( x_t) dt + \sigma ( x_t) dw_t, \quad  x_0\sim \rho
\end{align*}
on said probability space. By Theorem \ref{thm: time reversal of diffusions} we know that under $\bar P$, $\left(x_{t}\right)_{t \in [0,T]}$ induces the path space measure of the time-reversed process.

Let $\left(C([0,T], \R^d), \mathscr{B}\right)$ denote the path space, where $\mathscr{B}$ is the Borel sigma-algebra generated by the sup norm $\|\cdot\|_\infty$. Denote trajectories of the process by $x_\bullet:=(x_t)_{t \in [0,T]}: \Omega \to C([0,T], \R^d)$.
In summary, we showed that, under $P, \bar P$, the distribution of $x_\bullet$ on $(C([0,T], \R^d),\mathscr{B})$ are $\p_{\left[0, T\right]}$ and $ \bar{\p}_{\left[0, T\right]}$, respectively.

By definition of Itô calculus, $Z_{T}$ is measurable with respect to $\langle x_{s}: 0 \leq t \leq T\rangle$, so there exists a positive measurable function $Z_T^C$ on the path space, such that $P$-a.s. $Z^C_T (x_\bullet(\omega))=Z_T(\omega)$ for $\omega \in \Omega$, i.e., the following diagram commutes
\begin{equation*}
  \begin{tikzcd}
    &C([0,T], \R^d) \arrow[dr, "Z_T^C"] &  \\
    \Omega \arrow[ur, "x_\bullet"] \arrow[rr, "Z_T"']&&\R_{>0}.
    \end{tikzcd}
\end{equation*}
Fix $A \in \mathscr{B}$. Then $x_\bullet^{-1}A \in \mathscr F$ as $x_\bullet$ is measurable. Obviously,
\begin{align*}
    \E_{\p}\left[Z_{T}^{C}  \mathds 1_{A}\right]=\E_{P}\left[ Z_{T} \mathds 1_{x_\bullet^{-1} A}\right]=\E_{\bar{P}} \left[\mathds 1_{x_\bullet^{-1}A}\right]=\E_{\bar{\p}} \left[\mathds 1_{A}\right].
\end{align*}
It follows that
\begin{align*}
    \frac{d \bar{\p}_{[0, T]}}{d \p_{[0, T]}}=Z_{T}^{C}.
\end{align*}

Through this, we obtain the relative entropy between the forward and backward path-space measures
\begin{align*}
\H\left[\p_{[0, T]}, \bar \p_{[0, T]}\right] &=\E_{\p}\left[\log \frac{d \p_{[0, T]}}{d \bar \p_{[0, T]}}(\gamma)\right] \\
&=\E_{P}\left[\log \frac{d \p_{[0, T]}}{d \bar \p_{[0, T]}}(x(\omega))\right] \\
&= \E_{P}\left[-\log Z_T^C(x(\omega))\right]=\E_{P}\left[-\log Z_{T}\right] \\
&= \E_{P}\left[2 \int_0^T \langle \sigma^- b_{\mathrm{irr}}(x_t), dw_t \rangle\right] + \E_P\left[2 \int_0^T |\sigma^- b_{\mathrm{irr}}(x_t) |^2 dt\right]\\
&=2T \int_{\R^d}  \left|\sigma^- b_{\mathrm{irr}}(x) \right|^2\rho(x) dx\\
&= T\int_{\R^d} b_{\mathrm{irr}}^\top D^- b_{\mathrm{irr}} \rho(x) dx.
\end{align*}
The penultimate equality follows from the fact that Itô stochastic integrals are martingales (and hence vanish in expectation), Tonelli's theorem and stationarity. Finally, by Theorem \ref{thm: ep is a rate over time}, we obtain the entropy production rate
\begin{align*}
    e_p = \frac 1 T \H\left[\p_{[0, T]}, \bar \p_{[0, T]}\right] = \int_{\R^d} b_{\mathrm{irr}}^\top D^- b_{\mathrm{irr}} \rho(x) dx. 
\end{align*}
\end{proof}

We prove Proposition \ref{prop: suff conds exp cond}:

\begin{proof}
\begin{enumerate}
    \item[\textit{\ref{item: suff cond martingale}.}] Follows as $Z_0 =1$ and by definition of a martingale. 
\end{enumerate}
We define the $\mathscr F_t$-adapted process $\psi_t =-2 \sigma^- b_{\mathrm{irr}}(x_t)$.
\begin{enumerate}
\setcounter{enumi}{1}
    \item[\textit{\ref{item: suff cond novikov}} $\Rightarrow$ \textit{\ref{item: suff cond martingale}.}] This follows from \cite[Theorem 1]{rufNewProofConditions2013}.
    \item[\textit{\ref{item: suff cond delta exp condition}.}] By \cite[Proposition 10.17 (ii)]{pratoStochasticEquationsInfinite2014}, a sufficient condition for \eqref{eq: exponential condition} is the existence of a $\delta>0$ such that $\sup _{t \in[0, T]} \mathbb{E}_P\left(e^{\delta|\psi_t|^{2}}\right)<+\infty$. By stationarity of $x_t$ at $\rho$, $\mathbb{E}_P\left(e^{\delta|\psi_t|^{2}}\right)= \mathbb{E}_{\rho}\left(e^{4\delta|\sigma^- b_{\mathrm{irr}}(x)|^{2}}\right)$, and so the result follows.
    \item[\textit{\ref{item: suff cond Kazamaki}} $\Rightarrow$ \textit{\ref{item: suff cond martingale}}.] Follows from \cite[Theorem 1]{rufNewProofConditions2013}.
    \item[\textit{\ref{item: suff cond Dellacherie meyer}} $\Rightarrow$ \textit{\ref{item: suff cond novikov}}.] $A_t:=\frac 1 2\int_0^t\left|\psi_s\right|^2 ds= 2\int_0^t\left|\sigma^-b_{\mathrm{irr}}(x_s)\right|^2 ds$ is a non-decreasing, $\mathcal F_t$-adapted process. By assumption, $\E_P\left[A_{T}-A_{t} \mid \mathcal{F}_{t}\right] \leq K$ for all $t \in [0,T]$. By \cite[Theorem 105 (b)]{dellacherieProbabilitiesPotentialTheory1982} $\E_P\left[\exp(A_{T})\right]<+\infty$.
    \item[\textit{\ref{item: suff cond tail}} $\Rightarrow$ \textit{\ref{item: suff cond delta exp condition}}.]
    Let $\delta \in (0 ,c) $. The first equality follows a standard fact about non-negative random variables:
    \begin{align*}
        \E_\rho\left [e^{\delta |\sigma^- b_{\mathrm{irr}}(x)|^{2}}\right] &= \int_0^\infty   P\left(e^{ \delta |\sigma^- b_{\mathrm{irr}}(x)|^{2}}>r\right)dr\\
        &\leq 1 + \int_{e^{cR}}^\infty P\left(e^{ \delta |\sigma^- b_{\mathrm{irr}}(x)|^{2}}>r\right) dr\\
        &=1 + \int_{e^{cR}}^\infty P\left(|\sigma^- b_{\mathrm{irr}}(x)|^{2}> \delta^{-1}\log r\right)dr\\
        &\leq 1 + C \int_{e^{cR}}^\infty r^{- c \delta^{-1}} dr  \quad \text{(by \ref{item: suff cond tail} as $\delta^{-1}\log r> R$)} \\
        &<+\infty.
    \end{align*}
\end{enumerate}
\end{proof}

\subsubsection{Singularity}
\label{app: epr singular}

We prove Theorem \ref{thm: epr singular}:

\begin{proof}
Under the assumption that $b_{\mathrm{irr}}(x) \in \im \sigma(x)$ does not hold for $\rho$-a.e. $x \in \R^d$, we will show that there are paths taken by the forward process that are not taken by the backward process---and vice-versa---resulting in the mutual singularity of forward and backward path space measures.


Recall from Theorem \ref{thm: time reversal of diffusions} that any solution to the following SDE
\begin{align}
d\bar x_t= \bar b(\bar x_t) dt + \sigma(\bar x_t) dw_t, \quad \bar x_0 \sim \rho,
\end{align}
induces the path space measure of the time-reversed process.

We rewrite the forward and backward SDEs into their equivalent Stratonovich SDEs \cite[eq. 3.31]{pavliotisStochasticProcessesApplications2014}
\begin{align*}
    dx_t &= b^s(x_t) dt + \sigma(x_t) \circ dw_t, \quad  x_0 \sim \rho,\\
    d\bar x_t&= \bar b^s(\bar x_t) dt + \sigma(\bar x_t)\circ  dw_t, \quad \bar x_0 \sim \rho,
\end{align*}
By Remark \ref{rem: stratonovich Helmholtz}, time-reversal and the transformation from Itô to Stratonovich commute so $\bar b^s$ is unambiguous, and $b_{\mathrm{irr}} = b^s(x)-\bar b^s(x)$. The volatility and Stratonovich drifts are locally Lipschitz as $\sigma \in C^2$.

Consider an initial condition $x= x_0= \bar x_0$ to the trajectories, with $\rho(x)>0$. Consider trajectories in the Cameron-Martin space 
\begin{align*}
    \gamma \in \mathcal C:=\{\gamma \in A C\left([0, T] ; \mathbb{R}^m\right) \: \mid \: \gamma(0)=0 \text { and } \dot{\gamma} \in L^2\left([0, T] ; \mathbb{R}^m\right)\}
\end{align*}

Given such a trajectory, the approximating ODEs
\begin{align*}
    dx_t &= b^s(x_t) dt + \sigma(x_t) d\gamma_t, \quad  x_0 =x,\\
    d\bar x_t&= \bar b^s(\bar x_t) dt + \sigma(\bar x_t) d\gamma_t, \quad \bar x_0 =x,
\end{align*}
have a unique solution in $[0,T]$, with $T>0$ uniform in $\gamma$. 

We can thus apply the Stroock-Varadhan support theorem \cite[Theorem 3.10]{prokopTopologicalSupportSolutions2016} 
to state 
the possible paths under the forward and backward protocols. These are as follows
\begin{align*}
   \supp \p^{x}_{[0,T]}&=\overline{\left\{ x_t^\gamma = x + \int_0^t b^s(x_s^\gamma) ds + \int_0^t \sigma(x_s^\gamma) \gamma'(s) ds , t \in [0,T]\mid  \gamma \in \mathcal C \right\}}, \\
   \supp \bar \p^{x}_{[0,T]}&=\overline{\left\{  \bar x_t^\gamma = x + \int_0^t \bar b^s(\bar x_s^\gamma) ds + \int_0^t \sigma(\bar x_s^\gamma) \gamma'(s) ds, t \in [0,T] \mid \gamma \in \mathcal C \right\}}.
\end{align*}
where the closure is with respect to the supremum norm on $C\left([0, T] ; \mathbb{R}^d\right)$. The time derivatives of these paths at $t=0$ are
\begin{align*}
   \partial_{t}\supp \p^{x}_{[0,T]}\mid_{t=0}&=\overline{\{ \partial_t x_t^\gamma |_{t=0} \in \R^d \mid \gamma \in \mathcal C \}} = \{b^s(x) + \sigma(x) v \mid  v\in \R^d\}, \\
  \partial_{t}\supp \bar \p^{x}_{[0,T]}\mid_{t=0}&=\overline{\{ \partial_t \bar x_t^\gamma |_{t=0} \in \R^d \mid \gamma \in \mathcal C \}} = \{\bar b^s(x) + \sigma(x) v \mid  v\in \R^d\}.
\end{align*}
where the closure is with respect to the sup norm on $\R^d$.

Consider an initial condition $x$ with $b_{\mathrm{irr}}(x)\not \in \im \sigma(x)$. This implies that the forward and backward path space measures are mutually singular because the time derivatives of the possible paths differ at the origin 
\begin{align*}
    & 2b_{\mathrm{irr}}(x) = b^s(x)-\bar b^s(x) \not \in \im \sigma(x) \\
     \iff& b^s(x) + \im \sigma(x) \neq \bar b^s(x) + \im \sigma(x)\\
    \iff &\partial_{t}\supp \p^{x}_{[0,T]}\mid_{t=0}\not \subset \partial_{t}\supp \bar \p^{x}_{[0,T]}\mid_{t=0} \text{and vice-versa} \\
    \Rightarrow& \supp \p^{x}_{[0,T]} \not \subset \supp \bar \p^{x}_{[0,T]} \text{ and vice-versa}\\
    \Rightarrow& \p^{x}_{[0,T]} \perp \bar \p^{x}_{[0,T]}
\end{align*}

Finally, from Proposition \ref{prop: aggregating local ep}
\begin{align*}
     e_p 
     &= \E_{x \sim \rho}\left[\H\left[\p^{x}_{[0,T]}\mid \bar \p^{x}_{[0,T]}\right]\right]\\
     &\geq \rho\left(\{x \in \R^d \mid  \p^{x}_{[0,T]} \perp \bar \p^{x}_{[0,T]} \}\right)\cdot \infty\\
     &\geq \underbrace{\rho\left(\{x \in \R^d  : b_{\mathrm{irr}}(x) \not \in \im \sigma(x) \}\right)}_{>0} \cdot \infty\\
     &= +\infty.
\end{align*}
\end{proof}

\subsection{Entropy production rate of the linear diffusion process}
\label{app: exact simul OU process}

We require the following Lemma, which can be proved by adjusting the derivation of the relative entropy between non-degenerate Gaussian distributions, cf. \cite[Section 9]{duchiDerivationsLinearAlgebra}.
\begin{lemma}
\label{lemma: KL Gaussian}
On $\R^d$
    \begin{align*}
       2 \H[\mathcal N(\mu_0, \Sigma_0) \mid\mathcal N(\mu_1, \Sigma_1)] &= \operatorname{tr}\left(\Sigma_{1}^{-} \Sigma_{0}\right)-\rank \Sigma_0
       +\log \left(\frac{\operatorname{det}^* \Sigma_{1}}{\operatorname{det}^* \Sigma_{0}}\right) \\
       &+\left(\mu_{1}-\mu_{0}\right)^{\top} \Sigma_{1}^{-}\left(\mu_{1}-\mu_{0}\right).
    \end{align*}
where $\cdot^-$ is the Moore-Penrose pseudo-inverse and $\operatorname{det}^*$ is the pseudo-determinant. 
\end{lemma}

\begin{proof}[Proof of Lemma \ref{lemma: limit ep formula OU process}]
We insert the definitions of the transition kernels \eqref{eq: OU transition kernels} into Lemma \ref{lemma: KL Gaussian}.
\begin{equation*}
\begin{split}
    &\E_{x\sim \rho}[2\H[p_\e(\cdot,x)\mid \bar p_\e(\cdot,x)]] \\
    &=\tr(\bar S_\e^{-}S_\e)-\rank \sigma + \log \frac{\det^*(\bar S_\e)}{\det^*(S_\e)} \\
&+ \E_{x \sim\rho} \left[ x^\top(e^{-\e C}-e^{-\e B})^\top \bar S_\e^{-}(e^{-\e C}-e^{-\e B})x \right] \\
&=\tr(\bar S_\e^{-}S_\e)-\rank \sigma + \log \frac{\det^*(\bar S_\e)}{\det^*(S_\e)}\\
    &+\tr( \Pi^{-1} (e^{-\e C}-e^{-\e B})^\top \bar S_\e^{-}(e^{-\e C}-e^{-\e B}))
\end{split}
\end{equation*}
To obtain the last line, we used the trace trick for computing Gaussian integrals of bilinear forms. The proof follows by Proposition \ref{prop: epr transition kernels}.
\end{proof}

\small
\bibliographystyle{unsrt}
\bibliography{bib2}

\end{document}